\documentclass[11pt,a4paper]{article}
\pdfoutput=1
\usepackage{jcappub}
\usepackage{aas_macros_JCAP}

\newcommand{\hMpc}{h^{-1}\;{\rm Mpc}}

\newcommand{\hGpc}{h^{-1}{\rm Gpc}}
\newcommand{\hMsun}{h^{-1}M_{\odot}}
\newcommand{\void}{\mathrm{v}}

\newcommand{\matter}{\mathrm{m}}
\newcommand{\tracer}{\mathrm{t}}

\newcommand{\rs}{r_s}
\newcommand{\dc}{\delta_c}

\title{Probing cosmology and gravity with redshift-space distortions around voids}

\author[a,b,c]{Nico Hamaus,}
\author[a,b,d,e,f]{P. M. Sutter,}
\author[a,b]{Guilhem Lavaux}
\author[a,b,g]{\\ and Benjamin D. Wandelt}

\affiliation[a]{Sorbonne Universit\'es, UPMC Univ Paris 06, UMR 7095, Institut d'Astrophysique de Paris,\\ 98 bis boulevard Arago, F-75014, Paris, France}
\affiliation[b]{CNRS, UMR 7095, Institut d'Astrophysique de Paris,\\ 98 bis boulevard Arago, F-75014, Paris, France}
\affiliation[c]{Universit\"ats-Sternwarte M\"unchen, Fakult\"at f\"ur Physik, Ludwig-Maximilians Universit\"at,\\ Scheinerstr. 1, D-81679 M\"unchen, Germany}
\affiliation[d]{Center for Cosmology \& AstroParticle Physics, Ohio State University,\\ 191 West Woodruff Avenue, Columbus, OH 43210, U.S.A.}
\affiliation[e]{INFN - National Institute for Nuclear Physics,\\ via Valerio 2, I-34127 Trieste, Italy}
\affiliation[f]{INAF - Osservatorio Astronomico di Trieste,\\ via Tiepolo 11, I-34143 Trieste, Italy}
\affiliation[g]{Departments of Physics and Astronomy, University of Illinois at Urbana-Champaign,\\ 1110 West Green Street, Urbana, IL 61801, U.S.A.}

\emailAdd{hamaus@iap.fr}
\emailAdd{sutter@iap.fr}
\emailAdd{lavaux@iap.fr}
\emailAdd{wandelt@iap.fr}

\abstract{Cosmic voids in the large-scale structure of the Universe affect the peculiar motions of objects in their vicinity. Although these motions are difficult to observe directly, the clustering pattern of their surrounding tracers in redshift space is influenced in a unique way. This allows to investigate the interplay between densities and velocities around voids, which is solely dictated by the laws of gravity. With the help of $N$-body simulations and derived mock-galaxy catalogs we calculate the average density fluctuations around voids identified with a watershed algorithm in redshift space and compare the results with the expectation from general relativity and the $\Lambda$CDM model. We find linear theory to work remarkably well in describing the dynamics of voids. Adopting a Bayesian inference framework, we explore the full posterior of our model parameters and forecast the achievable accuracy on measurements of the growth rate of structure and the geometric distortion through the Alcock-Paczy\'nski effect. Systematic errors in the latter are reduced from $\sim15\%$ to $\sim5\%$ when peculiar velocities are taken into account. The relative parameter uncertainties in galaxy surveys with number densities comparable to the SDSS MAIN (CMASS) sample probing a volume of $1h^{-3}{\rm Gpc}^3$ yield $\sigma_{f/b}\left/(f/b)\right.\sim2\%$ ($20\%$) and $\sigma_{D_AH}/D_AH\sim0.2\%$ ($2\%$), respectively. At this level of precision the linear-theory model becomes systematics dominated, with parameter biases that fall beyond these values. Nevertheless, the presented method is highly model independent; its viability lies in the underlying assumption of statistical isotropy of the Universe.
}


\arxivnumber{1507.04363}

\begin{document}
\maketitle


\section{Introduction\label{sec:intro}}
With the advent of modern galaxy redshift surveys that map out significant fractions of the observable Universe in ever more detail (e.g.,~\cite{DES,EUCLID,GAMA,SDSS,VIPERS,WFIRST,WiggleZ}), its least luminous and yet most extended constituents have drawn an increasing amount of attention recently: \emph{cosmic voids}. In fact, cosmic voids are not only interesting objects in their own right. As they take up most of space in the Universe they play a major role in the formation of the web-like pattern of its large-scale structure. Despite its complexity, this large-scale structure contains a wealth of information on the fundamental properties of the Universe, such as its initial conditions, its matter and energy content, or the geometry of spacetime on cosmological scales, for example. Yet not only cosmology, but the very nature of gravity itself can be investigated with large-scale structures, as it is gravitation that gives rise to structure formation in the first place. This happens through gravitational collapse of initially over-dense regions in the mass distribution into filaments, sheets and clusters. Thereby, the constituents of this mass distribution acquire increasing velocities over time, yielding a complex dynamical behavior in 6-dimensional phase space. However, the dynamics of voids remains rather simple, as they are mainly composed of single-stream flows~\cite{Neyrinck2013,Leclercq2013,Falck2015}.

Unfortunately, a dominant fraction of the mass distribution in the Universe is made up of invisible \emph{dark matter} and we can only observe the tracers of this mass distribution, such as galaxies or Hydrogen gas clouds,\footnote{See ref.~\cite{Stark2015} for an interesting proposal to identify voids in the Lyman-alpha forest.} for example. The statistical spatial distribution of these tracers is typically different from the one of the dark matter, which goes under the name of \emph{bias}. Therefore, in order to reconstruct the mass distribution of large-scale structure one has to understand the relation between tracers and the dark matter, for instance with the help of theoretical models or numerical simulations. Furthermore, this reconstruction is complicated by peculiar motions. The peculiar velocity of a tracer is difficult to determine observationally, and is typically not known for distant objects. However, it affects the redshift of each tracer via the Doppler effect in addition to the redshift caused by the Hubble flow. Since the total redshift is used to determine each tracer's line-of-sight distance and the two sources of redshift cannot be distinguished, the apparent distribution of tracers in redshift space (as opposed to real space) is skewed, an effect referred to as \emph{redshift-space distortion}. On large scales this results in elongated structures perpendicular to the line of sight due to the coherent streaming motions of tracers towards overdensities in the matter distribution, the so-called \emph{Kaiser} effect~\cite{Kaiser1987}. In contrast, on small scales structures preferentially stretch along the line of sight, which is commonly referred to as the \emph{Finger-of-God} (FoG) effect. It is caused by nonlinear virial motions of tracers within gravitationally collapsed objects, such as dark matter halos.

Similarly, incorrect assumptions about the expansion history and the geometry of the Universe can result in a skewed distribution of tracers. The conversion between redshifts and angles on the sky into Cartesian coordinates requires knowledge of the spacetime metric that describes the Universe, which in turn depends on the latter's matter and energy content. Any deviation from the true underlying cosmological parameters results in a distortion of tracer coordinates, which is known as the \emph{Alcock-Paczy\'nski} (AP) effect~\cite{Alcock1979}. In cases where the underlying statistical arrangement of tracers is known, it can be used to constrain cosmology.

Luckily, a tracer's angular position on the sky is not affected by redshift-space distortions. Therefore, by statistically comparing its line-of-sight and angular clustering patterns, one can disentangle the impact of peculiar motions and geometric distortions on the coordinate distribution of a tracer. This method is particularly powerful for cosmic voids, as on average they exhibit spherical symmetry in real space, which allows to assign any type of anisotropy to a skewed coordinate system in redshift space~\cite{Lavaux2012,Sutter2012b,Pisani2014,Sutter2014b,Hamaus2014c}. Vice versa, these anisotropies can be used to infer the peculiar velocities of tracers around voids~\cite{Padilla2005,Ceccarelli2013,Paz2013,Ruiz2015}, opening up the possibility to determine the strength of gravity on void scales. The main underlying assumption to this method is the cosmological principle, i.e. statistical homogeneity and isotropy of the Universe. Observational evidence for the anisotropic nature of the void-galaxy cross-correlation function has been provided by Paz et al.~\cite{Paz2013} using SDSS data, and Micheletti et al.~\cite{Micheletti2014} in the VIPERS survey.

While the investigation of redshift-space distortions between pairs of galaxies has received enormous attention in the literature, both theoretical and observational (see e.g.,~\cite{Scoccimarro2004,Taruya2010,Reid2011,Seljak2011,Kwan2012,Yoo2012,Hamaus2012,Marulli2012,Vlah2013,Blazek2014,Bianchi2015,Ross2015,White2015,Uhlemann2015,Jeong2015,Li2015,Okumura2015} and~\cite{Guzzo2008,Okumura2008,Cabre2009,Blake2011,Reid2012,Kazin2013,DeLaTorre2013,Samushia2014,Sanchez2014,Anderson2014,Song2014,Beutler2014,Alam2015,Granett2015,Marin2015}, respectively), comparable studies concerning voids remain rare~\cite{Padilla2005,Paz2013,Hamaus2014c,Micheletti2014}. The aim of this paper is to assess the feasibility of measuring redshift-space distortions around voids using galaxies from a realistic mock galaxy catalog, and to compare this data with a physically motivated model. The values for the parameters that go into this model will be investigated in detail, including their uncertainty, in order to forecast the potential to test the predictions of General Relativity and the current standard model of cosmology with the help of modern surveys.

\section{Theoretical prerequisites \label{sec:theory}}

\subsection{Tracer coordinates in redshift space \label{sec:theory:coordinates}}
Let us denote the coordinates of a tracer in real space by the vector $\mathbf{x}$. We can only observe its angular position on the sky and its redshift $z$. If this redshift were only caused by Hubble expansion, we could compute the comoving line-of-sight distance to the tracer as
\begin{equation}
x_\parallel=\int_0^z \frac{c}{H(z')}\;\mathrm{d}z'\;, \label{x_par}
\end{equation}
where $c$ is the speed of light and $H(z)$ describes the Hubble expansion as a function of redshift. The tracer's peculiar velocity component along the line of sight $v_\parallel$ causes a small additional Doppler shift of magnitude $v_\parallel/c$, which is indistinguishable from the cosmological redshift. Thus, the apparent comoving line-of-sight coordinate of a tracer in redshift space becomes
\begin{equation}
 \tilde{x}_\parallel = \int_0^{z+\frac{v_\parallel}{c}(1+z)} \frac{c}{H(z')}\;\mathrm{d}z' \simeq x_\parallel + \frac{v_\parallel}{H(z)}(1+z)\;. \label{x_s}
\end{equation}
Furthermore, in cosmological redshift surveys the angles $\vartheta$ between tracers observed on the sky are converted to comoving distances $x_\perp$ using the angular diameter distance $D_A(z)$,
\begin{equation}
 x_\perp = D_A(z)\,\vartheta\;. \label{x_per}
\end{equation}
In a Friedmann-Lema\^itre-Robertson-Walker metric of curvature $k$ the comoving angular diameter distance is given by
\begin{equation}
 D_A(z) = \frac{c}{H_0\sqrt{-\Omega_k}}\sin\left(H_0\sqrt{-\Omega_k}\int_0^z \frac{1}{H(z')}\;\mathrm{d}z'\right)\;.
\end{equation}
It is sensitive to the curvature parameter $\Omega_k=1-\Omega_\mathrm{m}-\Omega_\Lambda$, while the Hubble rate itself also depends on the matter and energy content in the Universe,
\begin{equation}
 H(z) = H_0\sqrt{\Omega_\mathrm{m}(1+z)^3+\Omega_k(1+z)^2+\Omega_\Lambda}\;.
\end{equation}
It is hence necessary to assume fiducial values for today's Hubble constant $H_0$, matter fraction $\Omega_\mathrm{m}$, curvature parameter $\Omega_k$, and cosmological constant $\Omega_\Lambda$ in order to construct a three-dimensional map of the large-scale distribution of tracers\footnote{Here we consider a cosmological constant $\Lambda$, but this can straightforwardly be extended to an arbitrary equation of state for dark energy.}. If these fiducial parameters do not coincide with the true cosmological values, or the assumed $\Lambda$CDM-model is incomplete, the distances in eqs.~(\ref{x_par}) and (\ref{x_per}) will be incorrect. This effect is commonly referred to as the AP test~\cite{Alcock1979}, as it allows to determine the true cosmological parameter values in cases where the geometry of the observed structures is known, like it is the case for cosmic voids~\cite{Lavaux2012,Sutter2012b,Pisani2014,Sutter2014b,Hamaus2014c}. However, only the combination
\begin{equation}
\varepsilon = \frac{\Delta x_\parallel}{\Delta x_\perp} \propto \frac{1}{D_A(z)H(z)}\;, \label{AP}
\end{equation}
can be constrained with this test, as this is the quantity that determines the degree of anisotropy in the spatial distribution of the tracers. If we define $D_A(z)$ and $H(z)$ to adopt the fiducial (and possibly incorrect) values for our cosmology, $\varepsilon$ quantifies the deviation from isotropy resulting from analyzing the data with an incorrect fiducial cosmology.

\subsection{Void-tracer cross-correlation function \label{sec:theory:correlation}}
Now let us consider the average real-space radial number-density profile of tracers in distance $r$ from a void center at the origin, $\rho_{\void\tracer}(r)$, which can also be referred to as a one-dimensionally \emph{stacked void}. Labeling each of all $N_\void$ void centers with index $i$ and coordinates $\mathbf{X}_i$, and similarly each of all $N_\tracer$ tracers with index $j$ and coordinates $\mathbf{x}_j$, we have
\begin{equation}
 \frac{\rho_{\void\tracer}(r)}{\bar{\rho}_\tracer} = \frac{1}{N_\void} \sum_i \frac{\rho_{\void\tracer}^{(i)}(r)}{\bar{\rho}_\tracer} = \frac{1}{N_\void} \sum_i \frac{V}{N_\tracer} \sum_j \delta^\mathrm{D}(\mathbf{X}_i-\mathbf{x}_j+\mathbf{r})\;,
\end{equation}
where the first equality describes the ensemble average over all individual void density profiles and the second one represents a histogram of Dirac delta functions $\delta^\mathrm{D}$ for the tracer positions $\mathbf{x}_j$ at separation $\mathbf{r}$ from each void center $\mathbf{X}_i$. This expression can then be written as a convolution of the number density of void centers $\rho_\void$ with the number density of tracers $\rho_\tracer$,
\begin{equation}
 V \sum_{i,j} \int \frac{1}{N_\void}\delta^\mathrm{D}(\mathbf{X}_i-\mathbf{x}) \frac{1}{N_\tracer}\delta^\mathrm{D}(\mathbf{x}-\mathbf{x}_j+\mathbf{r}) \,\mathrm{d}^3x = \frac{1}{V} \int \frac{\rho_\void(\mathbf{x})}{\bar{\rho}_\void}\frac{\rho_\tracer(\mathbf{x}+\mathbf{r})}{\bar{\rho}_\tracer} \,\mathrm{d}^3x = 1+\xi_{\void\tracer}(r) \;, \label{rho-xi}
\end{equation}
where $V$ is the total observed volume, $\bar{\rho}_\void=N_\void/V$, $\bar{\rho}_\tracer=N_\tracer/V$, and $\xi_{\void\tracer}(r)$ denotes the void-tracer cross-correlation function in real space (see also ref.~\cite{Peebles1980}).

Now, in order to obtain the corresponding redshift-space correlation function, one needs to consider the \emph{pairwise velocity probability distribution function} $\mathcal{P}(\mathbf{v},\mathbf{r})$. This function maps all pairs of separation $\mathbf{r}$ to separation $\tilde{\mathbf{r}}=\left(\tilde{r}_\parallel,\tilde{r}_\perp\right)^\intercal$ due to their relative velocity $\mathbf{v}$, such that the correlation function in redshift space is calculated as~\cite{Peebles1980}
\begin{equation}
 1+\tilde{\xi}(\tilde{\mathbf{r}}) = \int\left[1+\xi(r)\right]\mathcal{P}(\mathbf{v},\mathbf{r})\,\mathrm{d}^3v\;. \label{xi_s}
\end{equation}
According to eq.~(\ref{x_s}) only the magnitude of the relative velocity component along the line of sight $\mathbf{v}_\parallel$ affects the coordinates $\tilde{\mathbf{r}}$, which reduces eq.~(\ref{xi_s}) to a one-dimensional integral via the replacements $\mathcal{P}(\mathbf{v},\mathbf{r})\rightarrow\mathcal{P}(v_\parallel,\mathbf{r})$ and $\mathrm{d}^3v\rightarrow\mathrm{d}v_\parallel$.
What remains to be specified is the explicit form of $\mathcal{P}(v_\parallel,\mathbf{r})$, itself a major topic of interest in the literature~(e.g., \cite{Peebles1976,Davis1983,Strauss1998,Scoccimarro2004}). Numerical studies of the pairwise velocity probability distribution function of dark matter particles in $N$-body simulations suggest an exponential form at small separations $\mathbf{r}$, with a non-negligible skewness. Towards large separations, the skewness gets weaker and $\mathcal{P}(v_\parallel,\mathbf{r})$ approaches a more Gaussian shape; its exponential wings however, remain~\cite{Scoccimarro2004}.

\subsection{Gaussian streaming model \label{sec:theory:GSM}}
Nevertheless, for the purpose of modeling observed tracers of the dark matter, such as galaxies, a Gaussian form for the pairwise velocity probability distribution function in many cases remains a reasonable approximation. Galaxies are found in dark matter halos, which move according to the net velocities of their constituent particles. Therefore, the small-scale nonlinear dynamics of individual dark matter particles do not affect the bulk motions of halos and hence their hosted galaxies. This leads to a more Gaussian behavior of their velocity statistics, which has been confirmed with the help of simulations~\cite{Tinker2007,Reid2011,Wang2014,Bianchi2015,White2015,Uhlemann2015}. Moreover, as the average tracer-density fluctuations inside and around voids are rather moderate~\cite{Hamaus2014b}, extreme nonlinearities (e.g. from virial motions of galaxies inside clusters) are less severe in this case. The fact that we model a cross-correlation between void centers and their surrounding tracers also mitigates nonlinear effects, since the main velocity contribution comes from the tracer, while the void center is practically at rest~\cite{Sutter2014c}. In contrast, in tracer auto-correlations the velocities at two locations are being correlated with each other, which effectively squares their individual nonlinearities. We will therefore assume a Gaussian pairwise velocity probability distribution function of the form
\begin{equation}
 \mathcal{P}(v_\parallel,\mathbf{r}) = \frac{1}{\sqrt{2\pi}\sigma_v(\mathbf{r})} \exp\left[-\frac{\left(v_\parallel-v_\void(r)\frac{r_\parallel}{r}\right)^2}{2\sigma_v^2(\mathbf{r})}\right]\;, \label{pdf}
\end{equation}
with a mean of $v_\void(r)\frac{r_\parallel}{r}$ and dispersion $\sigma_v(\mathbf{r})$. This model is referred to as the \emph{Gaussian streaming model}, and was first introduced by Fisher~\cite{Fisher1995} as a description for the galaxy auto-correlation function in redshift space. Here, the comoving real-space coordinates $r_\parallel$ and $r$ can be obtained from their measured redshift-space counterparts via
\begin{equation}
 r_\parallel = \tilde{r}_\parallel - \frac{v_\parallel}{H(z)}(1+z)\;, \qquad r_\perp = \varepsilon \tilde{r}_\perp \;, \qquad r = \sqrt{r_\parallel^2 + r_\perp^2}\;.
\end{equation}
Further, $v_\void(r)$ is the mean streaming velocity of tracers in distance $r$ from the void center. Using the conservation of mass and following linear theory it is given by~\cite{Peebles1980}
\begin{equation}
 v_\void(r) \simeq -\frac{1}{3} \frac{f(z) H(z)}{1+z} r \Delta_{\void\matter}(r) \;, \label{v_lin}
\end{equation}
where $f(z)\simeq\Omega_\mathrm{m}^\gamma(z)$ is the logarithmic growth rate of density perturbations with growth index $\gamma\simeq0.55$ in the standard $\Lambda$CDM model~\cite{Linder2005}, and
\begin{equation}
 \Delta_{\void\matter}(r) = \frac{3}{r^3}\int_0^r \left(\frac{\rho_{\void\matter}(q)}{\bar{\rho}_\matter} - 1\right)q^2 \,\mathrm{d}q\;, \label{Delta_vm}
\end{equation}
is the cumulative average mass-density fluctuation of matter around voids, compared to the background density $\bar{\rho}_\matter$. With the help of $N$-body simulations, ref.~\cite{Hamaus2014b} have empirically found a universal fitting function for the nonlinear mass-density profile of voids,
\begin{equation}
 \frac{\rho_{\void\matter}(r)}{\bar{\rho}_\matter} - 1 = \dc\,\frac{1-(r/\rs)^\alpha}{1+(r/r_\void)^\beta}\;, \label{HSW}
\end{equation}
containing four free parameters: a scale radius $\rs$, a central under-density $\dc$, and two slopes $\alpha$ and $\beta$. Here, $r_\void$ denotes the \emph{effective void radius}, which is determined from the data itself as described in sec.~\ref{sec:analysis:voids}. Given the density profile of eq.~(\ref{HSW}) one can integrate eq.~(\ref{Delta_vm}) analytically, which yields
\begin{multline}
 \Delta_{\void\matter}(r) = \dc\;{}_2F_1\!\!\left[1,\frac{3}{\beta},\frac{3}{\beta}+1,-\left(\frac{r}{r_\void}\right)^\beta\right]\\
 -\dc\left(\frac{r}{\rs}\right)^\alpha\frac{3}{\alpha+3}\;{}_2F_1\!\!\left[1,\frac{\alpha+3}{\beta},\frac{\alpha+3}{\beta}+1,-\left(\frac{r}{r_\void}\right)^\beta\right]\;, \label{intprofile}
\end{multline}
where ${}_2F_1$ is the Gauss hypergeometric function. For biased tracers of the mass and following linear theory, one can relate eq.~(\ref{Delta_vm}) to the fluctuation of tracers $\Delta_{\void\tracer}(r)$ with the linear tracer-bias parameter $b$,
\begin{equation}
 \Delta_{\void\tracer}(r) \simeq b \Delta_{\void\matter}(r) \;. \label{bias}
\end{equation}
Finally, note that the velocity dispersion of eq.~(\ref{pdf}) depends both on magnitude and direction of the void-tracer separation $\mathbf{r}$. It can be decomposed into its components along and perpendicular to the line of sight,
\begin{equation}
 \sigma_v^2(\mathbf{r}) = \sigma_\parallel^2(r)\left(r_\parallel/r\right)^2 + \sigma_\perp^2(r)\left[1-\left(r_\parallel/r\right)^2\right] \;. \label{sv}
\end{equation}
Following ref.~\cite{Peebles1993} (page 515), linear theory allows to estimate $\sigma_\parallel$ and $\sigma_\perp$ via
\begin{eqnarray}
 \sigma_\parallel^2(r) &\simeq& H_0^2f^2\left[\frac{J_5(r)}{3r^3}+\frac{K_2(r)}{3}\right]\;, \label{sv_par} \\
 \sigma_\perp^2(r) &\simeq& H_0^2f^2\left[\frac{J_3(r)}{2r}-\frac{J_5(r)}{6r^3}+\frac{K_2(r)}{3}\right]\;, \label{sv_per}
\end{eqnarray}
where $J_n$ and $K_n$ are defined through integrals over the correlation function,
\begin{equation}
 J_n(r) = \int_0^r \xi(q)r^{n-1}\,\mathrm{d}q\;, \qquad K_n(r) = \int_r^\infty \xi(q)r^{n-1}\,\mathrm{d}q\;. \label{J_K}
\end{equation}
In our case we can identify $\xi(r)$ as the void-matter cross-correlation function $\xi_{\void\matter}(r)$ and thus, according to eq.~(\ref{rho-xi}), with the mass-density profile around voids, eq.~(\ref{HSW}). Plugged into eq.~(\ref{J_K}), this can be expressed analytically as
\begin{multline}
 J_n(r) = \dc r^{n} \left\{\frac{1}{n}\;{}_2F_1\!\! \left[1,\frac{n}{\beta},\frac{n}{\beta}+1,-\left(\frac{r}{r_\void}\right)^\beta\right]\right. \\
 \left. - \left(\frac{r}{\rs}\right)^\alpha\frac{1}{\alpha+n}\;{}_2F_1\!\! \left[1,\frac{\alpha+n}{\beta},\frac{\alpha+n}{\beta}+1,-\left(\frac{r}{r_\void}\right)^\beta\right]\right\}\;, \label{J_n}
\end{multline}
\begin{multline}
 K_n(r) = \dc r^{n}\left(\frac{r_\void}{r}\right)^\beta \left\{\frac{1}{\beta-n}\;{}_2F_1\!\! \left[1,\frac{\beta-n}{\beta},\frac{\beta-n}{\beta}+1,-\left(\frac{r_\void}{r}\right)^\beta\right]\right. \\
 \left. - \left(\frac{r}{\rs}\right)^\alpha\frac{1}{\beta-\alpha-n}\;{}_2F_1\!\! \left[1,\frac{\beta-\alpha-n}{\beta},\frac{\beta-\alpha-n}{\beta}+1,-\left(\frac{r_\void}{r}\right)^\beta\right]\right\}\;. \label{K_n}
\end{multline}
However, although linear theory is able to describe the dynamics of voids remarkably well, it is important to point out that it does not apply in all cases. Especially the deeply under-dense interior of small voids is subject to nonlinear evolution, which was shown to spoil the validity of eq.~(\ref{v_lin}) at distances $r\lesssim10\hMpc$ from the void center~\cite{Hamaus2014b}. Moreover, the empirical fitting function of eq.~(\ref{HSW}) for the void-density profile only describes the immediate void environment, but not the large-scale clustering regime out to arbitrarily large scales~\cite{Hamaus2014a,ChuenChan2014}, such as the baryonic acoustic oscillation (BAO) scale, for instance. Therefore, quantities that depend on integrals over the correlation function out to infinity, such as $K_n(r)$ in eq.~(\ref{J_K}), cannot be determined accurately in this case.

\section{Numerical analysis \label{sec:analysis}}

\subsection{Simulation setup \label{sec:analysis:setup}}
We analyze a large simulation that evolved $N=2048^3$ cold dark matter particles in a box of $1\hGpc$ side length with the adaptive $N$-body tree-code 2HOT~\cite{Warren2013} (this simulation is identical to the one used in refs.~\cite{Hamaus2014a,Hamaus2014b}). It adopts the cosmological parameters of the Planck 2013 measurement~\cite{Planck2013} with $\Omega_\mathrm{m}=0.318$, $\Omega_\mathrm{b}=0.049$, $\Omega_\Lambda=0.682$, $h=0.67$, $\sigma_8=0.834$, $n_s=0.962$. The power spectrum of initial density perturbations is calculated with the Boltzmann code CLASS~\cite{Blas2011}, a realization of this power spectrum is then generated with a modified version of 2LPTic~\cite{Crocce2006} using second-order Lagrangian perturbation theory. Catalogs of dark matter halos and sub-halos are created using the ROCKSTAR halo finder~\cite{Behroozi2013}. This algorithm is based on an adaptive hierarchical refinement of friends-of-friends groups in phase space to define virialized objects. In our simulation we find about $2\times10^7$ halos with masses ranging from $\sim2\times10^{11}\hMsun$ to $\sim5\times10^{15}\hMsun$ at redshift $z=0$.

\subsection{Mock-galaxy catalogs \label{sec:analysis:mocks}}
In order to generate realistic mock galaxy samples we refer to a standard halo occupation distribution (HOD) model using the code developed by ref.~\cite{Tinker2006}. It assigns central and satellite galaxies to each dark matter halo of mass $M$ according to a parametrized distribution. We use the parametrization of refs.~\cite{Zheng2007,Zehavi2011}, where the mean numbers of centrals and satellites per host halo are given by
\begin{eqnarray}
\langle N_\mathrm{cen}(M)\rangle &=& \frac{1}{2}\left[1 + \mathrm{erf}\left(\frac{\log M - \log M_\mathrm{min}}{\sigma_{\log M}}\right)\right]\;,\\
\langle N_\mathrm{sat}(M)\rangle &=& \langle N_\mathrm{cen}(M)\rangle\left(\frac{M-M_0}{M_1}\right)^\alpha \;.
\end{eqnarray}
The probability distribution of central galaxies is a nearest-integer distribution (i.e., all halos above a given mass threshold host a central galaxy), and satellite galaxies follow Poisson statistics. Central galaxies are assigned peculiar velocities of their host halo, and satellites are given an additional random velocity drawn from a Maxwell-Boltzmann distribution with a standard deviation that matches the velocity dispersion of their host halo's dark matter particles. Moreover, we apply eq.~(\ref{x_s}) to shift all galaxies from real to redshift space. Without loss of generality, we pick one axis of the simulation box as the line-of-sight direction. We adopt the the so-called \emph{plane-parallel approximation}, where changes in the line-of-sight direction with the observed angles on the sky are neglected. On the scales of voids and their immediate surroundings, as considered in this paper, this approximation is accurate.

We construct two different catalogs to investigate the impact of survey properties on our final results. For the first mock catalog we model a \emph{dense} galaxy survey with the parameters $M_\mathrm{min}\simeq2.0\times10^{11}\hMsun$, $M_0\simeq6.9\times10^{11}\hMsun$, $M_1\simeq3.8\times10^{13}\hMsun$, $\sigma_{\log M}\simeq0.21$, and $\alpha\simeq1.12$ adapted to the SDSS DR7 MAIN sample~\cite{Zheng2007} at redshift $z=0$. This results in a distribution of about $2\times10^7$ galaxies of mean number density $\bar{n}\simeq2.0\times10^{-2}h^3\;{\rm Mpc}^{-3}$, a mean separation of roughly $3.7\hMpc$, and a linear bias parameter of $b\simeq0.8$. The second mock catalog is adapted to the \emph{sparser} CMASS sample of the SDSS DR9~\cite{Manera2013} with parameters $M_\mathrm{min}\simeq1.2\times10^{13}\hMsun$, $M_0\simeq1.2\times10^{13}\hMsun$, $M_1\simeq1.0\times10^{14}\hMsun$, $\sigma_{\log M}\simeq0.60$, and $\alpha\simeq1.01$ at redshift $z=0.5$. It yields a distribution of roughly $3\times10^5$ galaxies of mean number density $\bar{n}\simeq3.0\times10^{-4}h^3\;{\rm Mpc}^{-3}$ that are separated by $15.0\hMpc$ on average, and have a linear bias parameter of $b\simeq1.8$.

\subsection{Void catalogs \label{sec:analysis:voids}}
These mock galaxies are then utilized to generate voids with the \emph{Void IDentification and Examination toolkit} VIDE~\cite{Sutter2014d}, which in its core is based on the ZOBOV~\cite{Neyrinck2008} algorithm. Moreover, we use a subsampled version of our original dark matter simulation in real space that matches our dense mock-galaxy catalog's sampling density of $\bar{n}\simeq2.0\times10^{-2}h^3\;{\rm Mpc}^{-3}$ to define a dark matter void catalog as a reference. VIDE finds density minima in a Voronoi tessellation of the tracer particles and grows basins around them applying the watershed transform~\cite{Platen2007}. It gives rise to a nested hierarchy of voids and sub-voids, all of which we consider in our analysis. We cease to merge basins with each other if the minimum ridge density between them is larger than $20\%$ of the mean tracer density, which prevents voids of growing too deep into over-dense structures~\cite{Neyrinck2008}. This value is inspired by the spherical evolution model~\cite{Sheth2004}, which predicts a minimum under-density of $0.2\bar{\rho}_\matter$ for voids whose surrounding ridges begin to undergo shell crossing. However, this is only a reference value, as the watershed algorithm identifies basins irrespective of whether shell crossing has occurred. We further define each void center $\mathbf{X}_i$ as the mean of all its member particle positions $\mathbf{x}_j$, weighted by their individual Voronoi cell-volume $V_j$~\cite{Lavaux2012},
\begin{equation}
 \mathbf{X}_i = \frac{\sum_j \mathbf{x}_j V_j}{\sum_j V_j}\;.
\end{equation}
The advantage of this definition over the location of the minimum density (largest Voronoi cell) is its insensitivity to Poisson noise and the fact that it makes use of the watershed geometry which defines the void. Further, an effective void radius $r_\void$ can be defined as the radius of a sphere comprising the same volume as the sum of all Voronoi cells that each void is composed of,
\begin{equation}
 r_\void = \left(\frac{3}{4\pi}\sum\nolimits_j V_j\right)^{1/3}\;.
\end{equation}
We consider a range of $7\hMpc \lesssim r_\void \lesssim 85\hMpc$ for the voids identified in our dense tracer samples, and a range of $30\hMpc \lesssim r_\void \lesssim 107\hMpc$ for the ones found in our sparse mock catalog. This corresponds to about twice the mean particle separation as a lower bound, and the largest void found as an upper bound, respectively. The lower bound is imposed to avoid contamination of the sample by misidentified voids arising through either random Poisson fluctuations~\cite{Neyrinck2008}, or the effects of redshift-space distortions~\cite{Pisani2015b}. It amounts to a total number of $\sim9.9\times10^4$ voids in the dark matter sample, $\sim6.4\times10^4$ voids in our dense mock-galaxy catalog, and $\sim2.0\times10^3$ in the sparse one.

\subsection{Void stacks \label{sec:analysis:stacks}}
As described by eq.~(\ref{xi_s}), the tracer-density profile around voids (respectively the void-tracer cross-correlation function) is no longer isotropic in redshift space. The spherical symmetry of voids is thus reduced to a cylindrical symmetry along the line-of-sight direction. Therefore, it is instructive to visualize this symmetry as a two-dimensional stack with axes showing the void-centric distances of tracers along and perpendicular to the line of sight. To this end we determine the relative separation vector
\begin{equation}
 \hat{\mathbf{r}}_{ij}=\frac{\mathbf{X}_i-\mathbf{x}_j}{r_\void}
\end{equation}
between every void position $\mathbf{X}_i$ and every tracer position $\mathbf{x}_j$ in units of $r_\void$, within a narrow range of the effective void radius with mean value $\bar{r}_\void$. This is done only up to a maximum distance of $\left|\hat{\mathbf{r}}_{ij}\right| = 3$. Further, we extract its line-of-sight component $\hat{r}_{ij}^\parallel$, as well as its angular component $\hat{r}_{ij}^\perp$, and histogram the number of tracers in bins of width $\delta \hat{r}_\parallel$ and $\delta \hat{r}_\perp$ to estimate the void-centric tracer density according to a Poisson-counts estimator~\cite{Nadathur2015a},
\begin{equation}
 \rho_{\void\tracer}\left(r_\parallel,r_\perp\right) = \frac{\sum_{i,j}\Theta(\hat{r}_{ij}^\parallel)\Theta(\hat{r}_{ij}^\perp)+1}{4\pi \hat{r}_\perp \delta \hat{r}_\perp \delta \hat{r}_\parallel N_\void \bar{r}_\void^3}\;, \label{stack_rho}
\end{equation}
where $\Theta(\hat{r}_{ij}) \equiv \theta[\hat{r}_{ij}-(\hat{r}-\delta \hat{r}/2)]\theta[-\hat{r}_{ij}+(\hat{r}+\delta \hat{r}/2)]$ combines two Heaviside step functions $\theta$ to define each radial bin for $r_\parallel$ and $r_\perp$, respectively. The uncertainty in a given bin can then be estimated via the standard deviation in the number of tracers among all $N_\void$ voids. Similarly, the two components of the velocity dispersion $\sigma_\parallel(r)$ and $\sigma_\perp(r)$ in eq.~(\ref{sv}) can be estimated via
\begin{equation}
 \sigma^2_{\parallel,\perp}(r) = \frac{\sum_{i,j}\Theta(\hat{r}_{ij})\left[v_j^{\parallel,\perp}(\hat{r}_{ij})-v_{\parallel,\perp}(\hat{r})\right]^2V_j}{\sum_{i,j}\Theta(\hat{r}_{ij})V_j}\;, \label{stack_sv}
\end{equation}
except that each of them only depends on the absolute distance $r$ to the void center and we weight by the Voronoi cell-volume $V_j$ of each tracer to obtain a volumetric and hence unbiased estimate of these velocity statistics. The mean velocities used in eq.~(\ref{stack_sv}) are estimated accordingly,
\begin{equation}
 v_{\parallel,\perp}(r) = \frac{\sum_{i,j}\Theta(\hat{r}_{ij})v_j^{\parallel,\perp}(\hat{r}_{ij})V_j}{\sum_{i,j}\Theta(\hat{r}_{ij})V_j}\;. \label{stack_v}
\end{equation}
Note that we neglected the velocities of the void centers, which have been shown to be vanishingly small in simulations~\cite{Sutter2014c}. 

\subsubsection{Dark matter \label{sec:analysis:stacks:matter}}
\begin{figure*}[!p]
\centering
\vspace{-38pt}
\resizebox{.78\hsize}{!}{
\includegraphics[trim=0 35 78 0,clip]{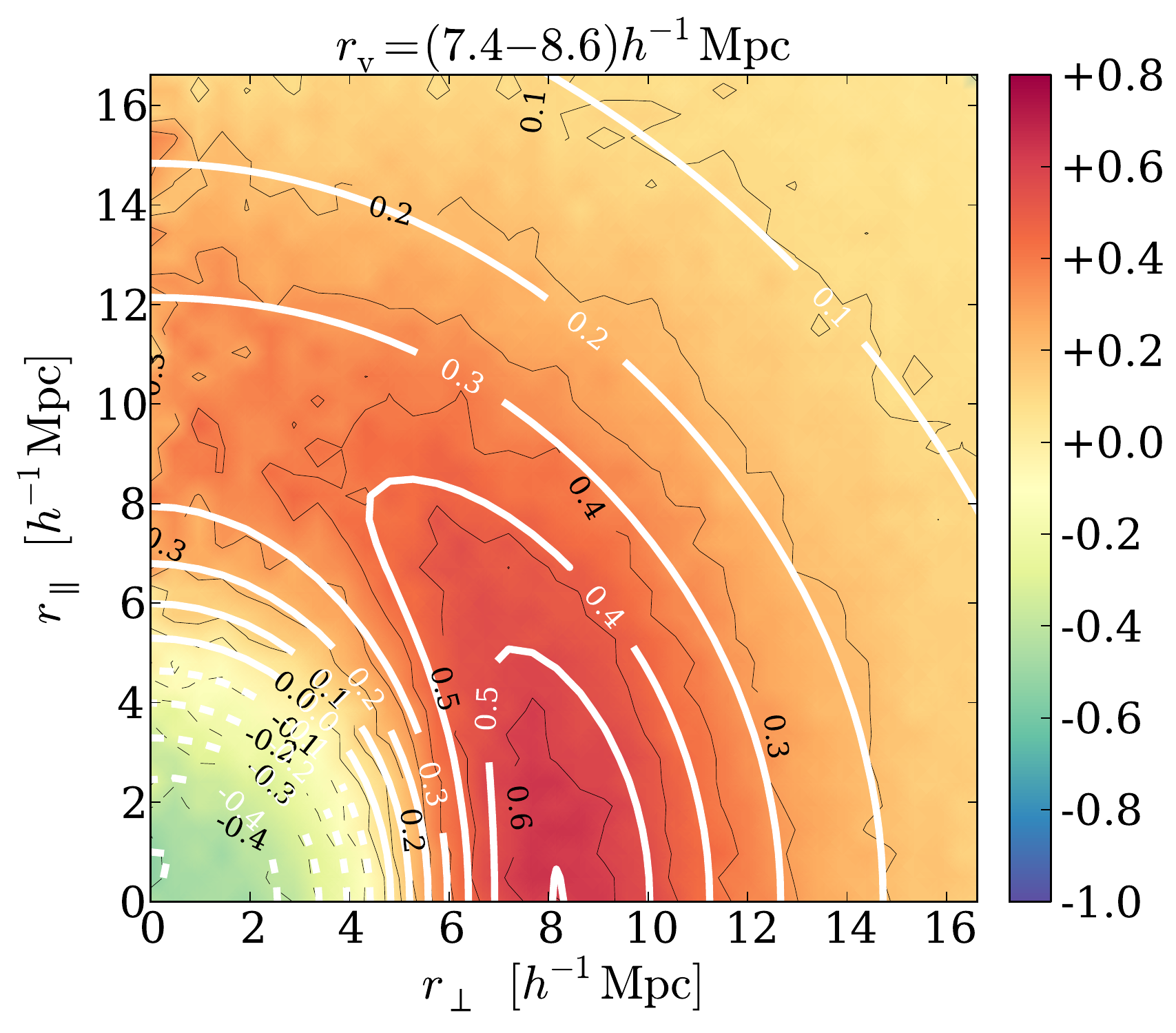}
\includegraphics[trim=40 35 0 0,clip]{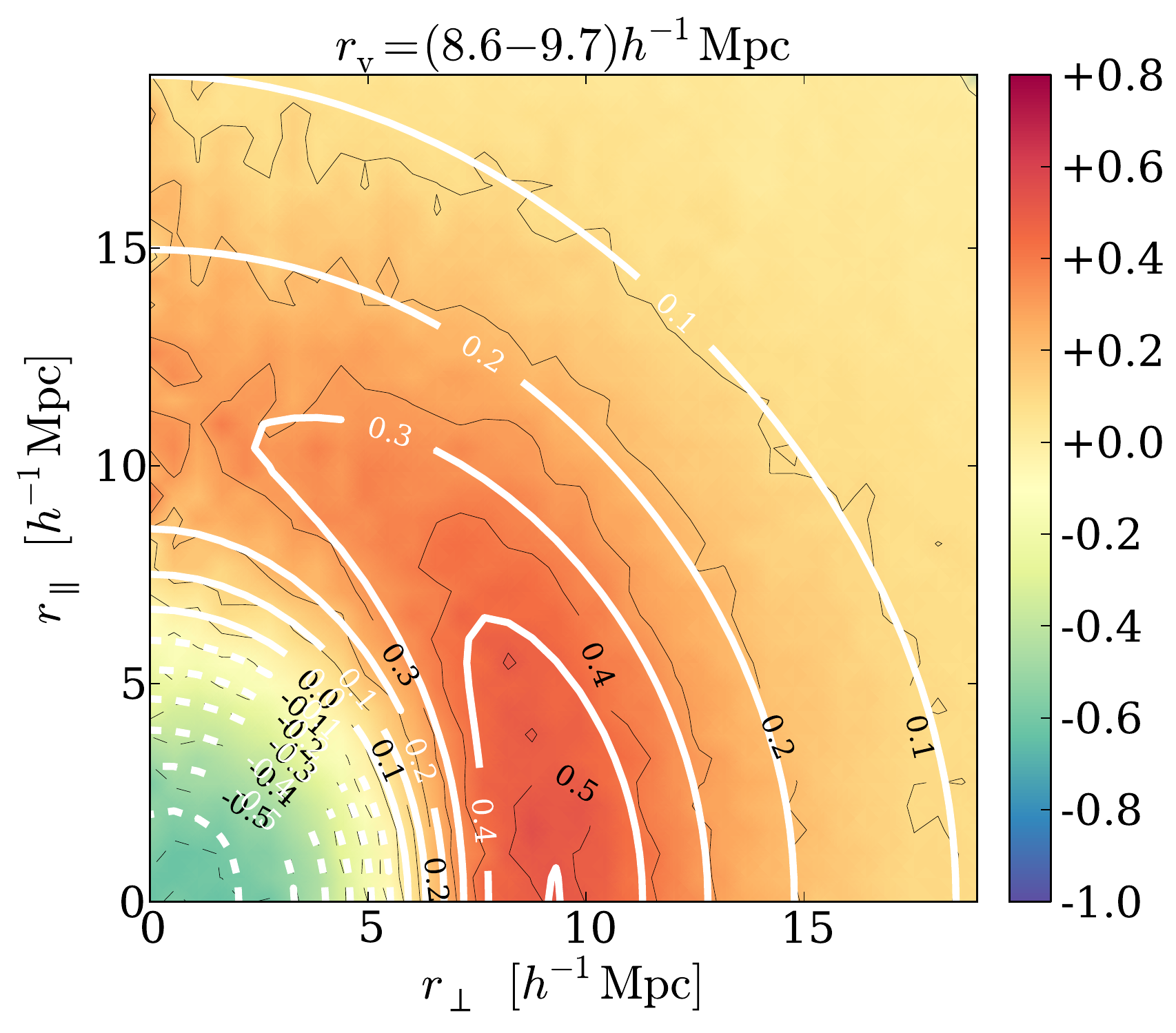}}
\resizebox{.78\hsize}{!}{
\includegraphics[trim=0 35 78 0,clip]{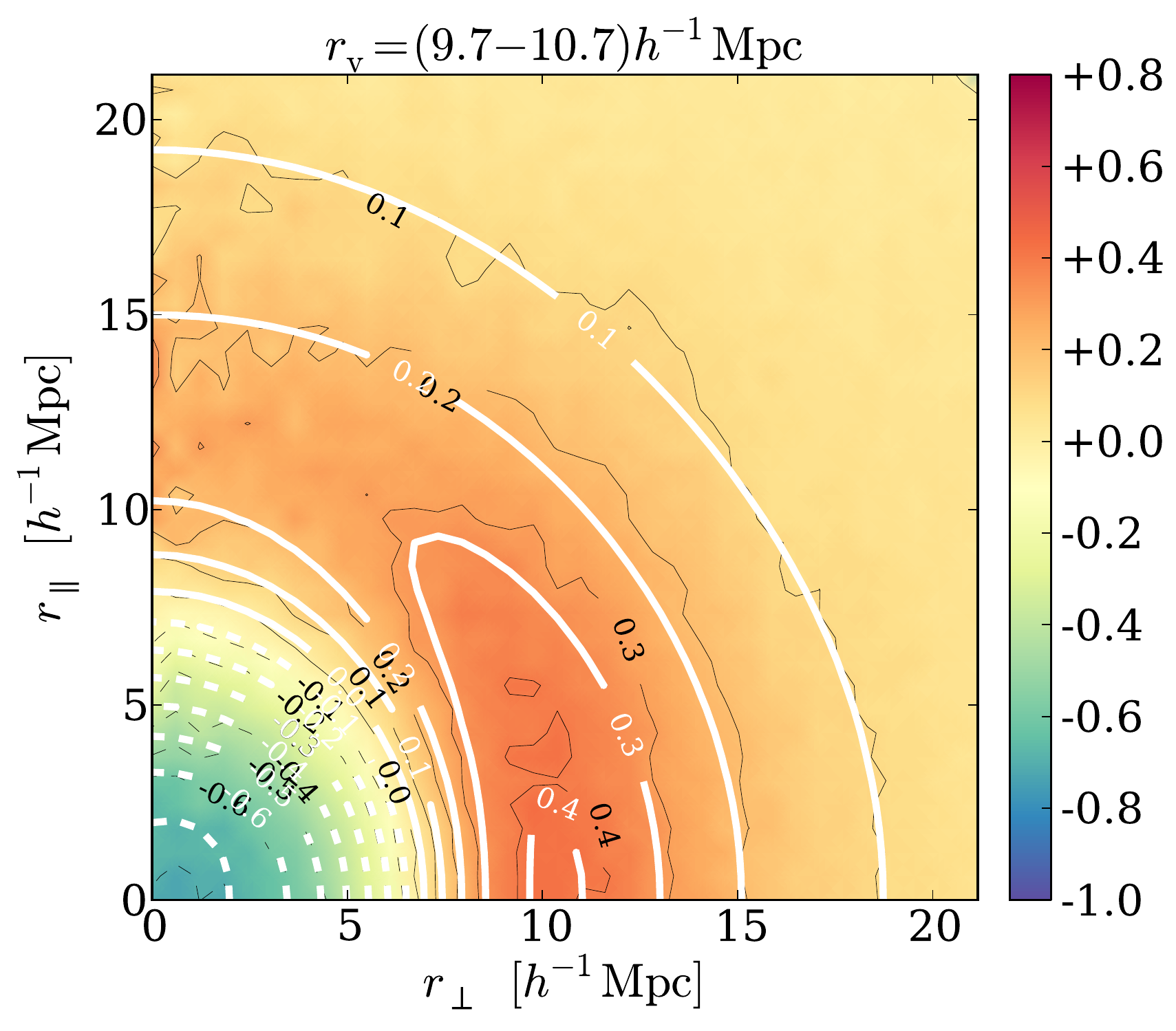}
\includegraphics[trim=40 35 0 0,clip]{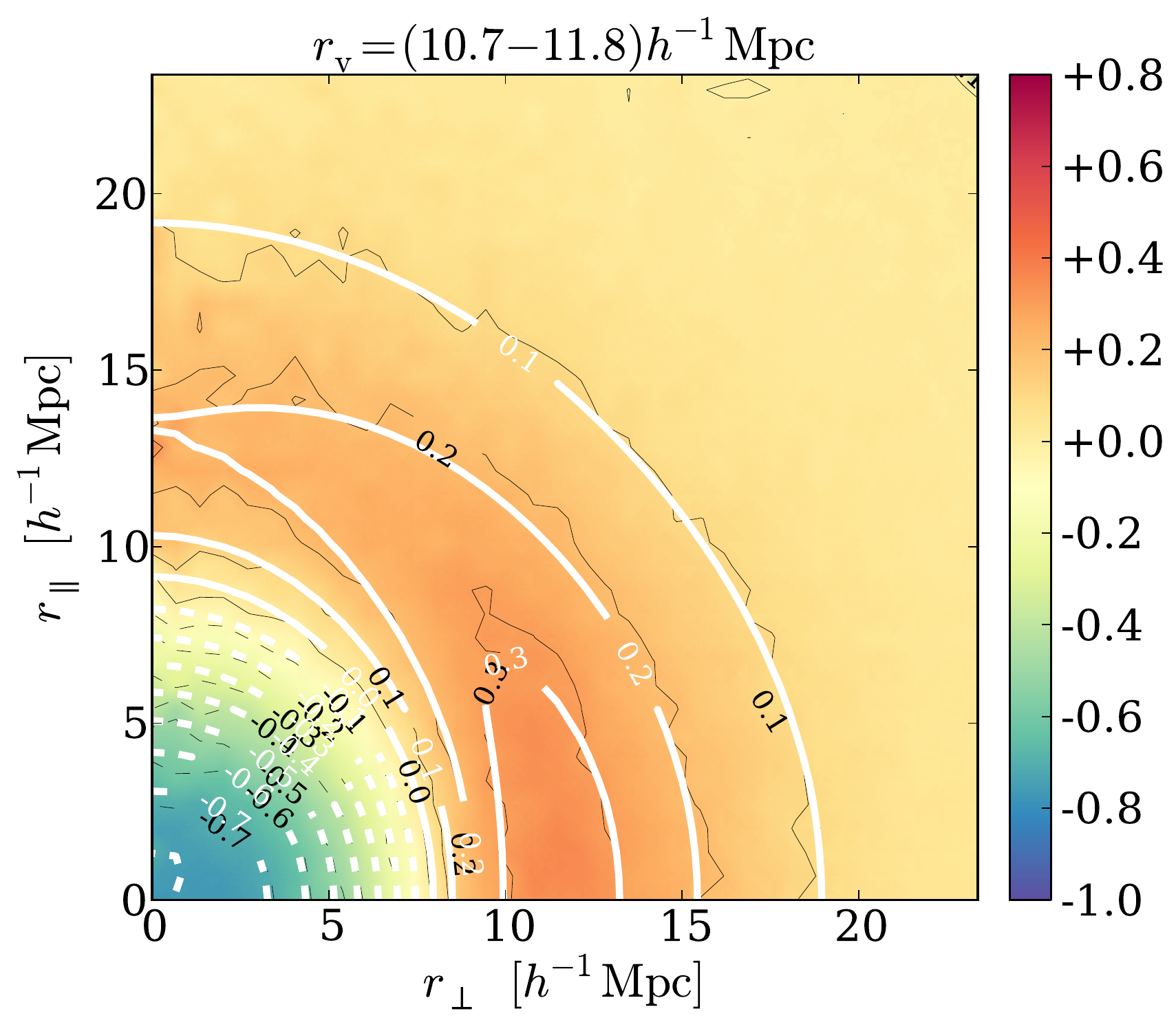}}
\resizebox{.78\hsize}{!}{
\includegraphics[trim=0 35 78 0,clip]{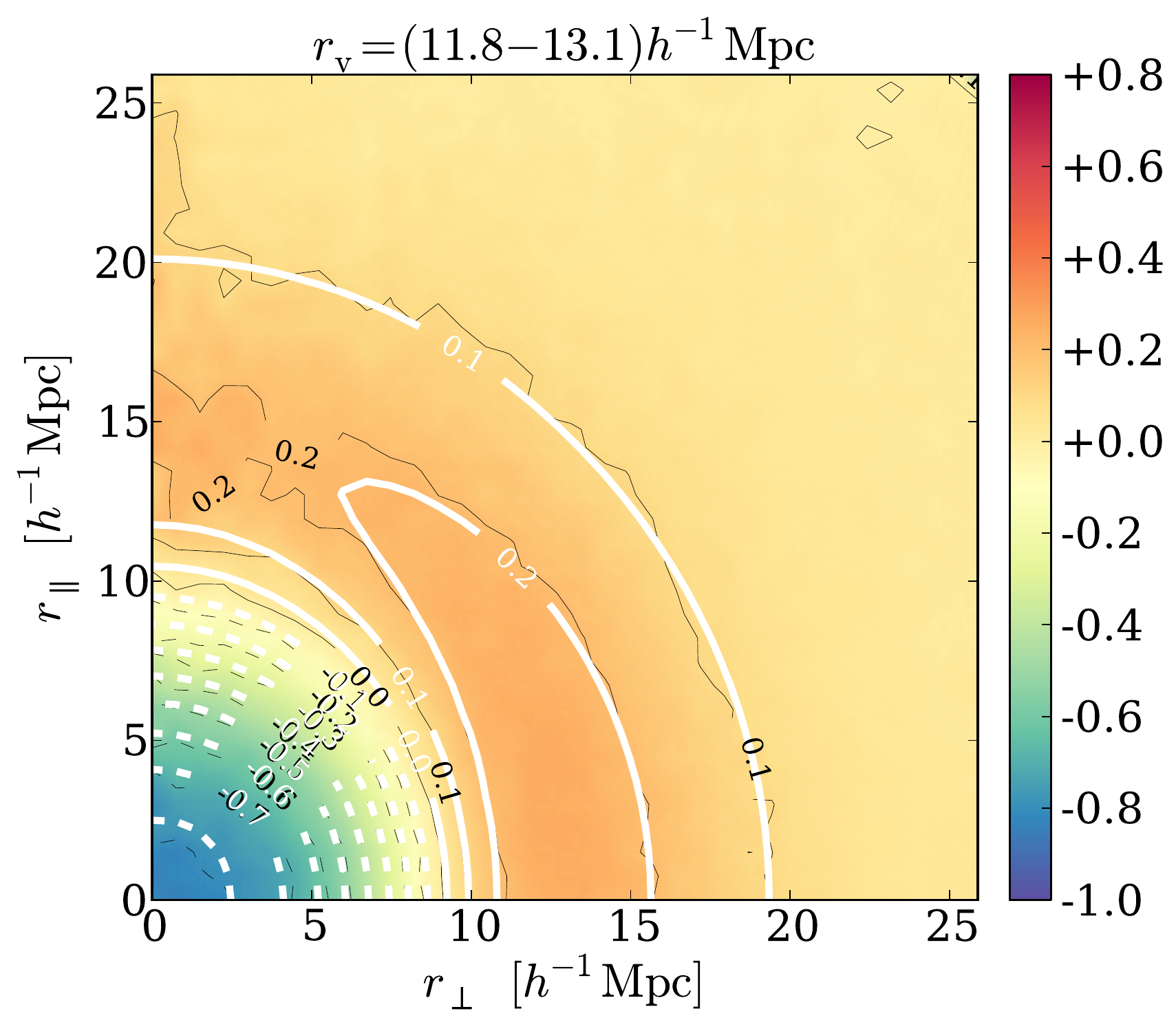}
\includegraphics[trim=40 35 0 0,clip]{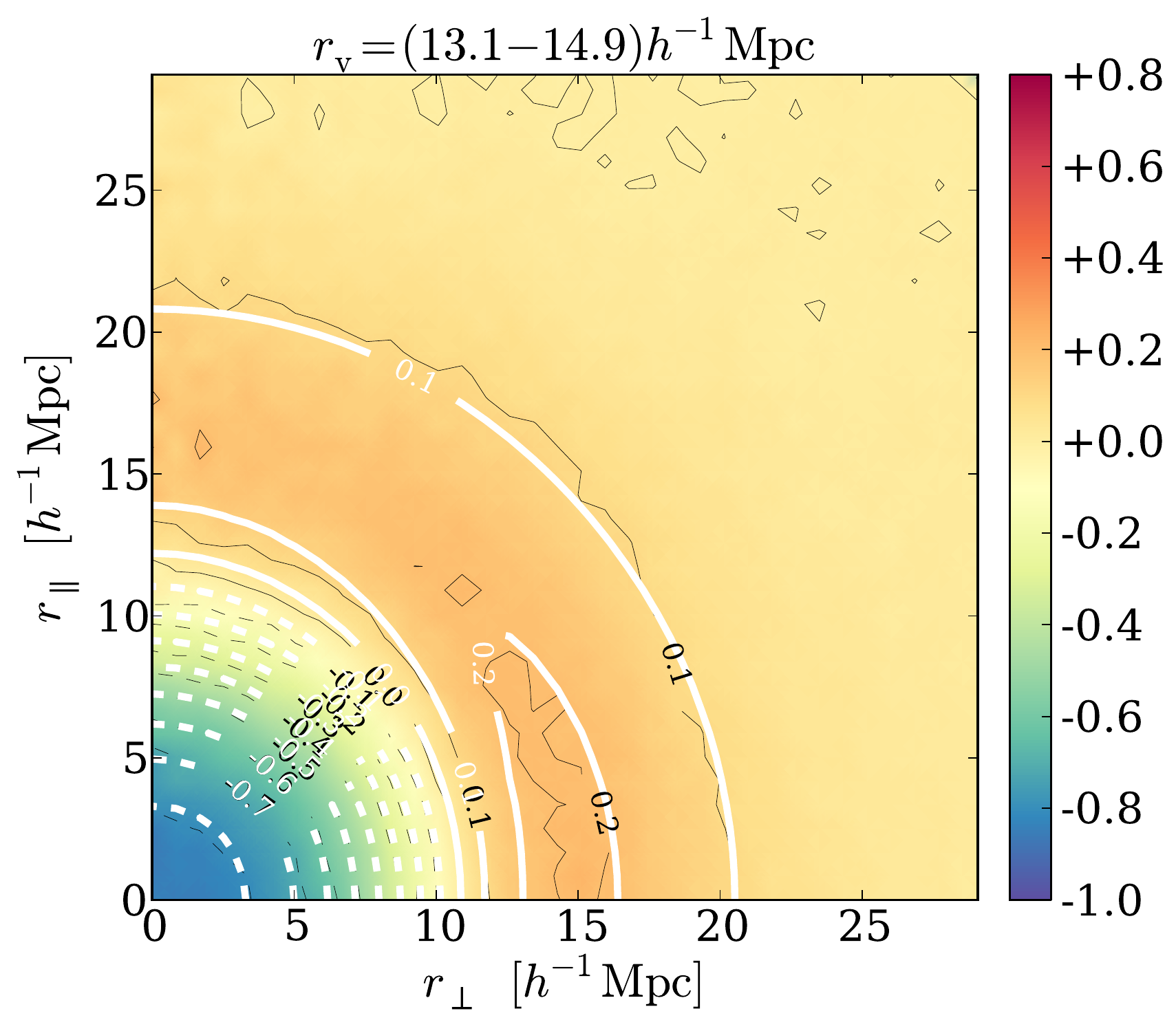}}
\resizebox{.78\hsize}{!}{
\includegraphics[trim=0 9 78 0,clip]{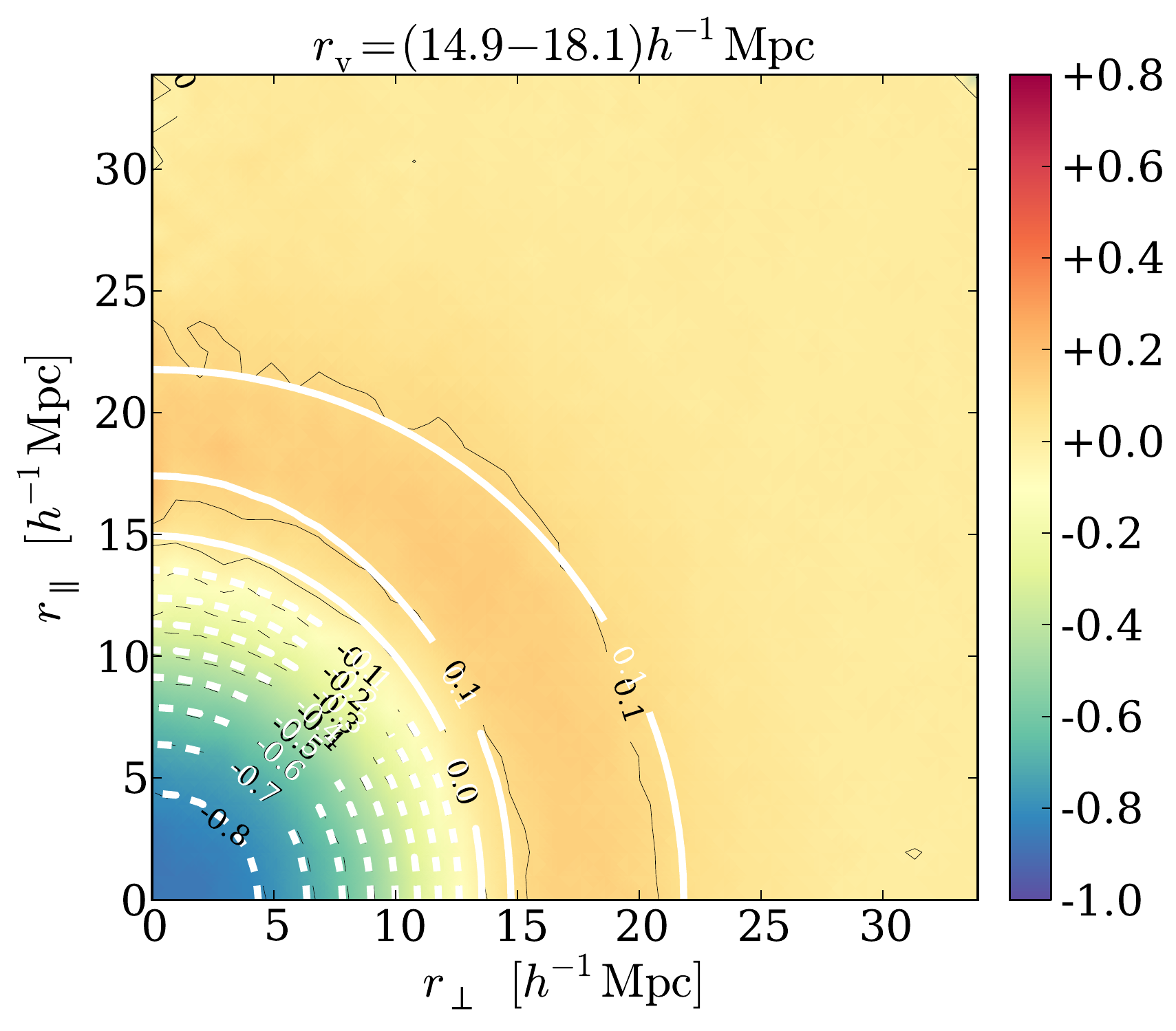}
\includegraphics[trim=40 9 0 0,clip]{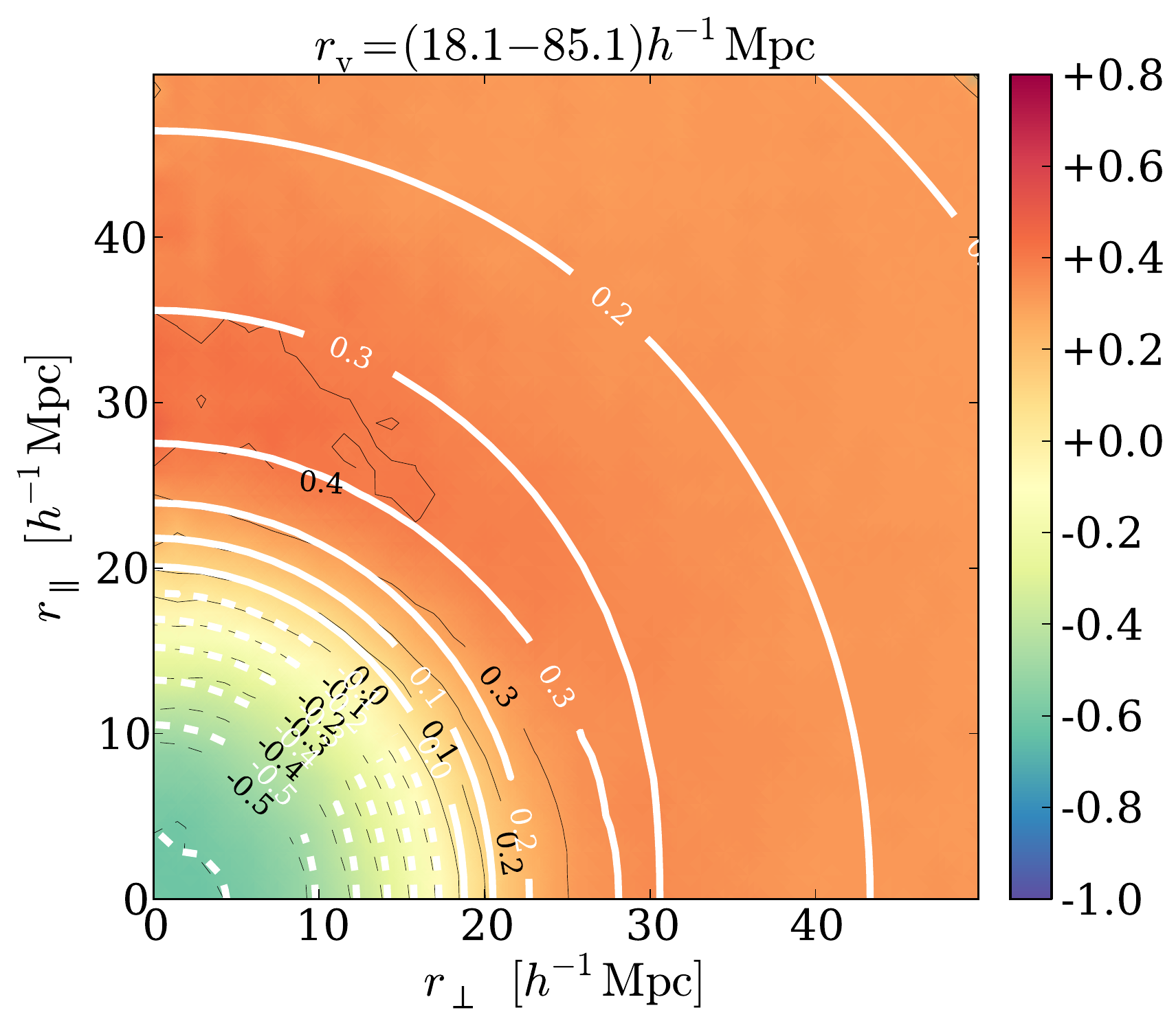}}
\caption{Dark matter void stacks $\rho_{\void\matter}(r_\parallel,r_\perp)/\bar{\rho}_\matter - 1$ in redshift space at $z=0$. Black solid/dashed lines show positive/negative contours of the data, white lines show the best-fit model.}
\label{Xvm2d}
\end{figure*}

Figure~\ref{Xvm2d} shows two-dimensional void stacks in the dark matter for eight contiguous bins in effective void radius, resulting in about 12500 voids per stack. The bin edges are chosen to yield an equal number of voids per bin, ensuring the signal-to-noise ratio to remain similar across the bins. Although these dark matter stacks are not directly observable in practice, they shall serve as a reference to compare with our theoretical model. To this end, we identified physical voids in the real-space dark matter distribution, and only used the dark matter in redshift space for the stacking in eq.~(\ref{stack_rho}). From the figure we can distinguish two regimes that are affected differently by redshift-space distortions. Small and intermediate-size voids with $r_\void\lesssim15\hMpc$ exhibit a relatively high ridge that decays away towards decreasing angles to the line of sight (i.e. small $r_\perp$). This behavior is reversed for the larger voids, whose ridge appears more prominent along the line of sight. These regimes may be attributed to the two well-known features observed in the galaxy auto-correlation function in redshift space: the FoG effect and the Kaiser effect~\cite{Kaiser1987}. The former is caused by nonlinear virial motions of galaxies within highly over-dense structures (such as clusters) and results in a stretching of contour lines on small distances along the line of sight. Because small voids exhibit the highest ridge densities, the FoG effect is most prominent for them and smears out the ridge density along the line of sight. This smearing is described by the velocity dispersion in eq.~(\ref{pdf}).

In contrast, the Kaiser effect is caused by the coherent streaming motion of galaxies towards over-densities on larger scales, as described by eq.~(\ref{v_lin}). This results in a squashing of contours towards the void ridges, which leads to their enhancement along the line of sight. Hence, the two effects act against each other and can even balance out in their transition regime, which can be seen in the bottom left panel of fig.~\ref{Xvm2d} with $14.9\hMpc<r_\void<18.1\hMpc$. Towards larger voids the Kaiser effect is still visible not only in the amplification of ridges, but also in emptying out the interior of the void, resulting in elongated negative contours along the line of sight (see also~\cite{Padilla2005,Paz2013,Micheletti2014}).

The influence of the AP effect on the two-dimensional void stacks is simpler to understand, which is why we do not show it explicitly here. According to eq.~(\ref{AP}), the ratio of tracer coordinates along and perpendicular to the line of sight is scaled by some constant factor, which results in a stretching of the void stack along either $r_\parallel$ or $r_\perp$. Therefore, geometric distortions due to the AP effect are expected to be far less dependent on void size and tracer density when compared to the dynamic distortions discussed above.

\begin{figure*}[!t]
\centering
\resizebox{\hsize}{!}{
\includegraphics[trim=0 0 0 0,clip]{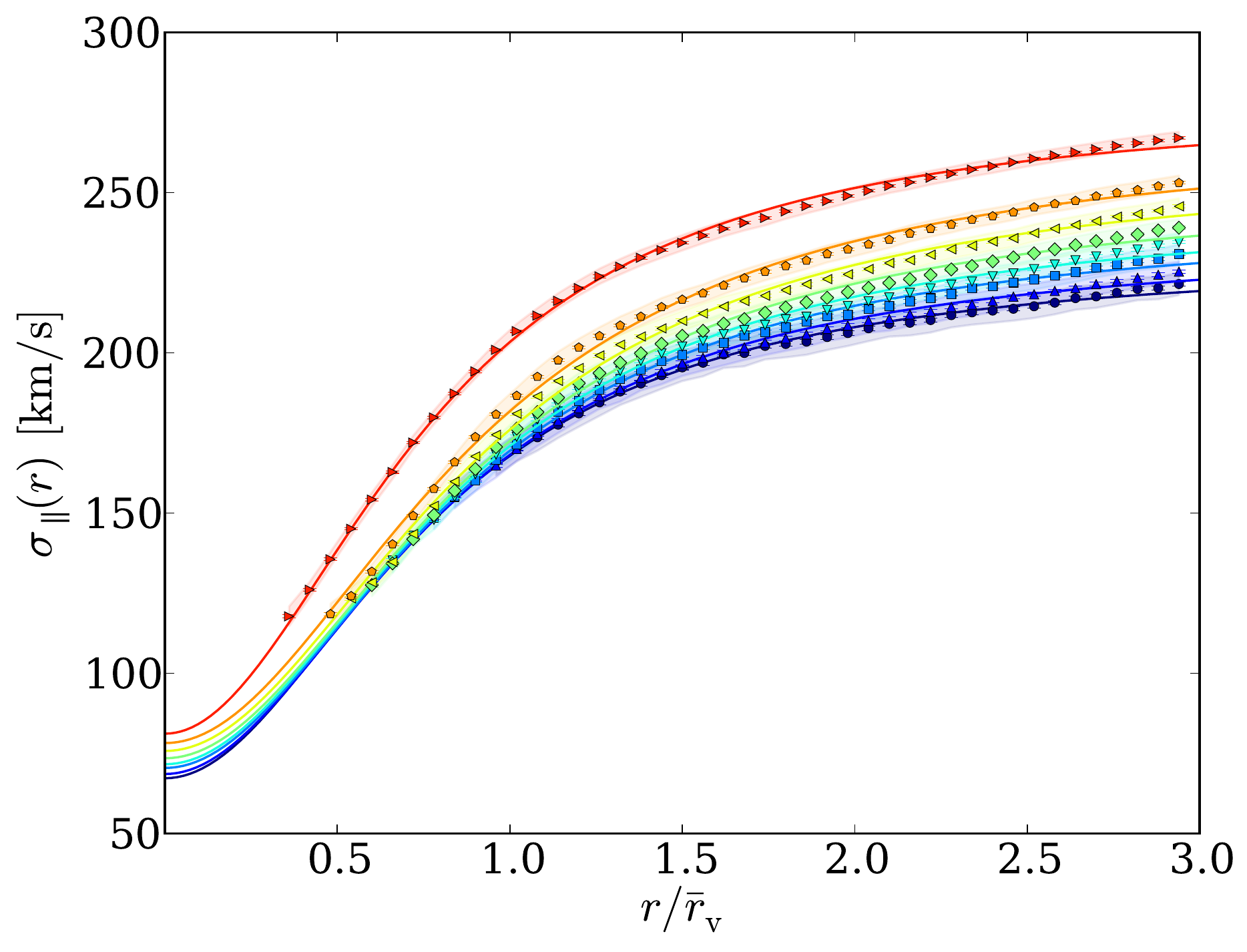}
\includegraphics[trim=0 0 0 0,clip]{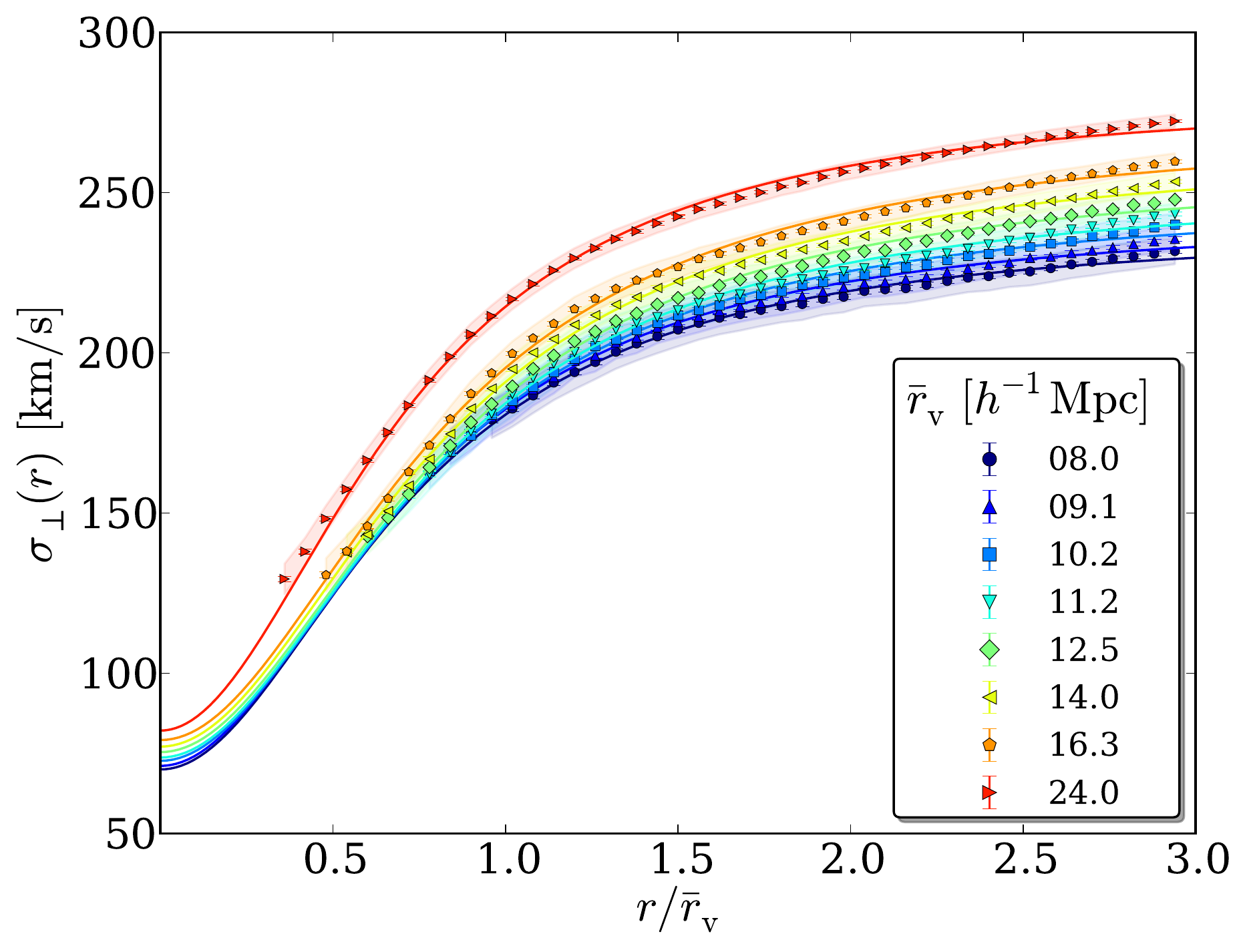}}
\resizebox{\hsize}{!}{
\includegraphics[trim=0 0 0 0,clip]{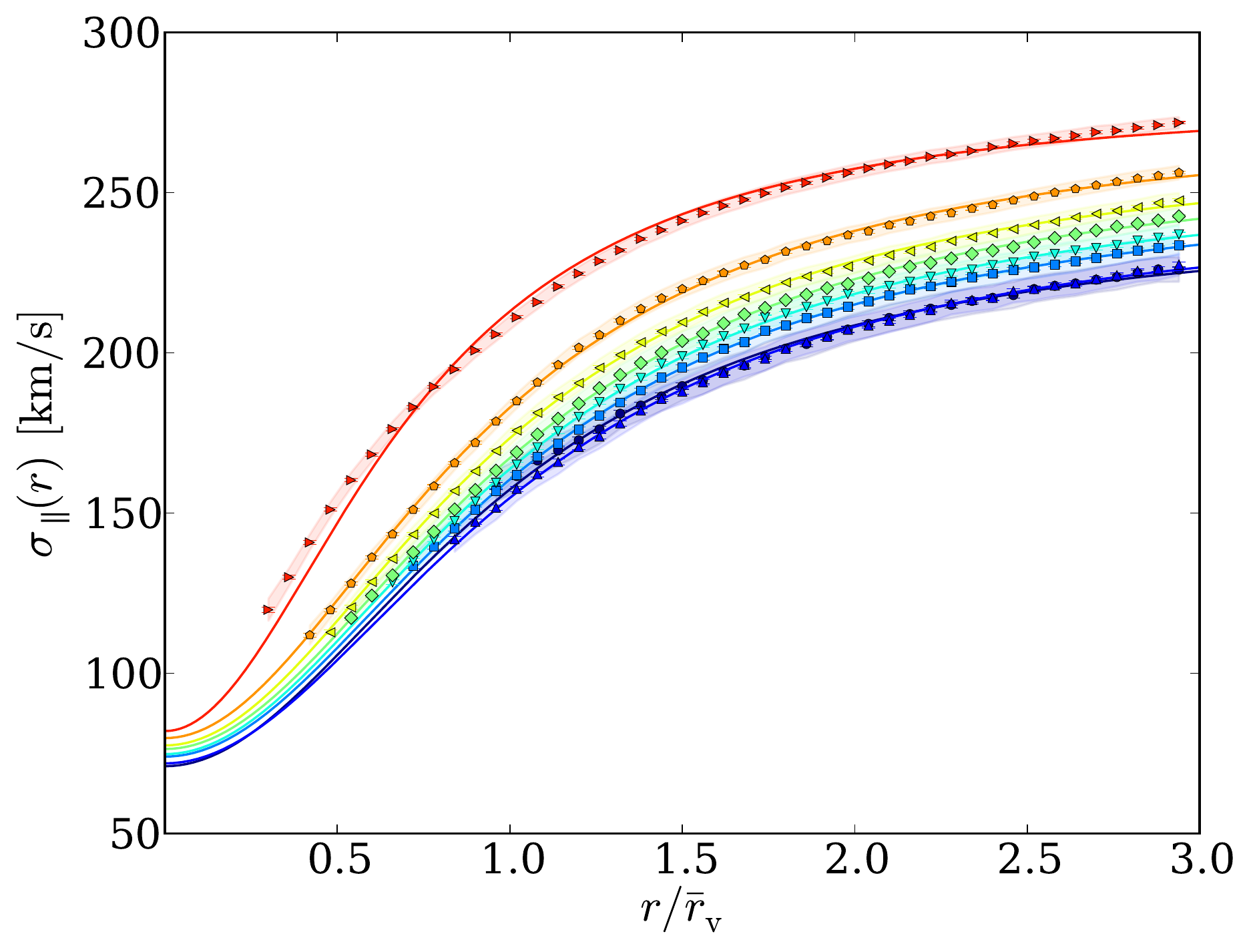}
\includegraphics[trim=0 0 0 0,clip]{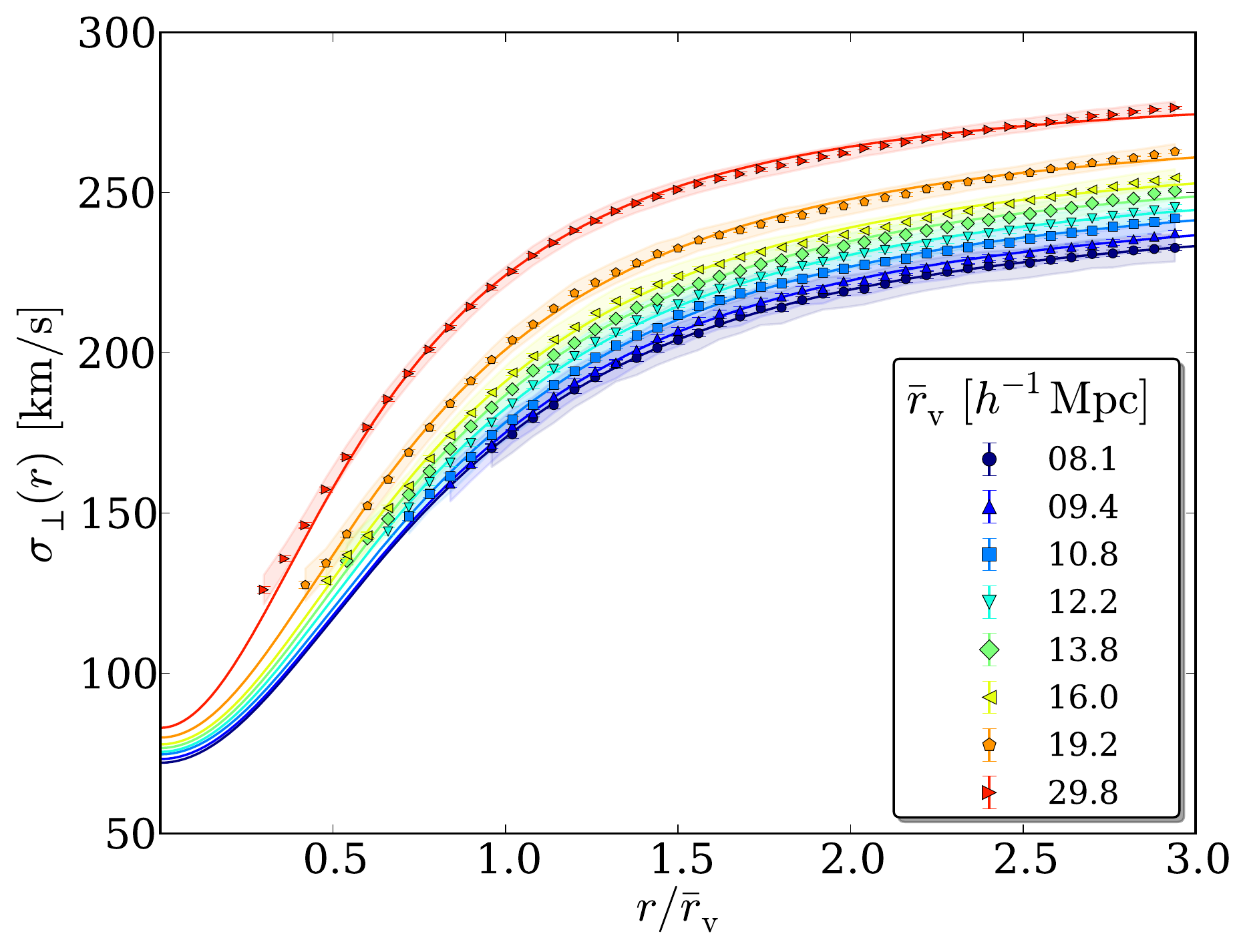}}
\resizebox{\hsize}{!}{
\includegraphics[trim=0 0 0 0,clip]{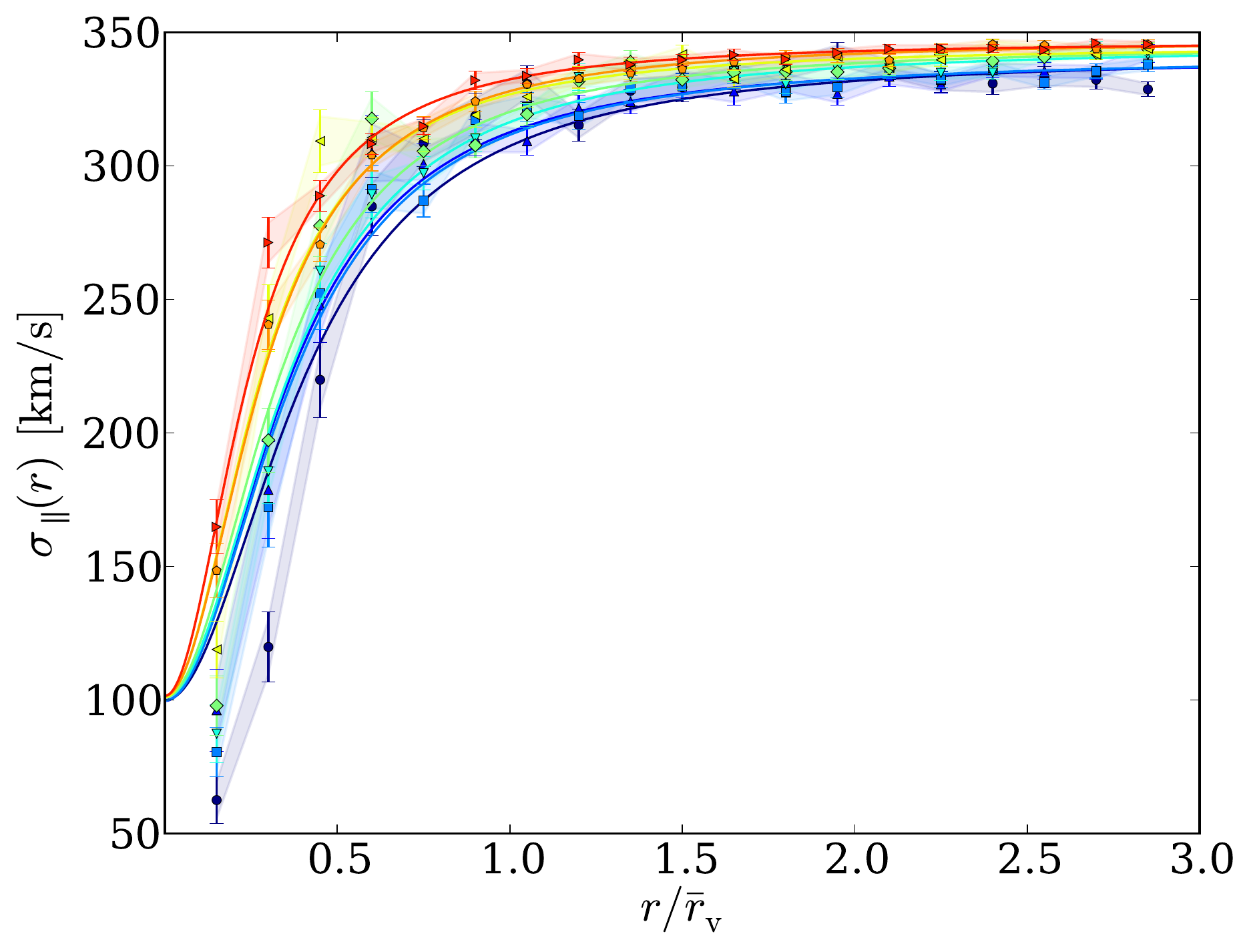}
\includegraphics[trim=0 0 0 0,clip]{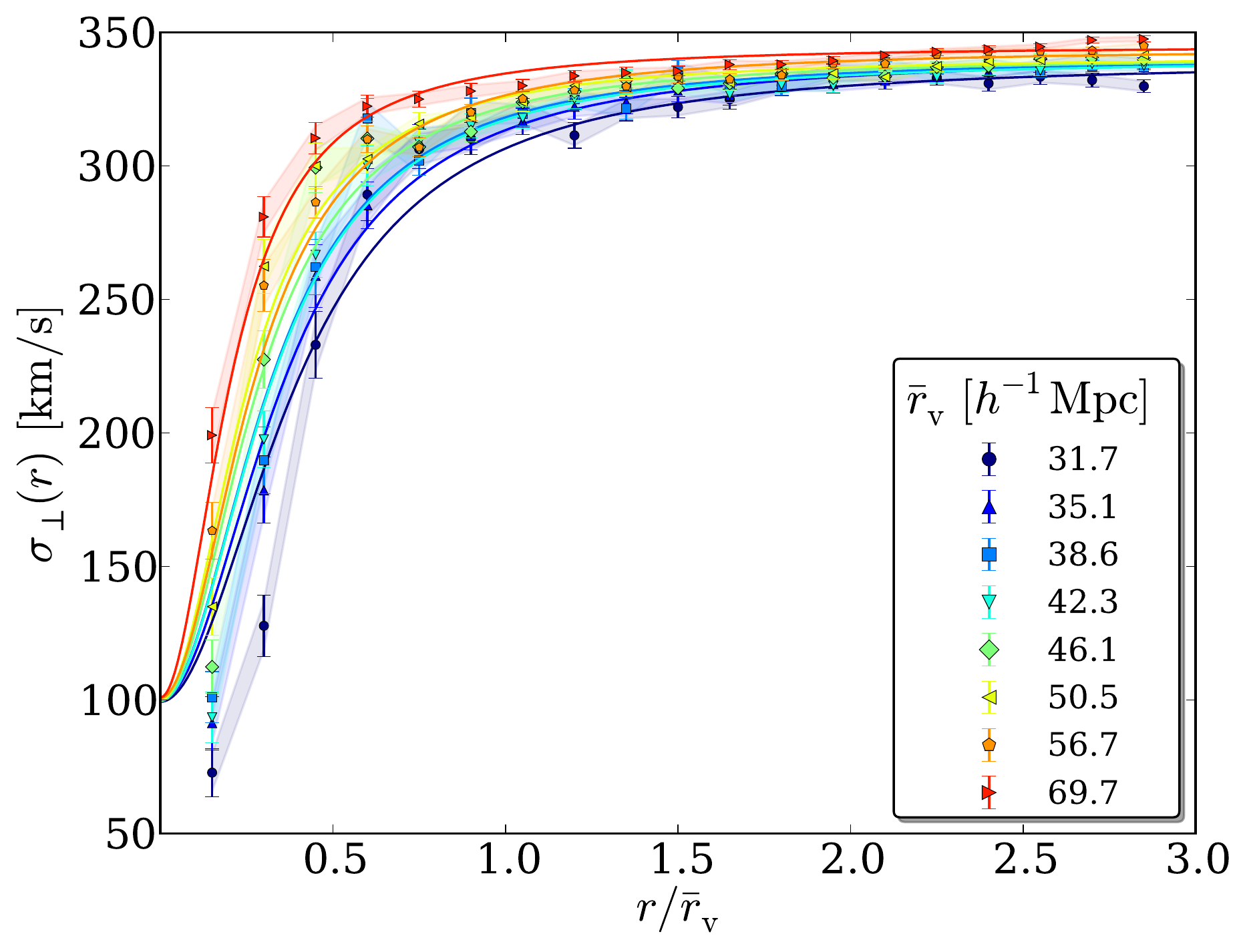}}
\caption{Velocity dispersion along (left) and perpendicular (right) to the line of sight around voids identified in dark matter particles at $z=0$ (top), dense mock galaxies at $z=0$ (middle), and sparse mock galaxies at $z=0.5$ (bottom). Shaded regions depict the void-to-void standard deviation within each bin (scaled down by 20 for visibility), error bars show standard errors on the mean, and solid lines show the best fits from eq.~(\ref{sv_model}). The insets list the mean effective void radius of each stack.}
\label{svm}
\end{figure*}

Let us now inspect the velocity dispersion in the dark matter around these voids. The upper panels in fig.~\ref{svm} depict its two components $\sigma_\parallel$ and $\sigma_\perp$ as a function of distance $r$ to the void center for the same bins in void effective radius. Both components increase rapidly up to the mean void radius $\bar{r}_\void$ of each stack, and then level off with a more moderate slope. Due to the rescaling of the abscissa by $\bar{r}_\void$, larger voids appear to have higher velocity dispersions in their surrounding, but at given physical distance these values are in fact remarkably similar. Likewise, the differences between $\sigma_\parallel(r)$ and $\sigma_\perp(r)$ are quite small. In order to model these velocity dispersions we tried fitting eqs.~(\ref{sv_par}) and (\ref{sv_per}) to our data. While this works well on large distances $r$ from the void center, the velocity dispersion in the void interior is not accurately reproduced with these formulae, which is likely due to the mentioned discrepancy of linear theory on those scales and long-range correlations in the velocity statistics. Instead, we find that a simple \emph{Lorentzian} of width $\omega$ empirically describes the observed behavior very well,
\begin{equation}
 \sigma_{\parallel,\perp}(r) = \sigma_v\left(1 - \frac{1/\sqrt{2}}{1+r^2/\omega^2}\right)\;. \label{sv_model}
\end{equation}
Here, $\sigma_v$ is simply a constant number and corresponds to the velocity dispersion at $r\to\infty$. Figure~\ref{svm} depicts the best fits of eq.~(\ref{sv_model}) to the data as solid lines. The free parameters in the fit are $\sigma_v$ and $\omega$, going from small to large voids they roughly range as $\sigma_v\in[230,300]\frac{\mathrm{km}}{\mathrm{s}}$ and $\omega\in[1.0,0.7]\bar{r}_\void$ for both $\sigma_\parallel(r)$ and $\sigma_\perp(r)$.

\subsubsection{Mock galaxies \label{sec:analysis:stacks:galaxies}}
\begin{figure*}[!p]
\centering
\vspace{-38pt}
\resizebox{.78\hsize}{!}{
\includegraphics[trim=0 35 78 0,clip]{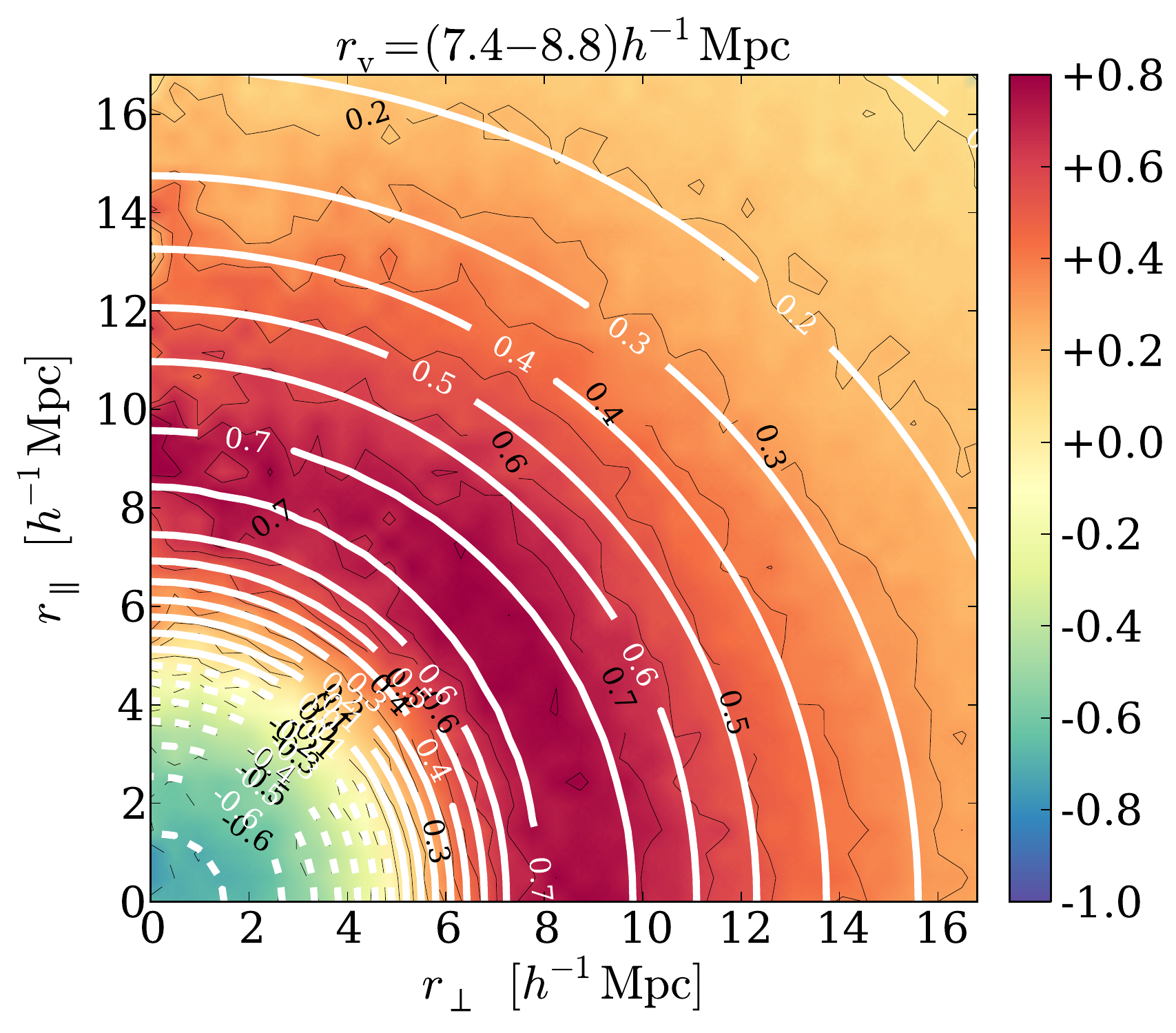}
\includegraphics[trim=40 35 0 0,clip]{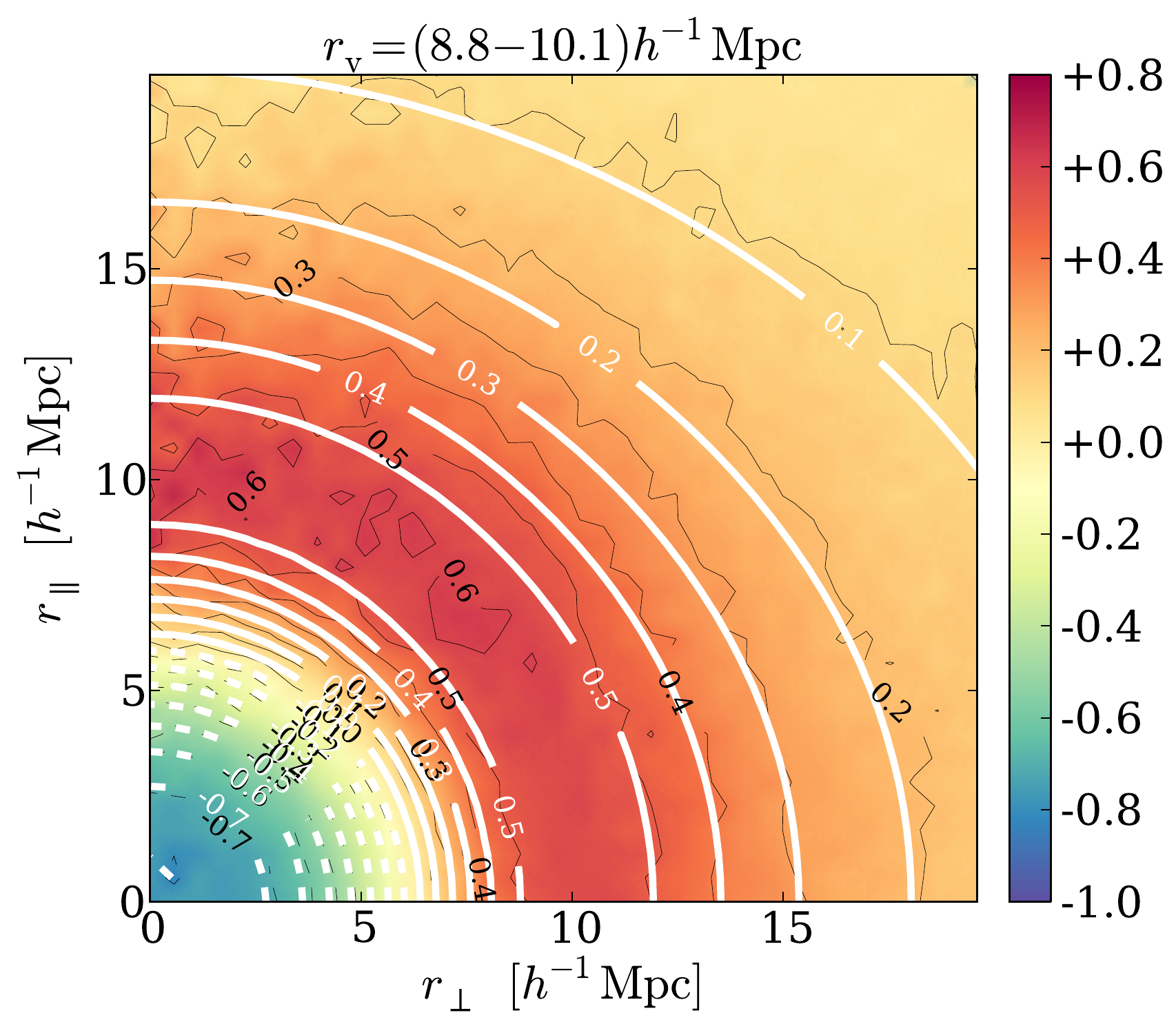}}
\resizebox{.78\hsize}{!}{
\includegraphics[trim=0 35 78 0,clip]{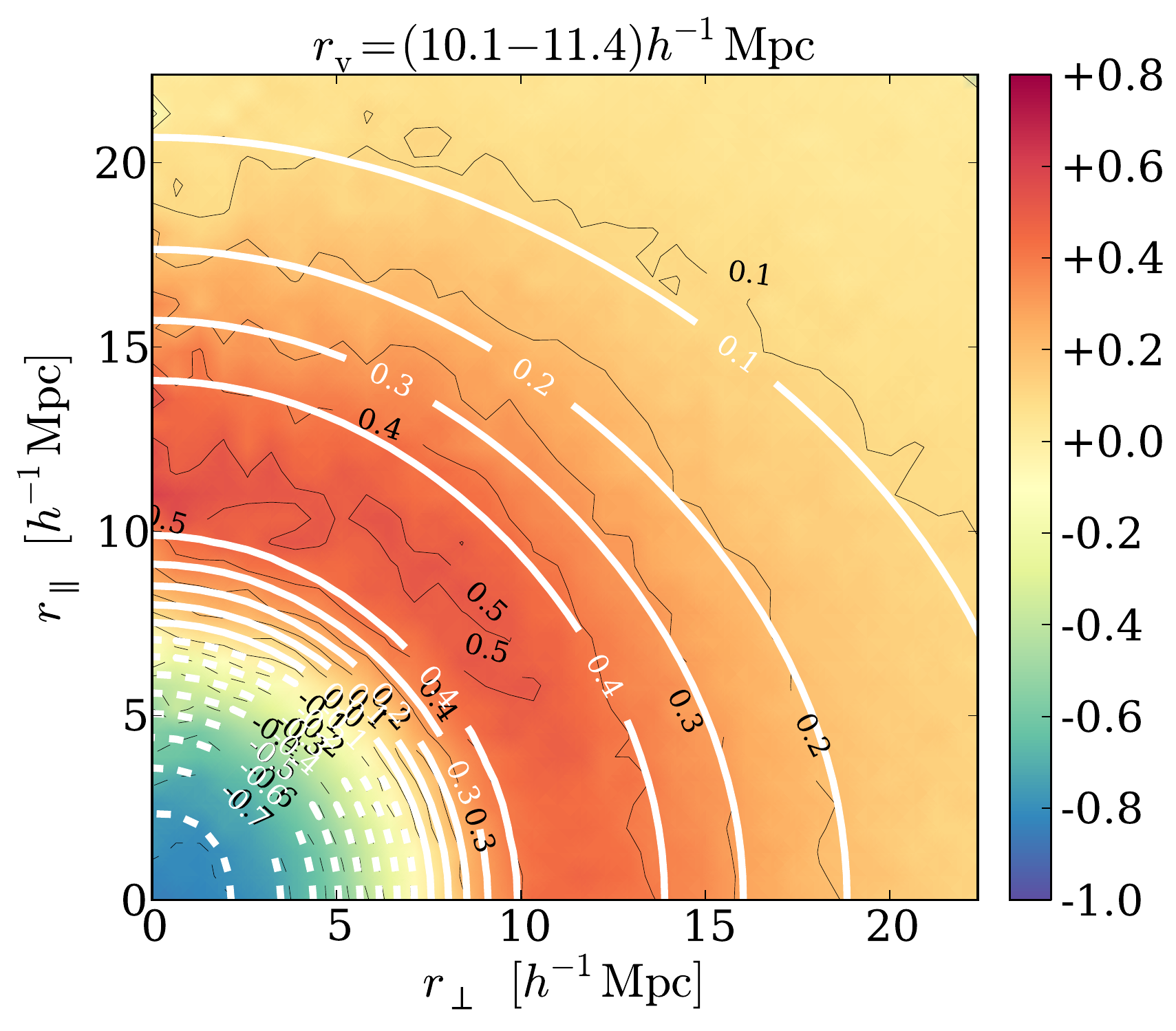}
\includegraphics[trim=40 35 0 0,clip]{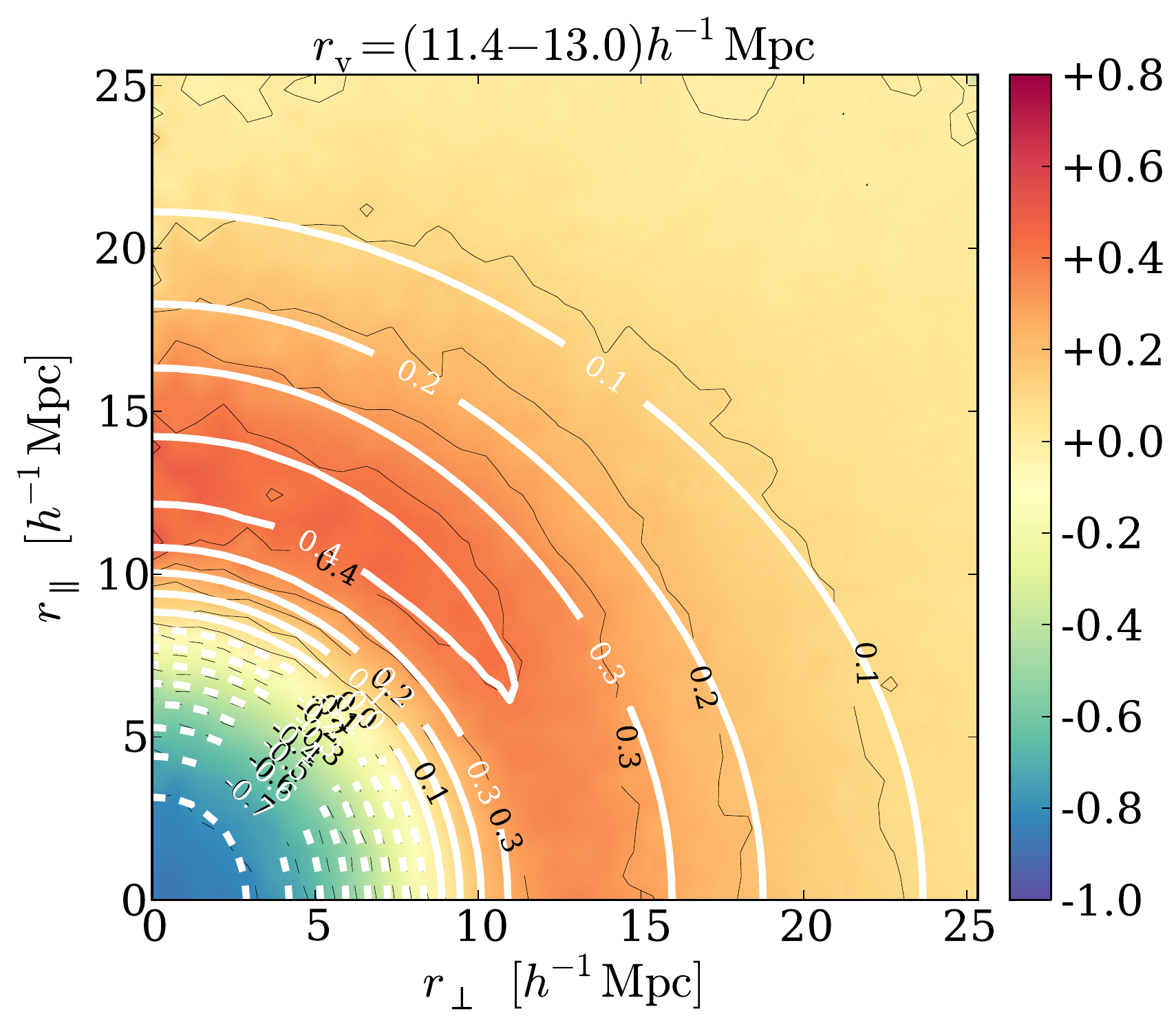}}
\resizebox{.78\hsize}{!}{
\includegraphics[trim=0 35 78 0,clip]{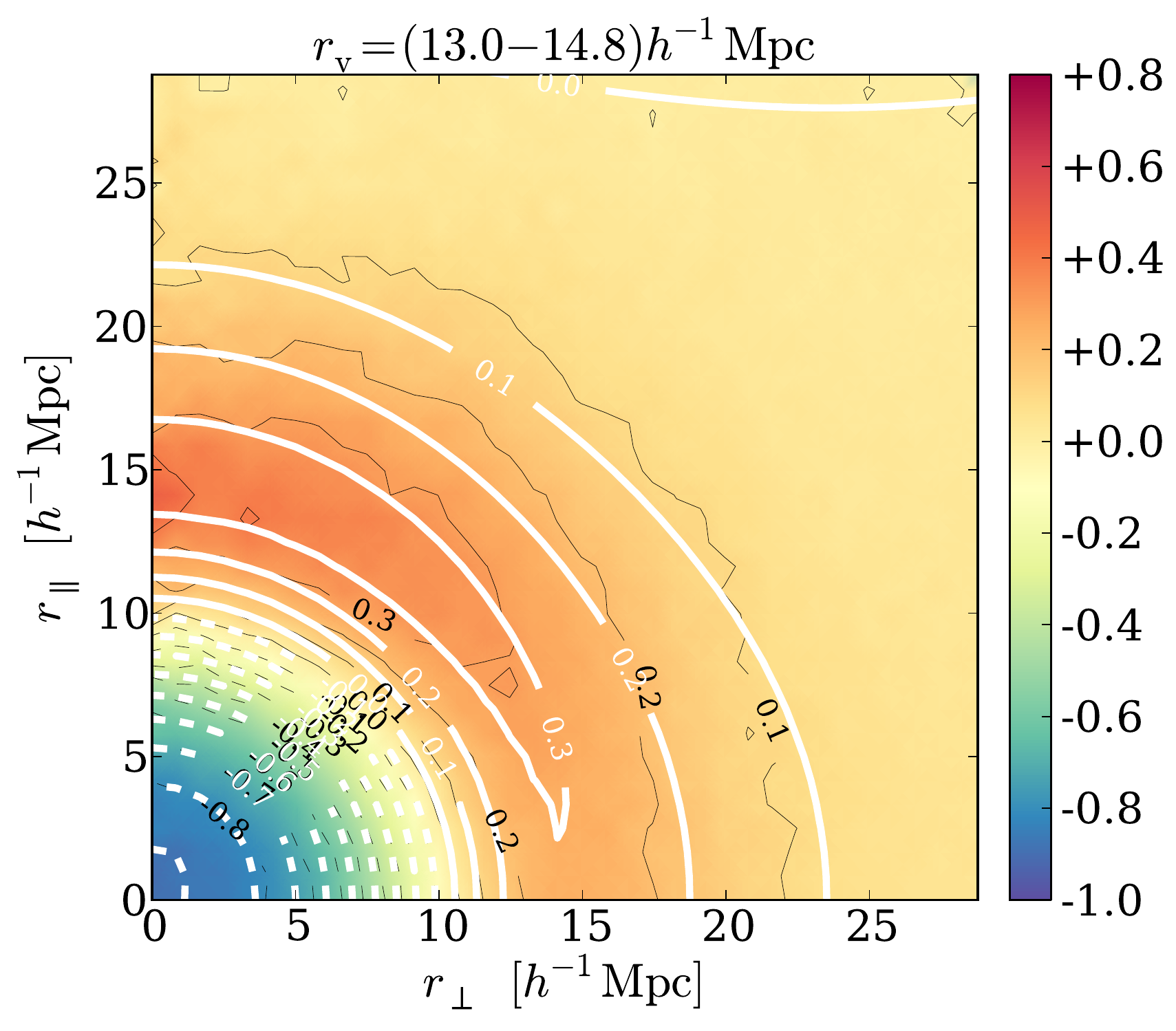}
\includegraphics[trim=40 35 0 0,clip]{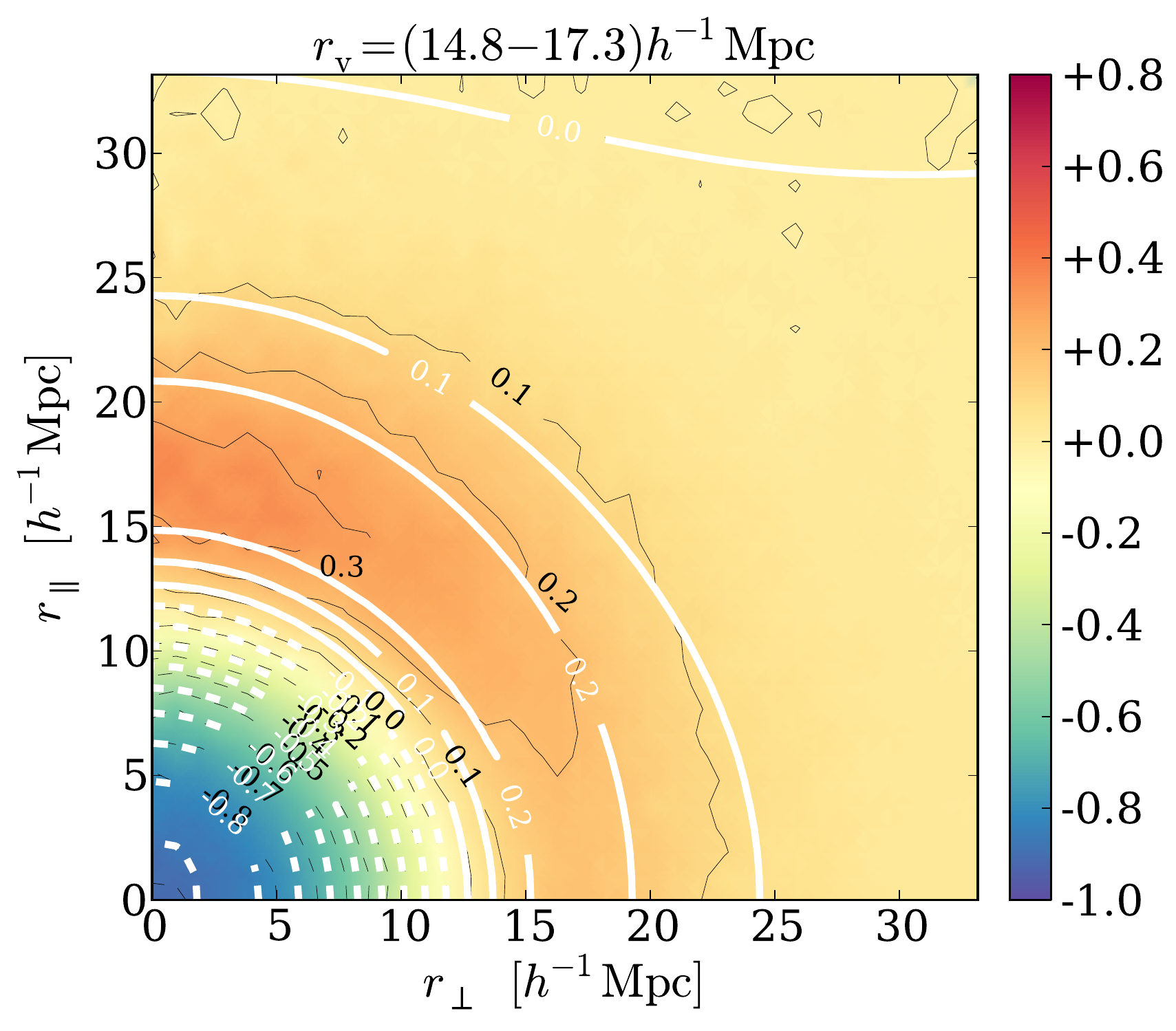}}
\resizebox{.78\hsize}{!}{
\includegraphics[trim=0 9 78 0,clip]{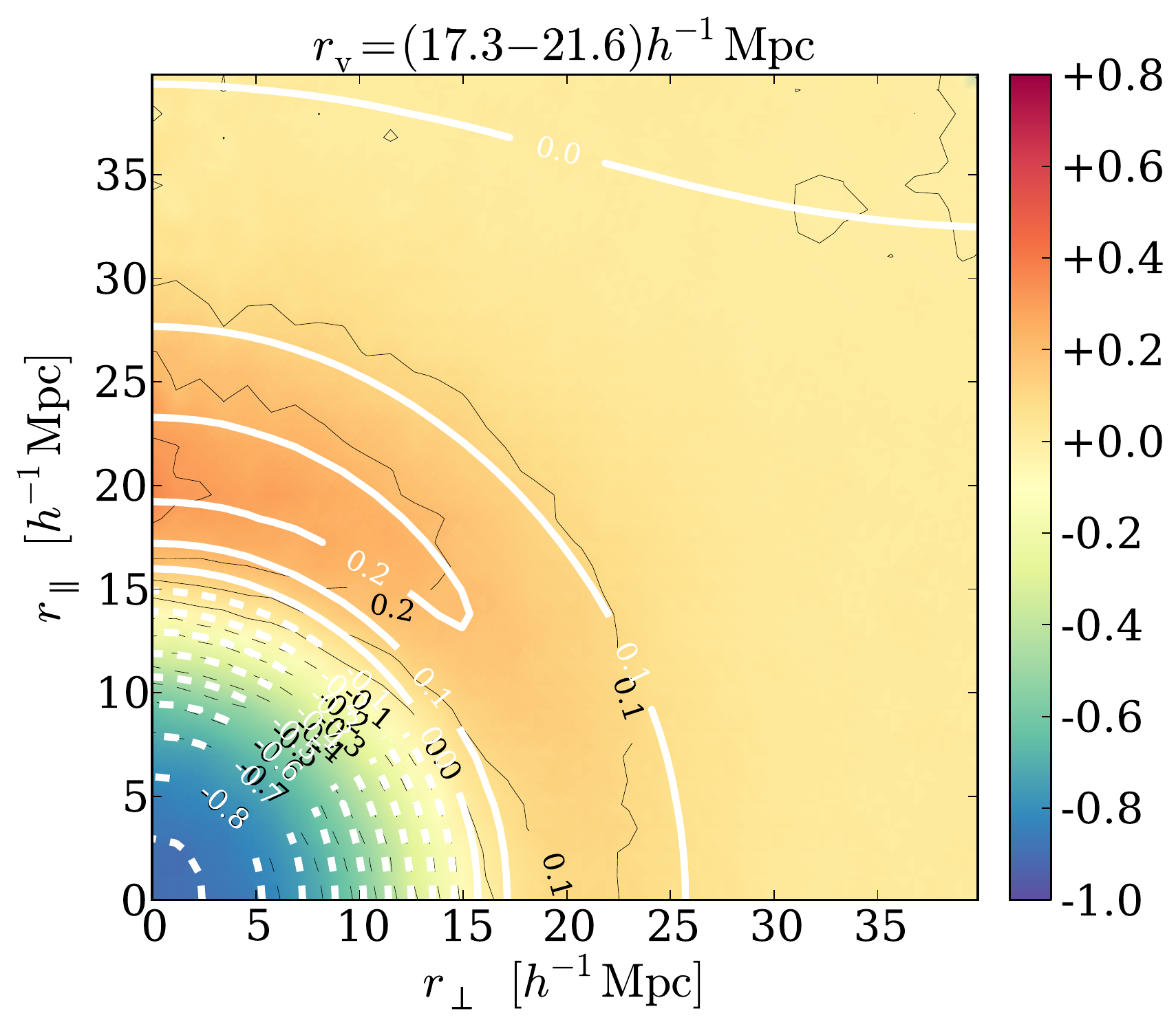}
\includegraphics[trim=40 9 0 0,clip]{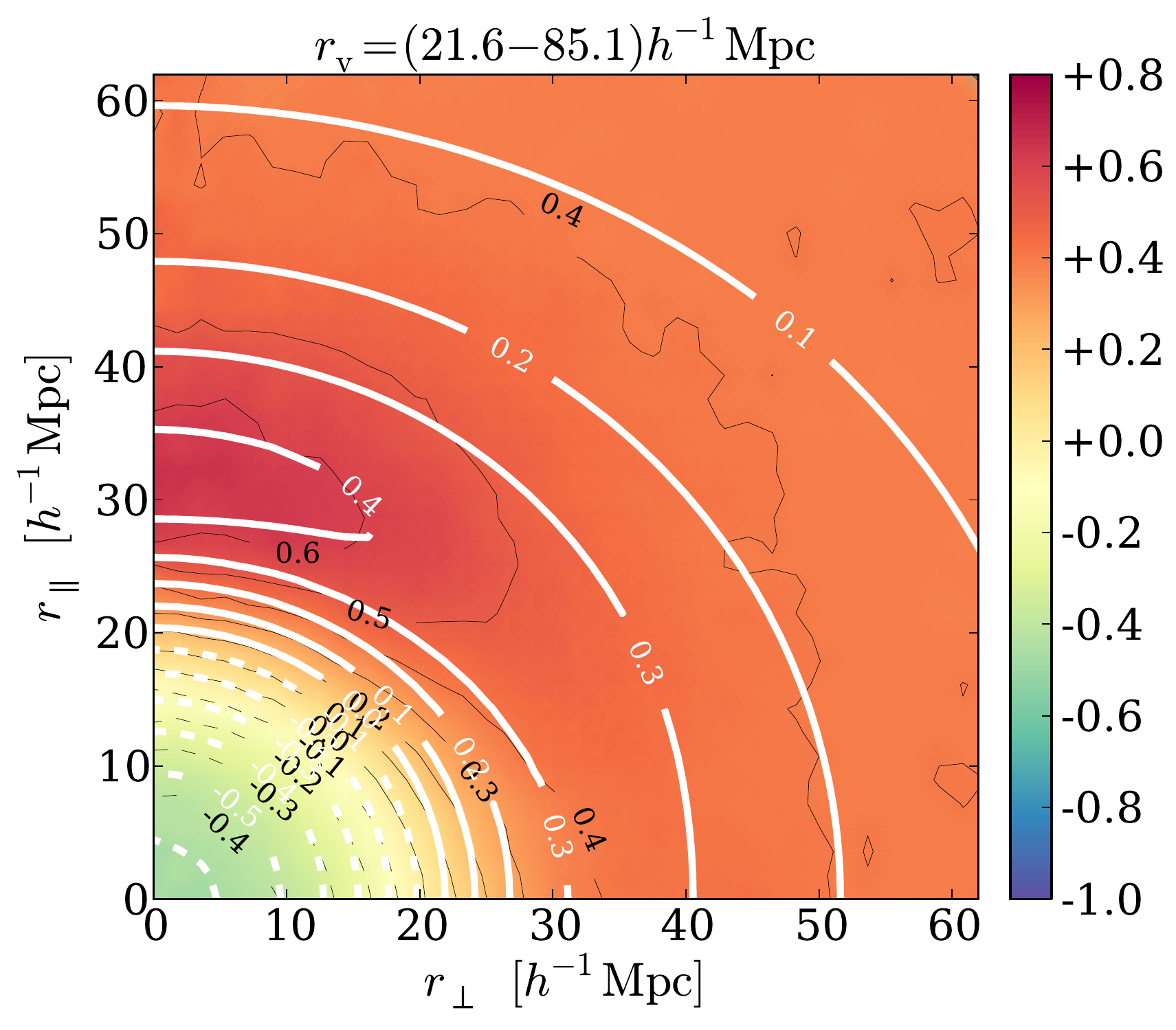}}
\caption{Dense mock-galaxy void stacks $\rho_{\void\tracer}(r_\parallel,r_\perp)/\bar{\rho}_\tracer - 1$ in redshift space at $z=0$. Black solid/dashed lines show positive/negative contours of the data, white lines show the best-fit model.}
\label{Xvt2d}
\end{figure*}

\begin{figure*}[!p]
\centering
\vspace{-38pt}
\resizebox{.79\hsize}{!}{
\includegraphics[trim=0 35 70 0,clip]{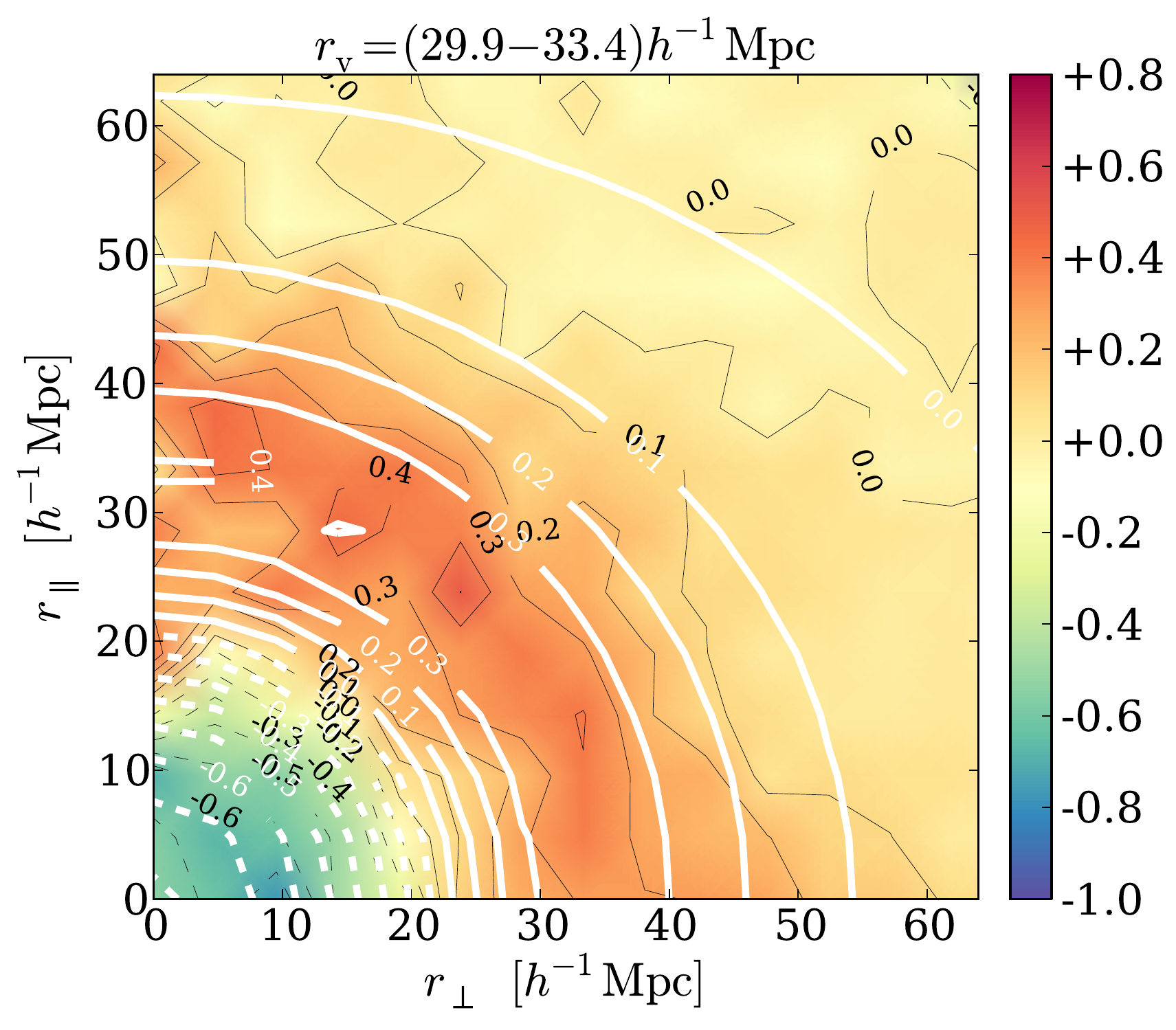}
\includegraphics[trim=40 35 0 0,clip]{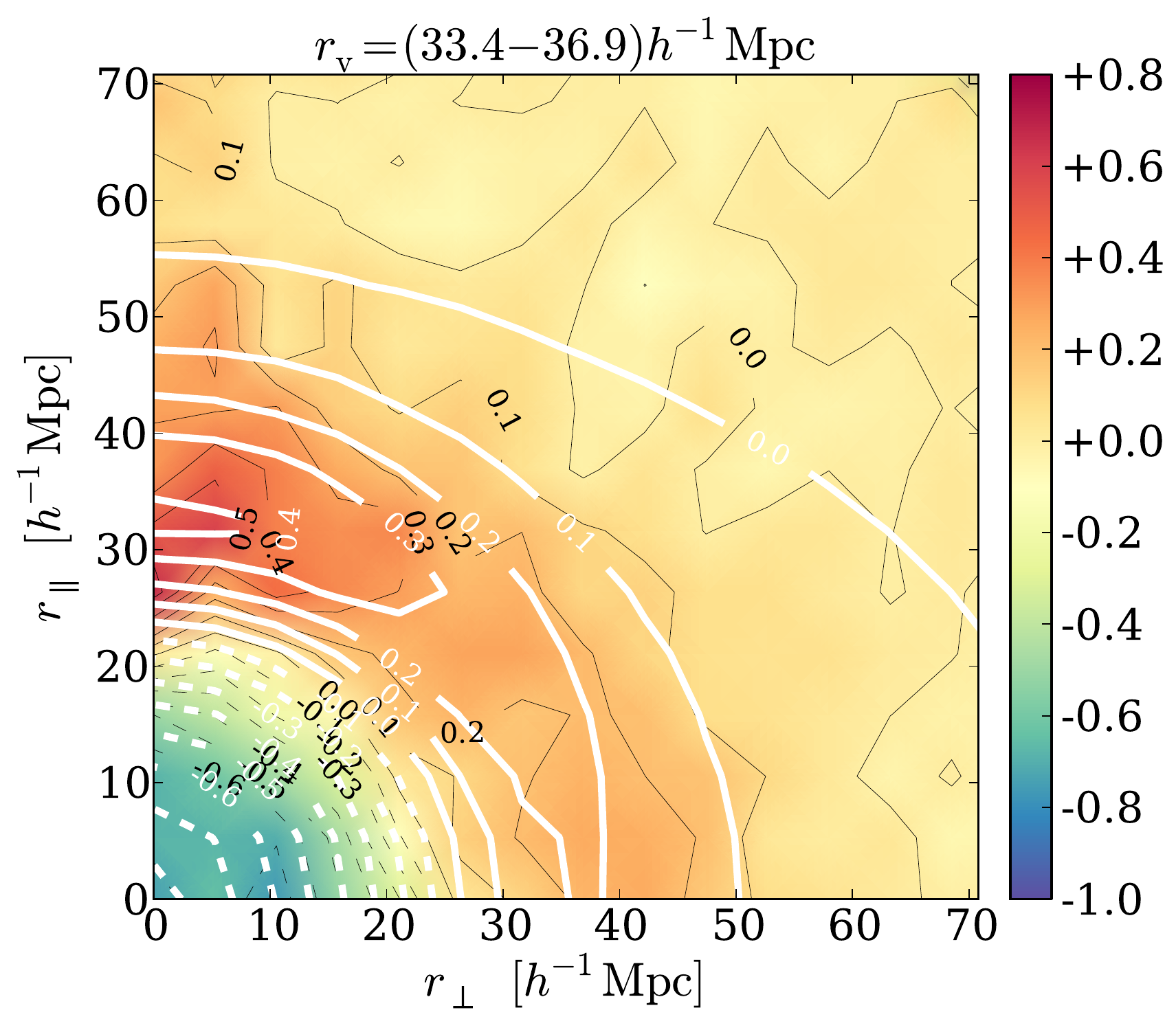}}
\resizebox{.79\hsize}{!}{
\includegraphics[trim=0 35 70 0,clip]{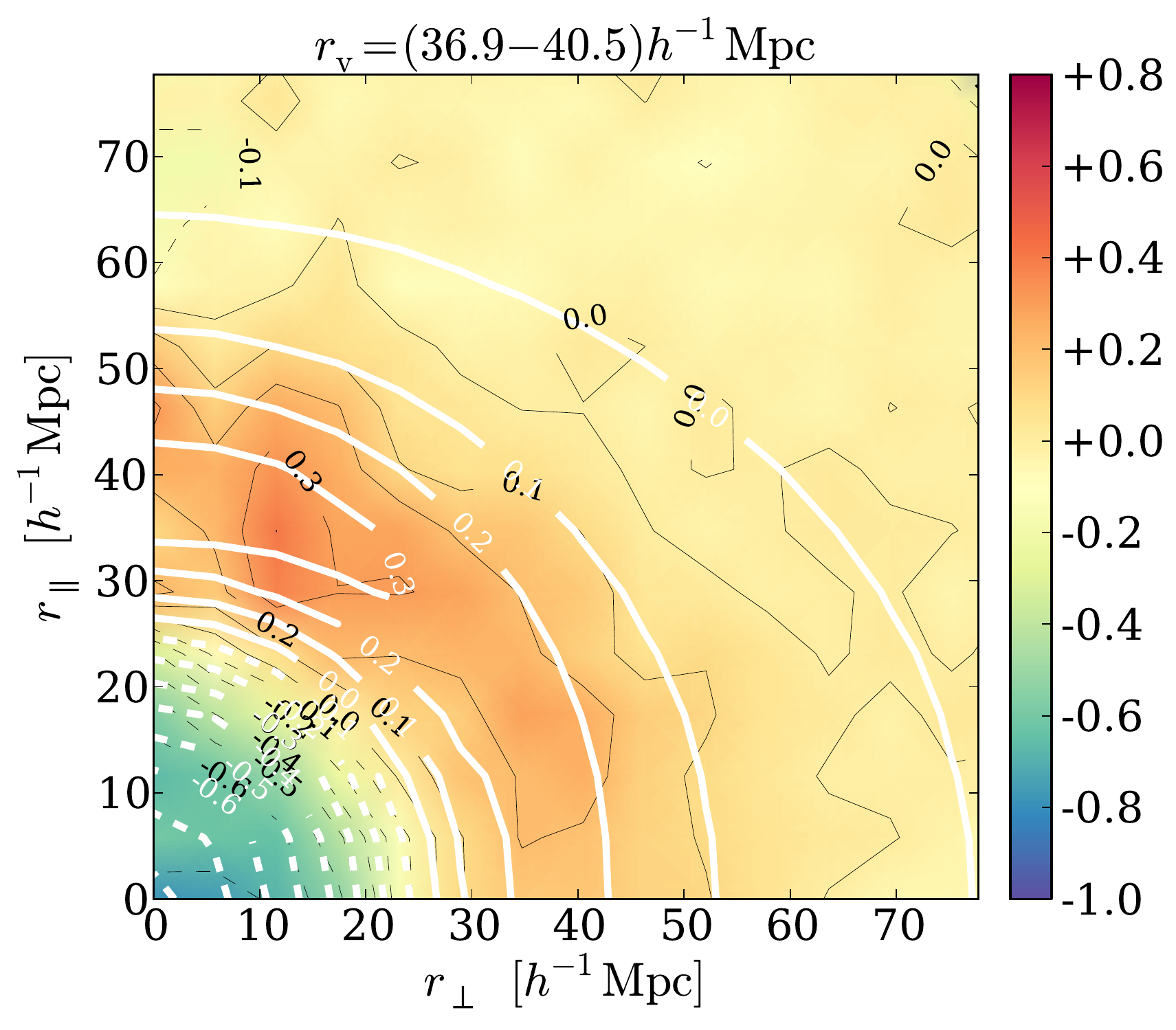}
\includegraphics[trim=40 35 0 0,clip]{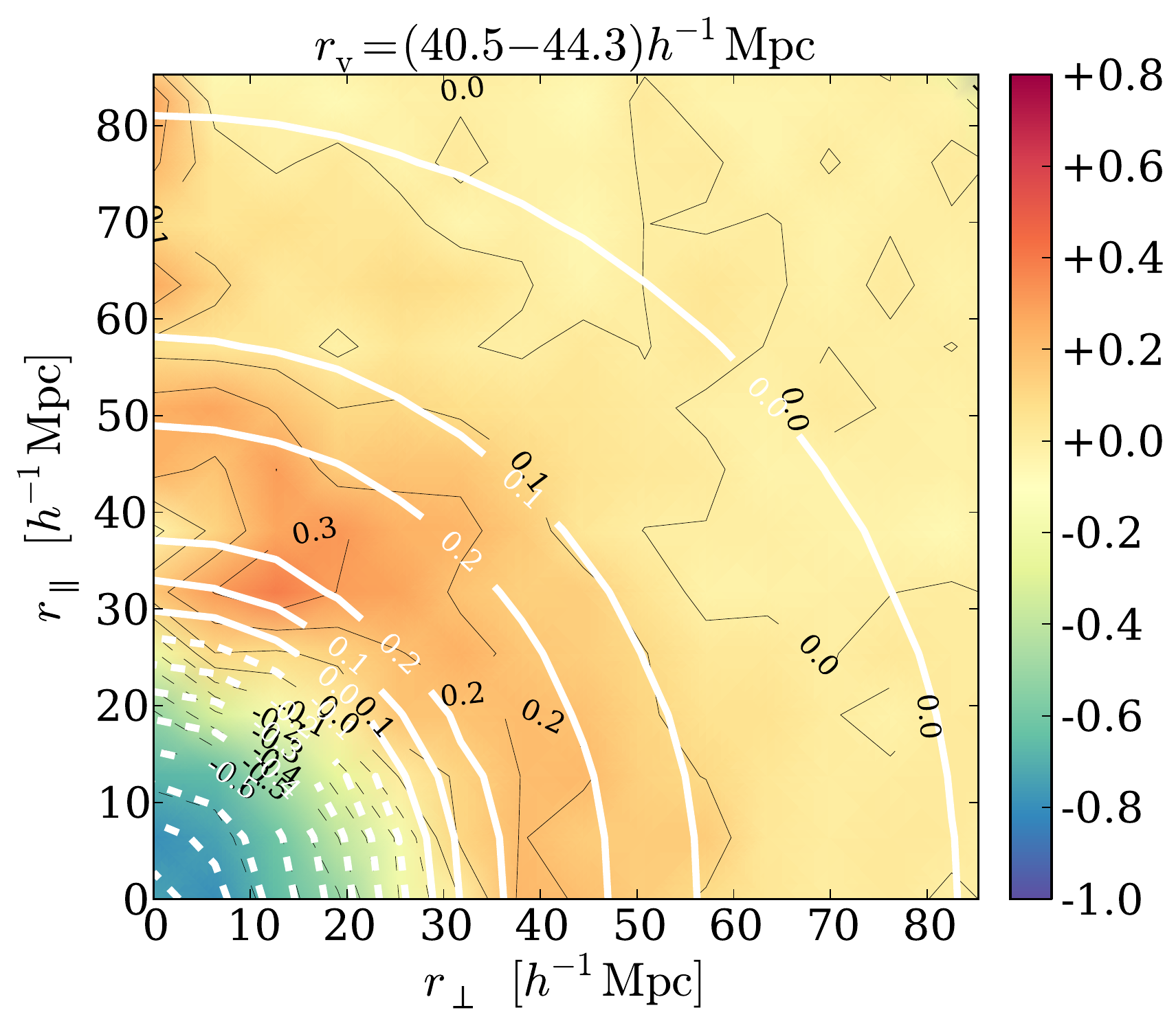}}
\resizebox{.79\hsize}{!}{
\includegraphics[trim=0 35 80 0,clip]{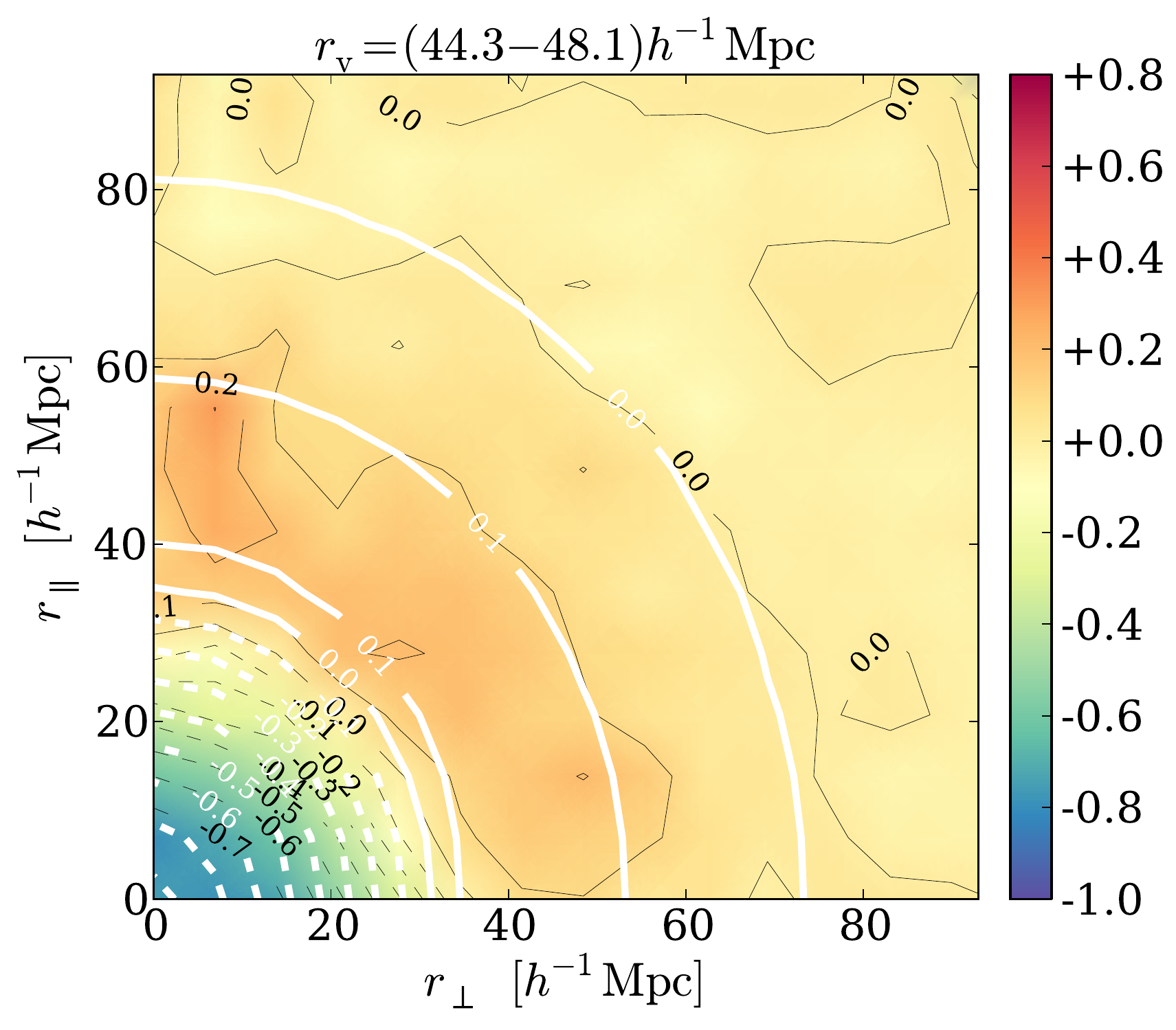}
\includegraphics[trim=40 35 0 0,clip]{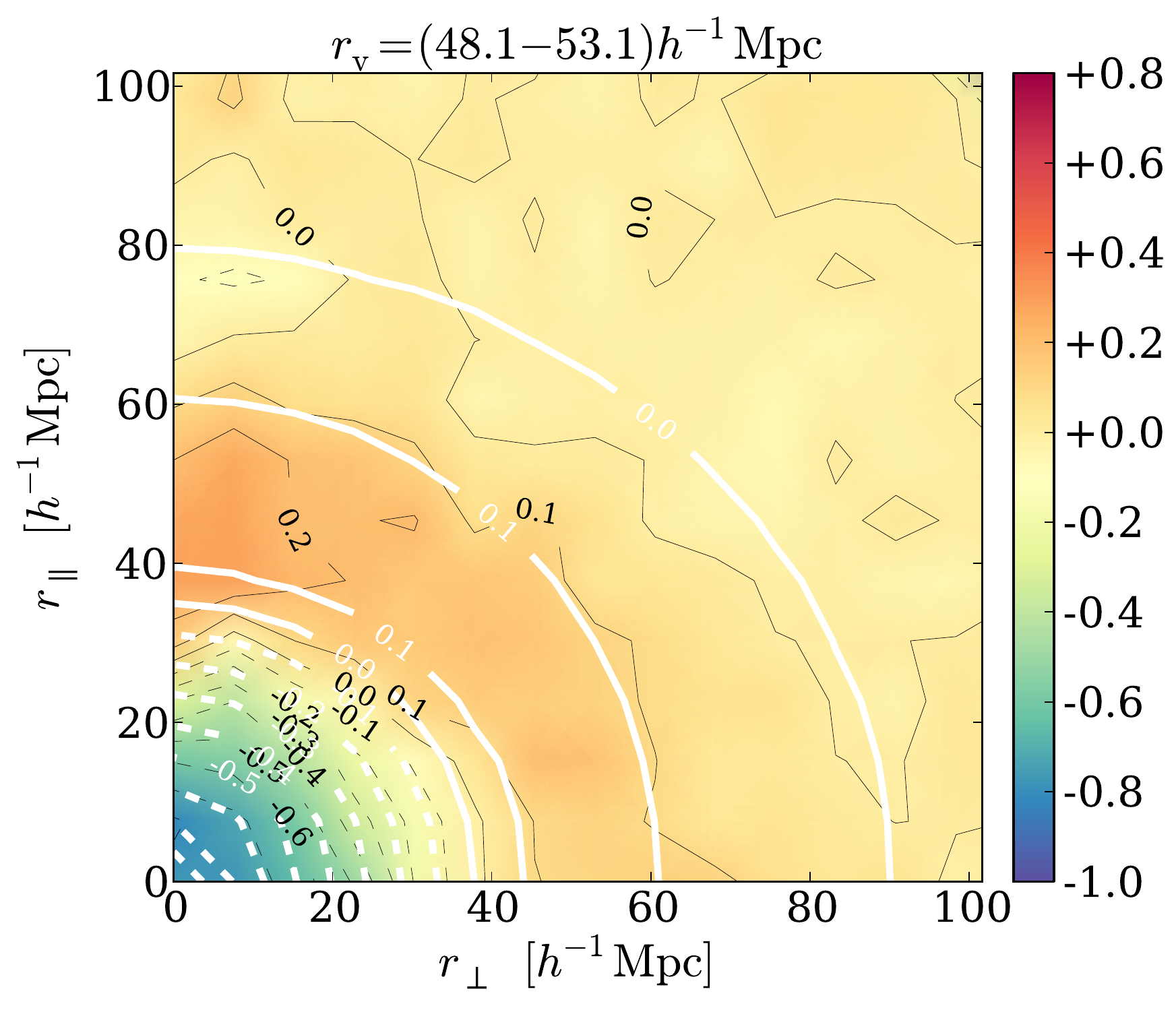}}
\resizebox{.79\hsize}{!}{
\includegraphics[trim=9 9 80 0,clip]{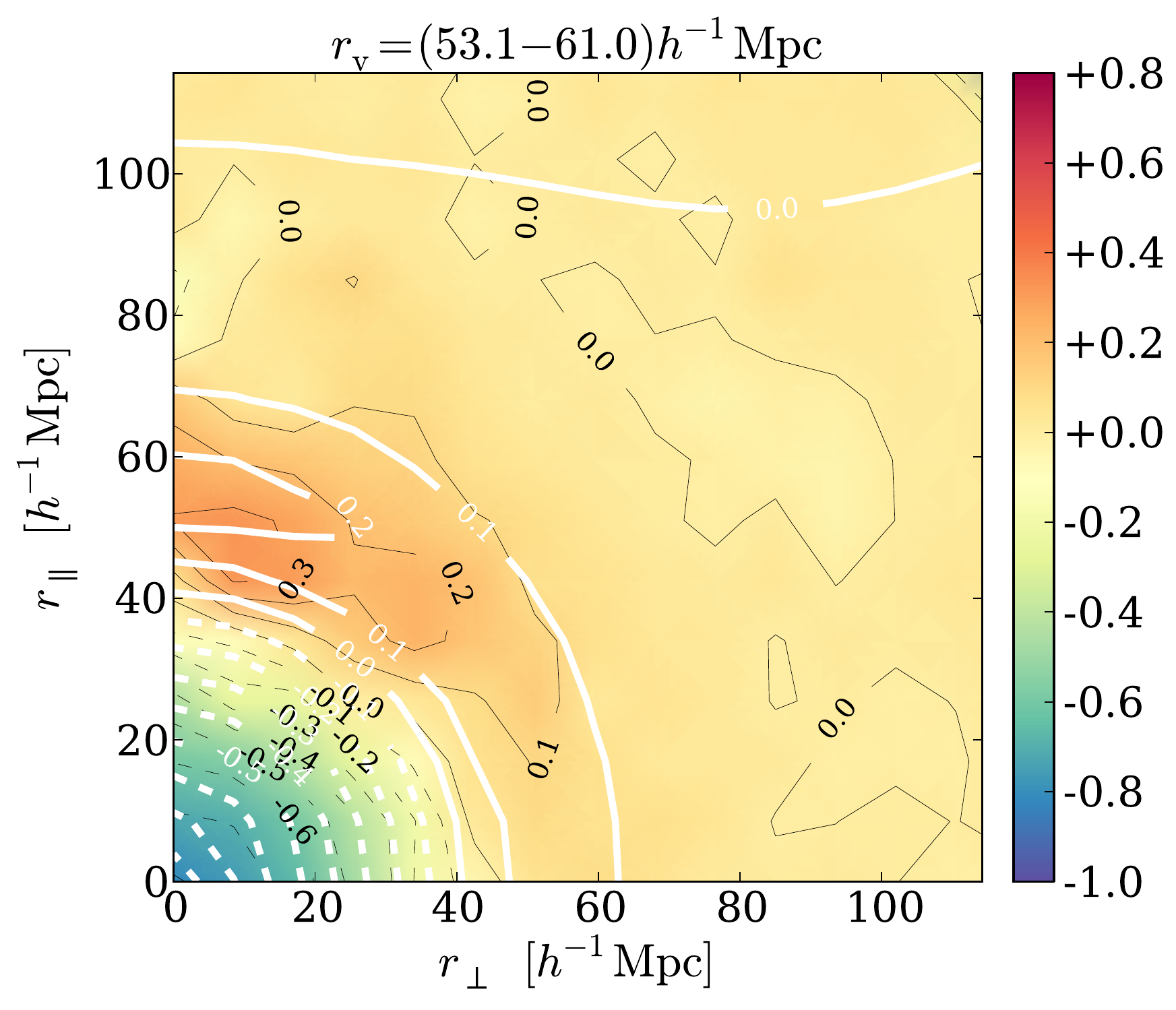}
\includegraphics[trim=40 9 0 0,clip]{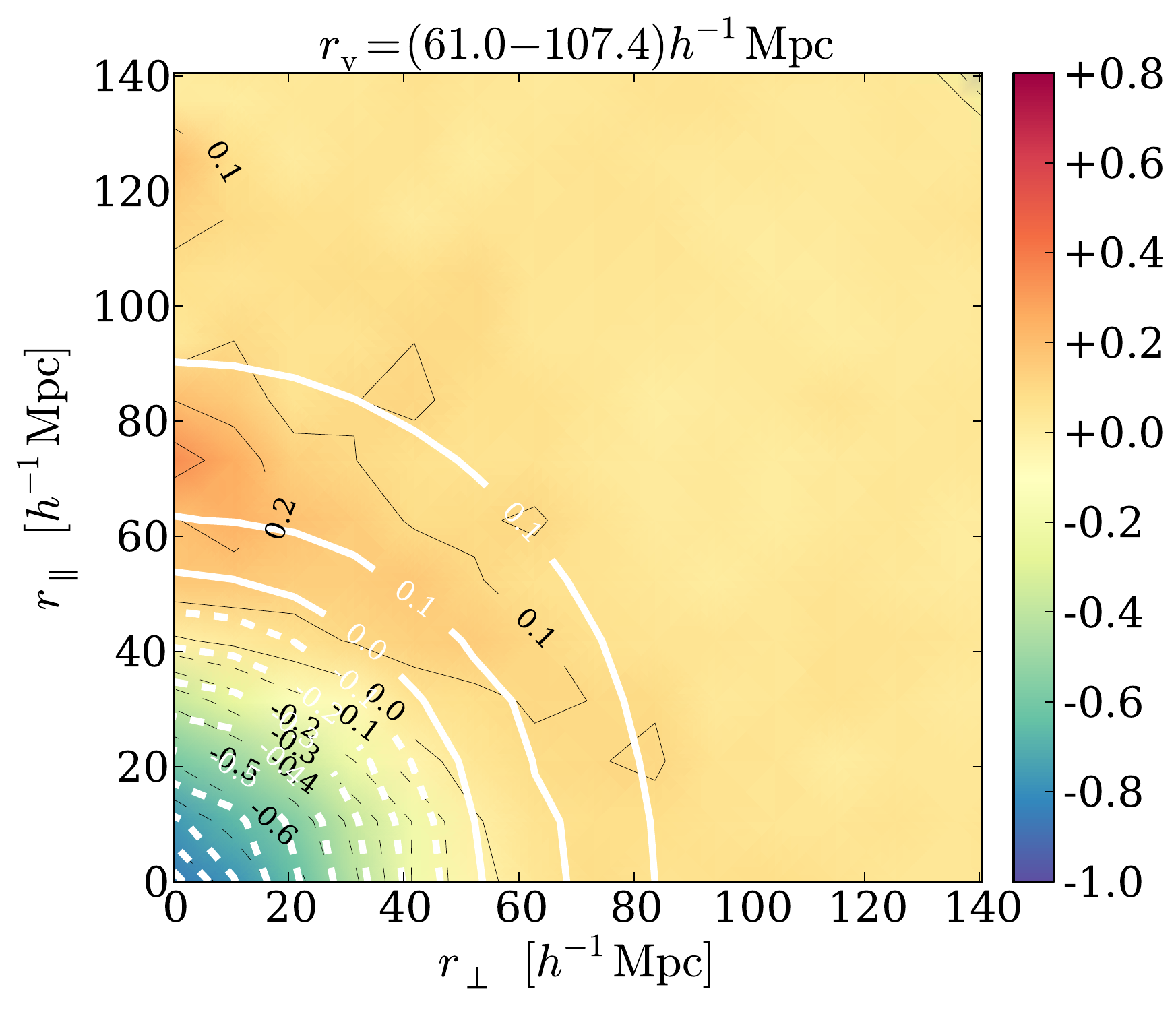}}
\caption{Sparse mock-galaxy void stacks $\rho_{\void\tracer}(r_\parallel,r_\perp)/\bar{\rho}_\tracer - 1$ in redshift space at $z=0.5$. Black solid/dashed lines show positive/negative contours of the data, white lines show the best-fit model.}
\label{Xvg2d}
\end{figure*}

The observationally more relevant scenario is the one where voids are identified from the distribution of galaxies in redshift space. Figure~\ref{Xvt2d} shows the corresponding void stacks in our dense mock-galaxy sample at redshift $z=0$ using a similar binning as before, yielding about 8000 voids per stack. On the one hand we notice the features in these stacks to appear more extreme, both the void ridges and their interiors. This can partly be attributed to the fact that galaxies are biased tracers of the mass, so fluctuations about the mean density are enhanced on void scales~\cite{Sutter2013a,Sutter2014a}. On the other hand, this affects the redshift-space distortions in favor of the Kaiser effect, because galaxies are less sensitive to small-scale velocity dispersion in the large-scale structure as compared to dark matter particles. We can already observe a balance between the FoG and the Kaiser effect in the first bin at $7.4\hMpc<r_\void<8.8\hMpc$. For all larger voids the Kaiser effect dominates and causes an enhanced ridge feature along the line of sight, which can be directly related to the logarithmic growth rate $f$ in eq.~(\ref{v_lin}). For our sparser mock catalog with roughly 250 voids per stack at redshift $z=0.5$ the FoG effect is already negligible in the smallest bin, as apparent from fig.~\ref{Xvg2d}. This is partly due to the fact that we can only resolve voids with $r_\void>30\hMpc$ in this sample, and partly due to the higher redshift. The Kaiser effect can even still be perceived for voids of effective radius $r_\void>100\hMpc$.

The velocity dispersion in the case of dense mock galaxies is shown in the middle panels of fig.~\ref{svm}. Compared to the dark matter shown in the upper panels, the behavior of $\sigma_\parallel(r)$ and $\sigma_\perp(r)$ is almost identical, both in magnitude and shape of these two curves. This is an encouraging result, insofar as it does not suggest any velocity bias between our dense mock galaxies and the underlying mass distribution. Moreover, fitting eq.~(\ref{sv_model}) to this mock-galaxy data yields roughly the same parameter values for $\sigma_v$ and $\omega$ as obtained from the dark matter. For the velocity dispersion around voids in our sparse catalog we observe slightly higher amplitudes and lower widths, as can be seen in the lower panels of fig.~\ref{svm}. This behavior is expected for a more biased sample of galaxies ($b\simeq1.8$), since these specifically trace out higher over-densities of the mass distribution that have evolved more nonlinear.

\subsection{MCMC analysis \label{sec:analysis:MCMC}}
The question of how well our model from sec.~\ref{sec:theory:GSM} can reproduce the data presented in sec.~\ref{sec:analysis:stacks} and how well its model parameters can be constrained will be the topic of this section. For this purpose we implemented a \emph{Markov Chain Monte Carlo} (MCMC) framework making use of the publicly available PyMC code package in Python~\cite{Patil2010}. Following Bayesian statistics, we formulate a likelihood function $\mathcal{L}$ that quantifies the probability of the data, given a model. In our case the data is given by the simulated void-tracer cross-correlation function $\tilde{\xi}_{\void\tracer}(\tilde{r}_\parallel,\tilde{r}_\perp)$ in redshift space, and the model is expressed by eq.~(\ref{xi_s}) with the ingredients from sec.~\ref{sec:theory:GSM}. Our model parameters can be summarized in the following parameter vector,
\begin{equation}
 \boldsymbol{\theta} = (\rs,\dc,\alpha,\beta,\sigma_v,f,\varepsilon)\;,
\end{equation}
with the four void density-profile parameters $\rs,\dc,\alpha,\beta$, the velocity-dispersion amplitude $\sigma_v$, the logarithmic growth rate $f$, and the AP parameter $\varepsilon$. For the width $\omega$ of the velocity dispersion in eq.~(\ref{sv_model}) we use the best-fit values obtained from the data beforehand and do not vary it in the MCMC. We also do not distinguish between the velocity dispersion along or perpendicular to the line of sight, as the two are found to be very similar in fig.~\ref{svm}. Further, we assume a multivariate Gaussian likelihood function for the data given a model of the form
\begin{equation}
 \mathcal{L}(\hat{\xi}_{\void\tracer}|\boldsymbol{\theta}) = (2\pi)^{-N_b^2/2}|\mathbf{C}|^{1/2}\exp\left[-\frac{1}{2}\left(\hat{\xi}_{\void\tracer}-\tilde{\xi}_{\void\tracer}\right)^\intercal\mathbf{C}^{-1}\left(\hat{\xi}_{\void\tracer}-\tilde{\xi}_{\void\tracer}\right)\right]\;, \label{likelihood}
\end{equation}
where $\hat{\xi}_{\void\tracer}$ is the measured void-tracer cross-correlation function in redshift space, $\tilde{\xi}_{\void\tracer}$ the corresponding model given by eq.~(\ref{xi_s}), $N_b$ its number of radial bins per dimension and $\mathbf{C}$ its covariance matrix. Note that the two-dimensional shape of $\xi_{\void\tracer}$ has been collapsed into the form of a vector in this notation, i.e. $\mathbf{C}$ is a matrix of dimensions $N_b^2 \times N_b^2$. We estimate the latter via Jackknife resampling of eq.~(\ref{stack_rho}), where the void stack is calculated $N_\void$ times after removing each individual void once. The resulting covariance matrix is then given by
\begin{equation}
 \hat{\mathbf{C}} = \frac{N_\void-1}{N_\void}\sum_{i=1}^{N_\void}\left(\hat{\xi}^i_{\void\tracer}-\hat{\xi}^*_{\void\tracer}\right)\left(\hat{\xi}^i_{\void\tracer}-\hat{\xi}^*_{\void\tracer}\right)^\intercal\;, \label{covariance}
\end{equation}
where $\hat{\xi}^i_{\void\tracer}$ is the $i$th Jackknife sample and $\hat{\xi}^*_{\void\tracer}=N_\void^{-1}\sum_i\hat{\xi}^i_{\void\tracer}$ the mean of all Jackknife samples. Since $\hat{\mathbf{C}}$ is only a statistical estimator of the true underlying covariance matrix $\mathbf{C}$ from the noisy data, it is subject to some degree of bias and uncertainty. The same is true for the inverse of this matrix, which in some cases may not even exist. A convenient solution to this issue can be achieved via \emph{tapering} of the covariance matrix, where its off-diagonal elements are increasingly diminished the more distant they are from the diagonal. An unbiased estimate of the inverse covariance matrix can be obtained by calculating~\cite{Kaufman2008,Paz2015}
\begin{equation}
 \hat{\mathbf{C}}^{-1} = \left(1-\frac{N_b^2+1}{N_\void-1}\right)\left(\hat{\mathbf{C}}\circ\mathbf{T}\right)^{-1}\circ\mathbf{T}\;, \label{inverse_covariance}
\end{equation}
where $\mathbf{T}$ is the so-called \emph{tapering matrix} and $\circ$ denotes an element-wise product. It can be shown that a specific type of functions yield a result that converges to the maximum-likelihood estimate of $\hat{\mathbf{C}}^{-1}$~\cite{Kaufman2008}. One such possible choice is the family of \emph{Wendland functions} $\psi_{d,k}(r)$~\cite{Wendland1995}, which in the case of $d=3$ dimensions and $k=1$ reads
\begin{equation}
 \psi_{3,1}(r) = \max\left(1-r, 0\right)^4\left(1+4r\right)\;.
\end{equation}
The elements of the tapering matrix can then be defined as $T_{ij} = \psi_{3,1}(\left|\mathbf{r}_i-\mathbf{r}_j\right|/r_t)$, where $\mathbf{r}=\left(r_\parallel,r_\perp\right)^\intercal$ and $r_t$ denotes the \emph{tapering scale}. Paz and S\'anchez~\cite{Paz2015} have investigated the covariance tapering method on BAO measurements with galaxy auto-correlations, and found that it yields more accurate (and in particular lower) uncertainties in the model parameters than the standard methods. In this case they report an optimal tapering scale of $r_t\simeq230\hMpc$, but point out that this value depends on the detailed structure of the covariance matrix. We have implemented the same value in our analysis and find it to yield reasonable results for void-galaxy cross-correlations, although it might not be the optimal choice. However, given this value to lie well within the linear regime of large-scale structure, where the two types of correlations are proportional to each other~\cite{Hamaus2014a}, this choice seems reasonable. The following uniform prior ranges for the values of our parameters are imposed,
\begin{eqnarray}
 &\rs\in[0.5,1.3]\bar{r}_\void\;,\quad \dc\in[-1,-0.6]\;,\quad \alpha\in[0,3]\;,\quad \beta\in[5,13]\;,& \nonumber \\
 &\sigma_v\in[0,600]\frac{\mathrm{km}}{\mathrm{s}}\;,\quad f\in[0,2]\;,\quad \varepsilon\in[0.7,1.3]\;.&
\end{eqnarray}
The selection of these ranges is partly motivated by earlier numerical studies of the void-density profile from eq.~(\ref{HSW})~\cite{Hamaus2014b}, and previous testing of our inference framework.

Eventually, we explore the full parameter space with a \emph{Metropolis-Hastings} algorithm provided in PyMC, and thereby sample the posterior probability distribution $\mathcal{P}(\boldsymbol{\theta}|\tilde{\xi}_{\void\tracer})$ of our model given the data with $\mathcal{O}(10^6)$ samples for each void stack separately. We also determine the maximum-likelihood model for each stack, which is represented by the white contours in figs.~\ref{Xvm2d}, \ref{Xvt2d} and \ref{Xvg2d}. As apparent from comparing the contour levels in these figures, the Gaussian streaming model yields an excellent fit to the data from the smallest to the largest voids in both dark matter and mock galaxies. In the following subsections, we will investigate the corresponding best-fit parameters and their uncertainties in more detail with the help of their full posterior distribution that resulted from our MCMC runs.

\subsubsection{Dark matter \label{sec:analysis:MCMC:matter}}
\begin{figure*}[!t]
\centering
\resizebox{\hsize}{!}{
\includegraphics[trim=0 0 3 0,clip]{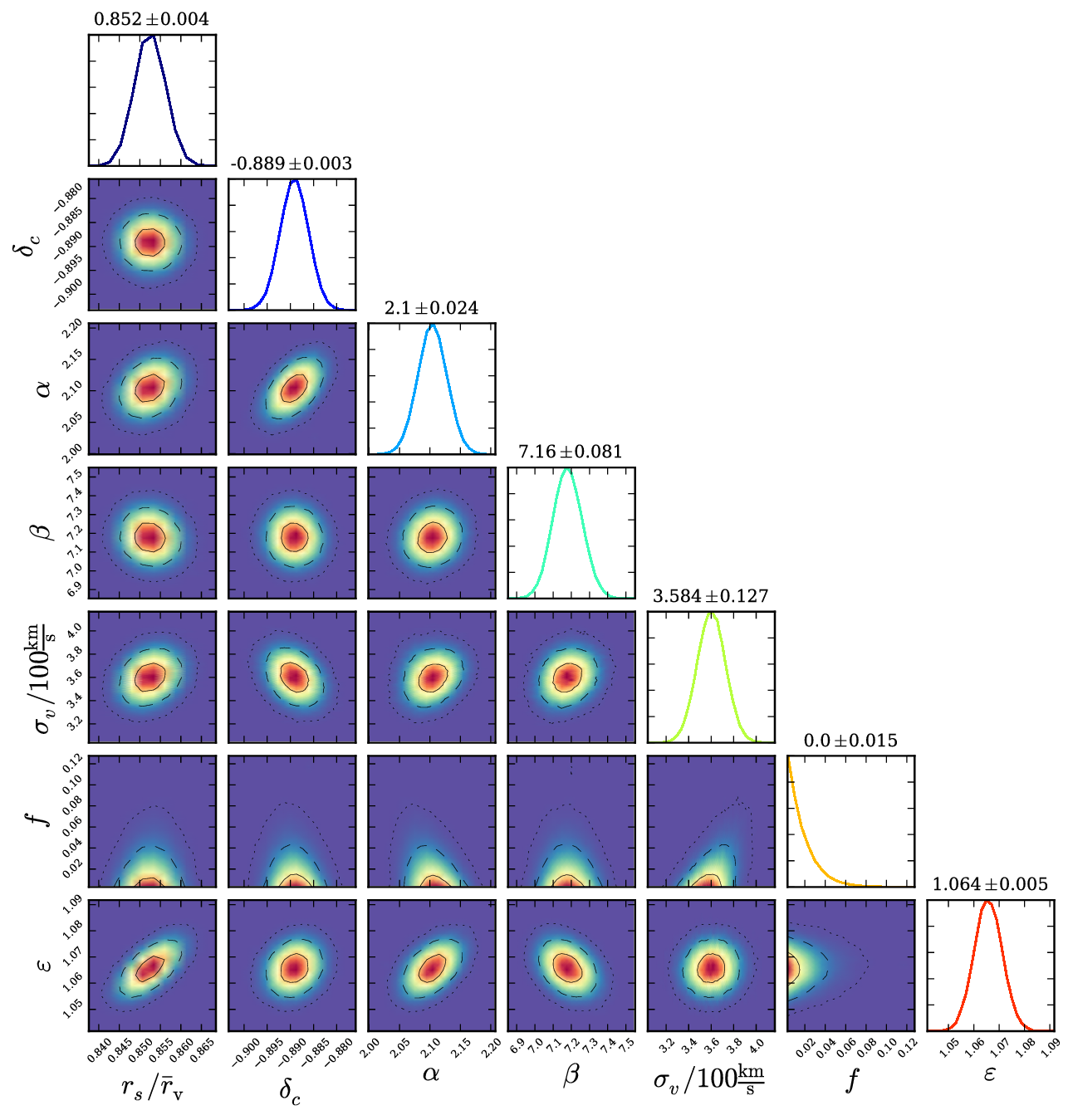}}
\caption{Posterior distribution of the model parameters for a dark matter void stack with effective radius range $r_\void=(13.1-14.9)\hMpc$ at $z=0$. Solid, dashed and dotted contours show the $68.3\%$, $95.5\%$ and $99.7\%$ confidence levels, respectively. Parameter values at the maximum of the marginal posterior distribution and its standard deviation are given at the top of each column.}
\label{pdfm}
\end{figure*}

Let us begin by investigating the full posterior parameter distribution for one of our dark matter void stacks of radius range $r_\void=(13.1-14.9)\hMpc$ shown in fig.~\ref{pdfm}. First of all, the void density-profile parameters $(\rs,\dc,\alpha,\beta)$ are very consistent with what has been found in a dedicated one-dimensional analysis in ref.~\cite{Hamaus2014b}. The constraints on these parameters are rather tight, showing no severe degeneracies amongst each other. Similarly, the posterior for the cosmological parameter set $(\sigma_v,f,\varepsilon)$ exhibits fairly symmetric and tight contours without strong degeneracies between any parameter pair. In particular, the preferred values for $\sigma_v\sim300\frac{\mathrm{km}}{\mathrm{s}}$ are roughly consistent with what we measured directly from the velocities of the dark matter particles in our simulation as shown in fig.~\ref{svm}. Nevertheless, the logarithmic growth rate is not consistent with the expected value from general relativity, $f=\Omega_\matter^{0.55}(z=0)\simeq0.53$, instead its posterior peaks at a value of zero. As already observed in fig.~\ref{Xvm2d}, the FoG effect dominates for void stacks in this radius range, leaving the Kaiser effect sub-dominant and hard to detect. On the other hand, the AP parameter more closely agrees with its fiducial value of $\varepsilon=1$ with a slight bias of about $6\%$. Hence, the $15\%$ systematic error in the latter, as originally reported by Lavaux and Wandelt~\cite{Lavaux2012}, can be reduced significantly when peculiar velocities are taken into account. Given the small relative $1\sigma$-uncertainty of only $0.5\%$ in this measurement, this serves as a useful consistency check that our model is able to treat all anisotropic distortions accurately enough to allow conducting an AP test on the few-percent level. However, we realize that the accuracy provided by linear theory is not sufficient to yield unbiased parameter constraints of the model.

Finally, it is interesting to note that the covariance between the two parameter sets discussed above is quite limited as well. This is expected, as the void density-profile parameters in eq.~(\ref{HSW}) only influence the isotropic part of the redshift-space correlation function, whereas the cosmological parameters modify its anisotropic behavior. In principle, this allows to separate the inference process into two independent steps that can be iterated one after each other to find the optimal complete parameter set. 

\begin{figure*}[!p]
\centering
\resizebox{.75\hsize}{!}{
\includegraphics[trim=0 0 0 0,clip]{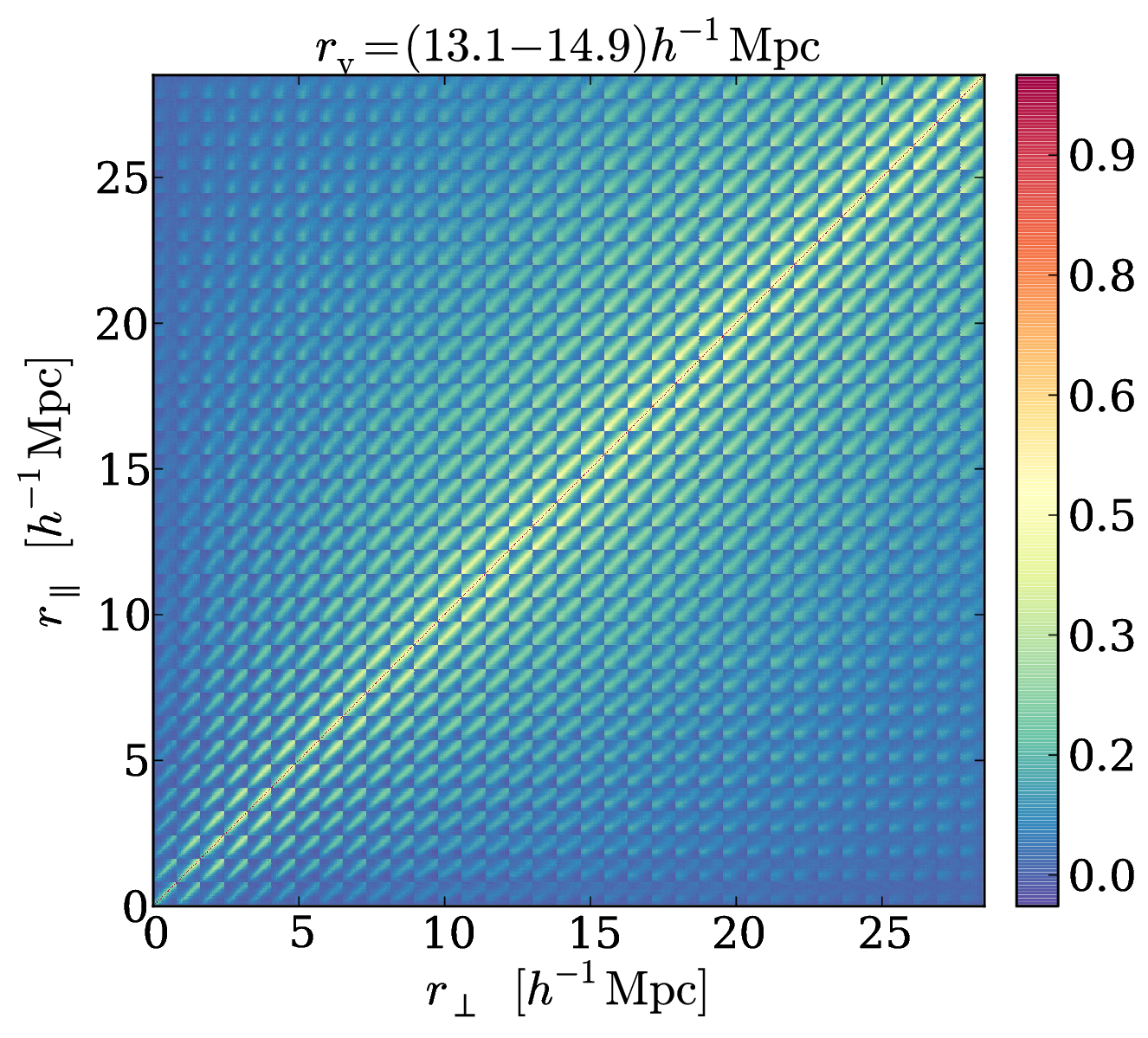}}
\resizebox{.75\hsize}{!}{
\includegraphics[trim=0 0 0 0,clip]{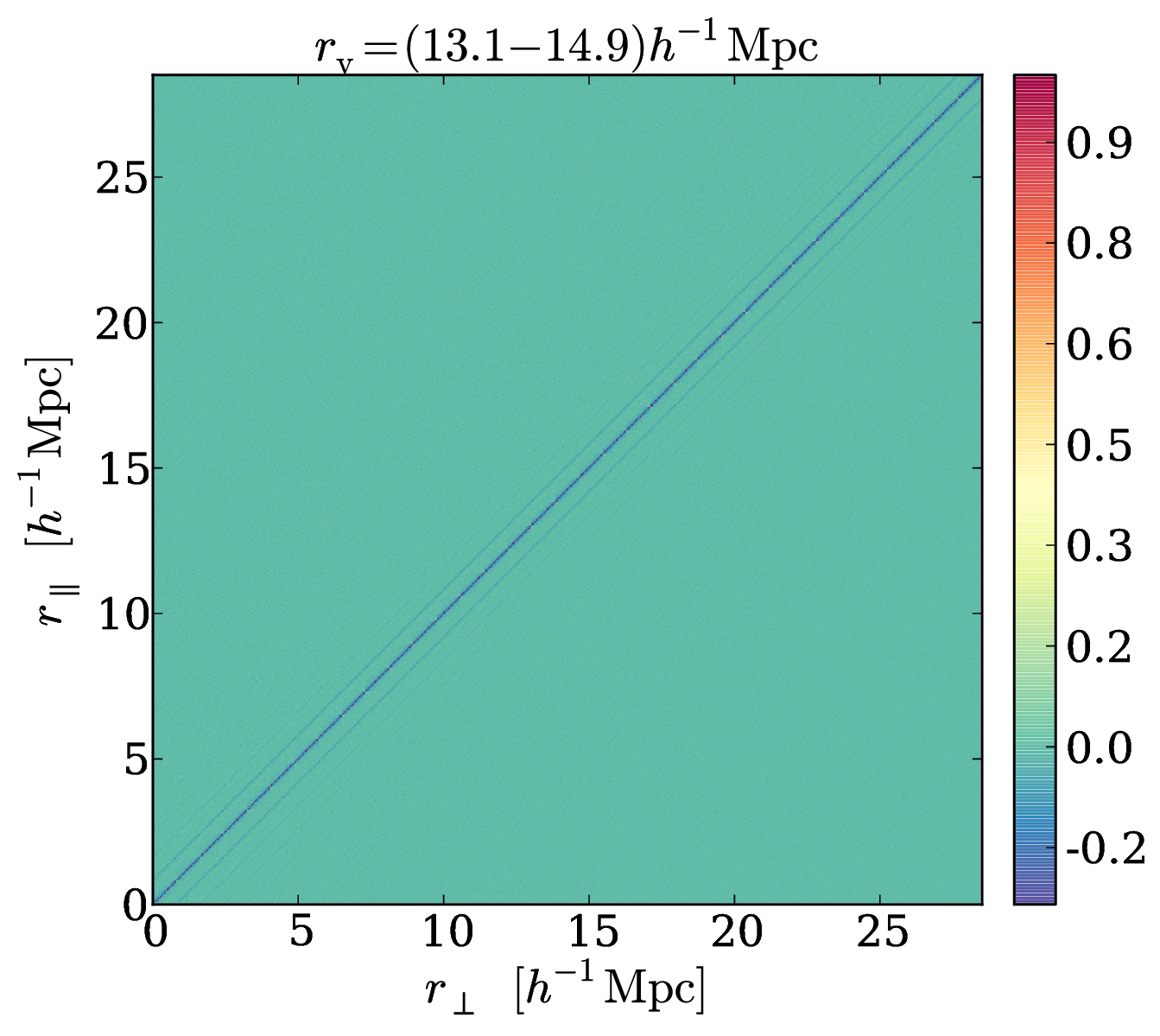}}
\caption{Covariance matrix (top) and its inverse (bottom) of a dark matter void stack with radius range $r_\void=(13.1-14.9)\hMpc$ at $z=0$. Each matrix is normalized by its diagonal, the lower triangular parts have been obtained using the tapering technique as described in the text.}
\label{Cvm}
\end{figure*}

The covariance matrix of the data is shown in the upper panel of fig.~\ref{Cvm}, normalized by its diagonal, which yields the correlation matrix $C_{ij}/\sqrt{C_{ii}C_{jj}}$. Because each two-dimensional void stack is represented by $N_b\times N_b$ bins, its covariance matrix exhibits $N_b^2\times N_b^2$ pixels, which can be thought of as a $N_b\times N_b$ matrix of $N_b^2$ sub-matrices with the same dimensions. Since the covariance matrix is symmetric by definition, in the upper triangular part we plot the Jackknife estimate from eq.~(\ref{covariance}), while in the lower triangle we show its tapered version $C_{ij}T_{ij}/\sqrt{C_{ii}C_{jj}}$. It can be seen that the tapering has almost no effect on these scales, only for the very largest voids in our sample it significantly downsizes the far off-diagonal elements of the covariance matrix. The lower panel of fig.~\ref{Cvm} depicts the inverse covariance matrix, again normalized by its diagonal. While the lower triangular part was obtained according to eq.~(\ref{inverse_covariance}), the upper part shows the direct inversion of eq.~(\ref{covariance}). Again, for the scales shown here and the high sampling density, the differences are marginal.

\subsubsection{Mock galaxies \label{sec:analysis:MCMC:galaxies}}
\begin{figure*}[!t]
\centering
\resizebox{\hsize}{!}{
\includegraphics[trim=0 0 0 0,clip]{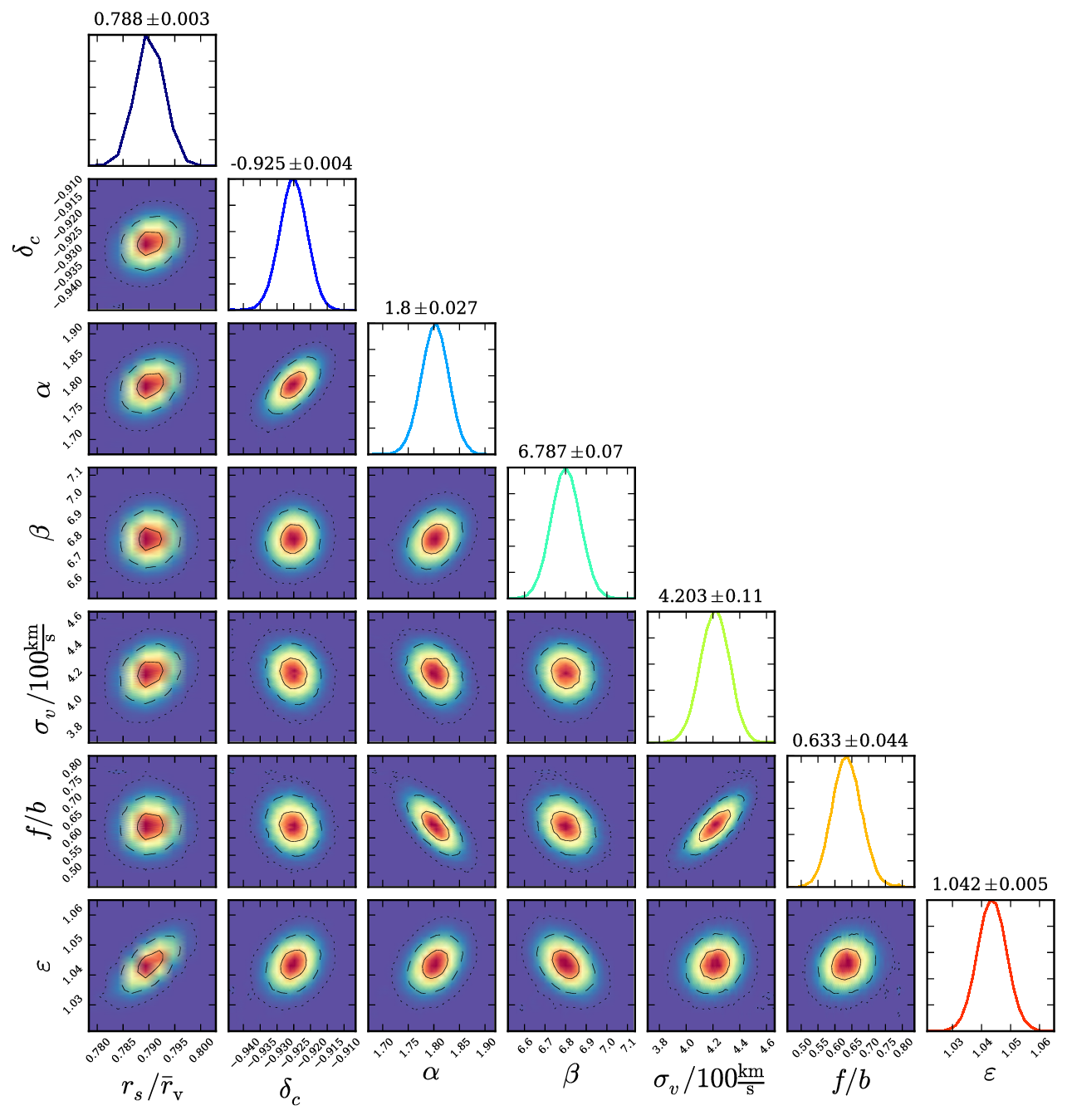}}
\caption{Posterior distribution of the model parameters for a dense mock-galaxy void stack with effective radius range $r_\void=(13.0-14.8)\hMpc$ at $z=0$. Solid, dashed and dotted contours show the $68.3\%$, $95.5\%$ and $99.7\%$ confidence levels, respectively. Parameter values at the maximum of the marginal posterior distribution and its standard deviation are given at the top of each column.}
\label{pdft}
\end{figure*}

The posterior parameter distribution obtained from our dense mock-galaxy void stack with roughly the same range of effective radii $r_\void=(13.0-14.8)\hMpc$ as for the dark matter voids above is presented in fig.~\ref{pdft}. First of all we notice a very similar precision on all parameter constraints. Moreover, there is a slight decrease in $\rs$, $\dc$, $\alpha$ and $\beta$, resulting in a steeper void density-profile with a more pronounced ridge feature. This confirms earlier studies~\cite{Sutter2013a,Sutter2014a} comparing void profiles in different tracer distributions. It can be explained by the biased clustering statistics of tracers of the mass distribution, resulting in an increased contrast in the density fluctuation of galaxies around voids.

By examining the difference between figs.~\ref{pdfm} and \ref{pdft} one can also perceive an improvement in the inferred value of $\varepsilon$, which is closer to its fiducial value in the dense mock-galaxy sample. While the constraining power on $f$ is degraded, its maximum-likelihood value significantly increases to a non-zero value. In fact, in the case of mock galaxies we are effectively constraining the parameter ratio $f/b$, which is often denoted $\beta$ in the literature. This can directly be seen when plugging eq.~(\ref{bias}) into eq.~(\ref{v_lin}). The linear bias parameter of our dense mock-galaxy sample amounts to $b\simeq0.8$, which explains the relatively high value of the effective growth rate $f/b\simeq0.67$ here, a value which is perfectly consistent with the inferred posterior distribution within $1\sigma$.

\begin{figure*}[!t]
\centering
\resizebox{\hsize}{!}{
\includegraphics[trim=-2 -2 0 0,clip]{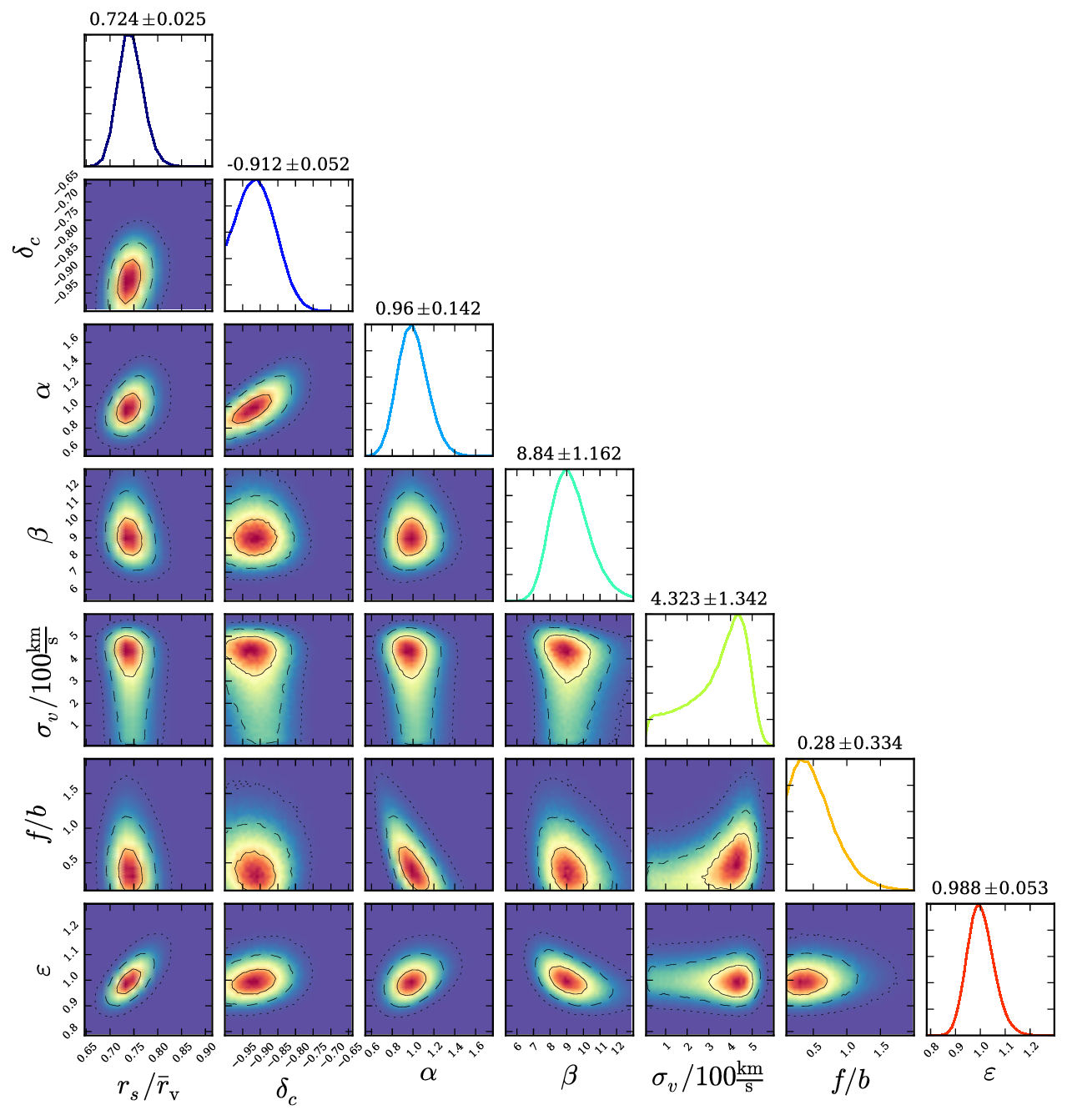}}
\caption{Posterior distribution of the model parameters for a sparse mock-galaxy void stack with effective radius range $r_\void=(48.1-53.1)\hMpc$ at $z=0.5$. Solid, dashed and dotted contours show the $68.3\%$, $95.5\%$ and $99.7\%$ confidence levels, respectively. Parameter values at the maximum of the marginal posterior distribution and its standard deviation are given at the top of each column.}
\label{pdfg}
\end{figure*}

\begin{figure*}[!p]
\centering
\resizebox{.75\hsize}{!}{
\includegraphics[trim=0 0 0 0,clip]{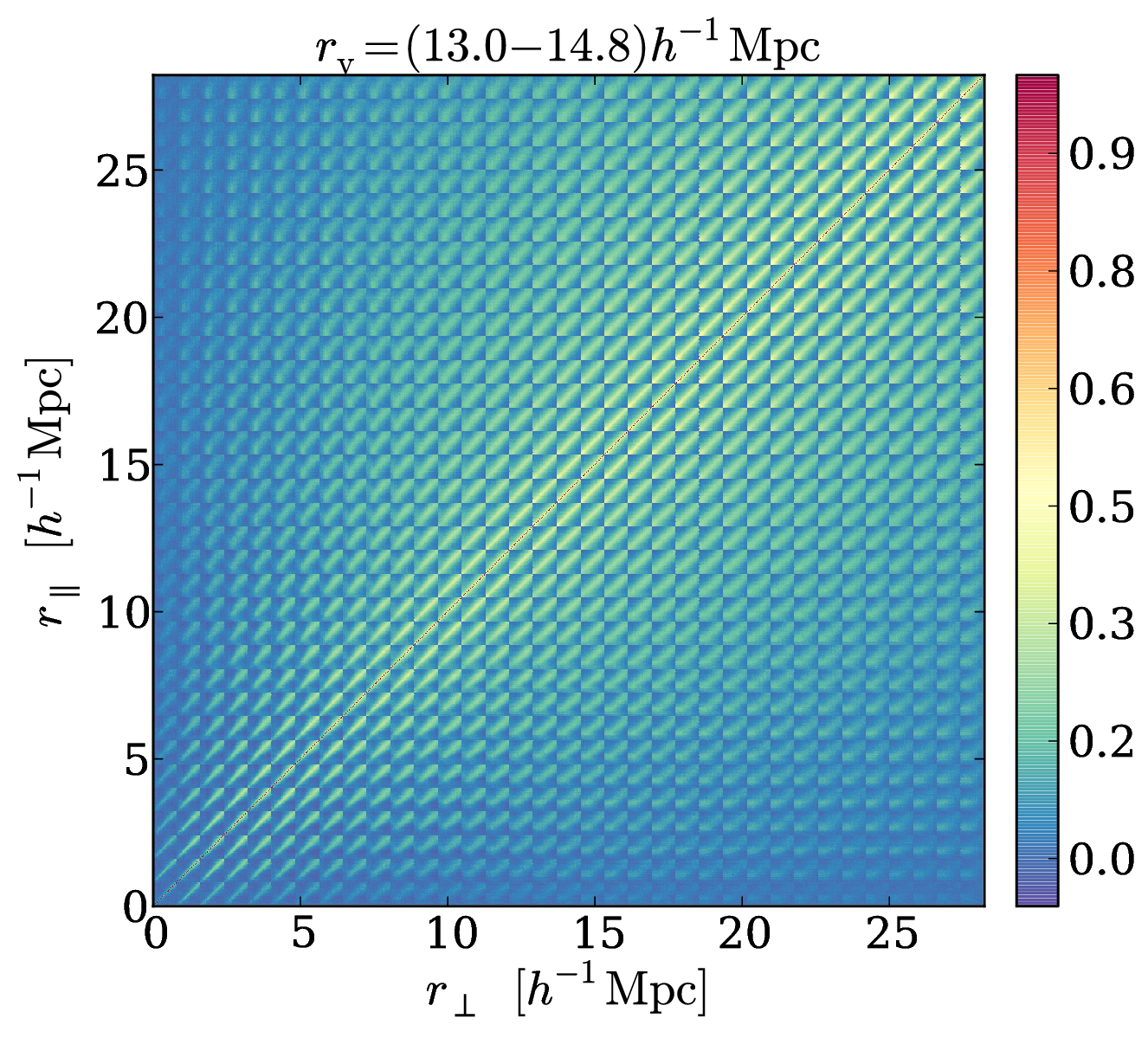}}
\resizebox{.75\hsize}{!}{
\includegraphics[trim=0 0 0 0,clip]{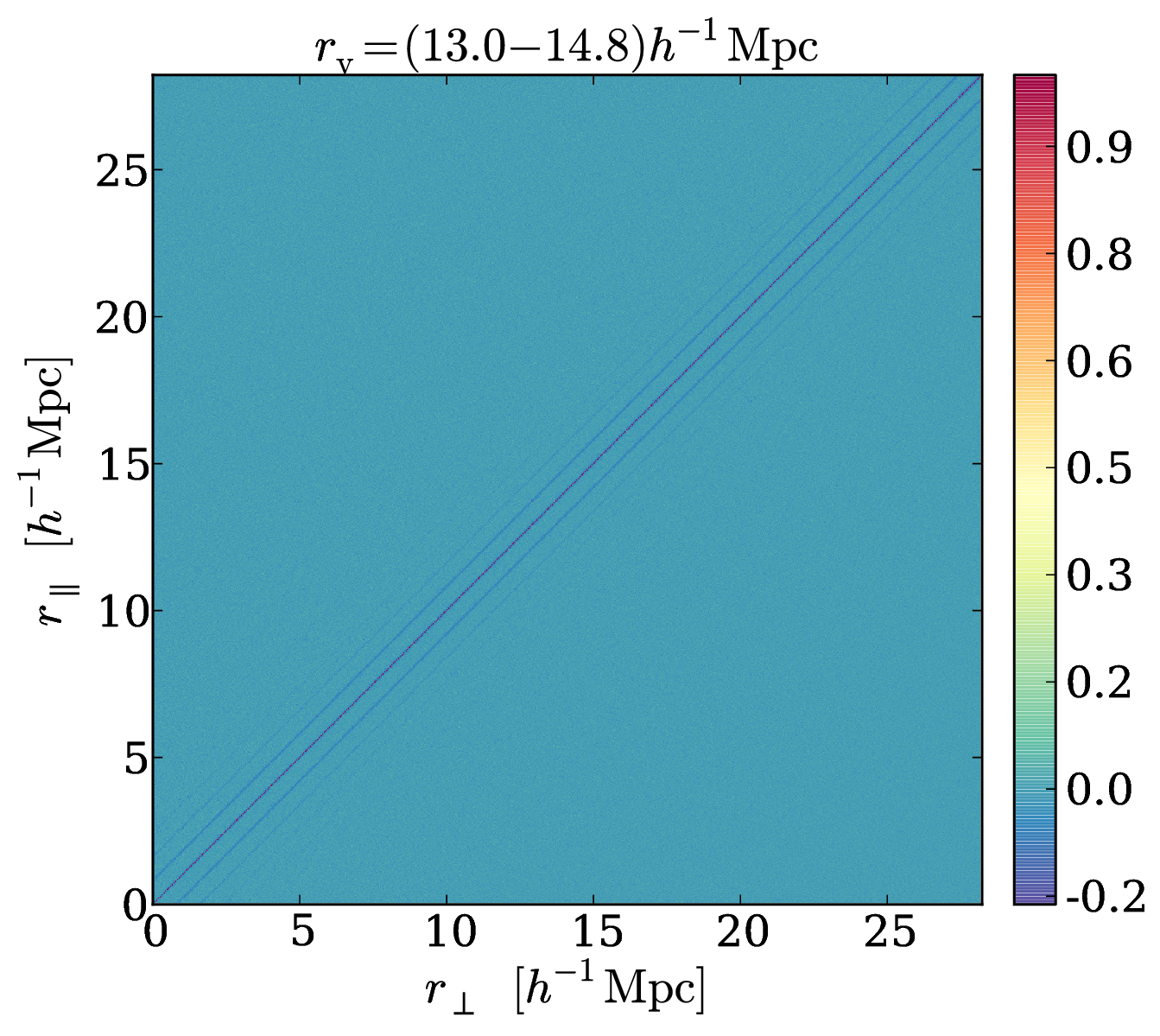}}
\caption{Covariance matrix (top) and its inverse (bottom) of a dense mock-galaxy void stack with radius range $r_\void=(13.0-14.8)\hMpc$ at $z=0$. Each matrix is normalized by its diagonal, the lower triangular parts have been obtained using the tapering technique as described in the text.}
\label{Cvt}
\end{figure*}

\begin{figure*}[!p]
\centering
\resizebox{.75\hsize}{!}{
\includegraphics[trim=0 0 0 0,clip]{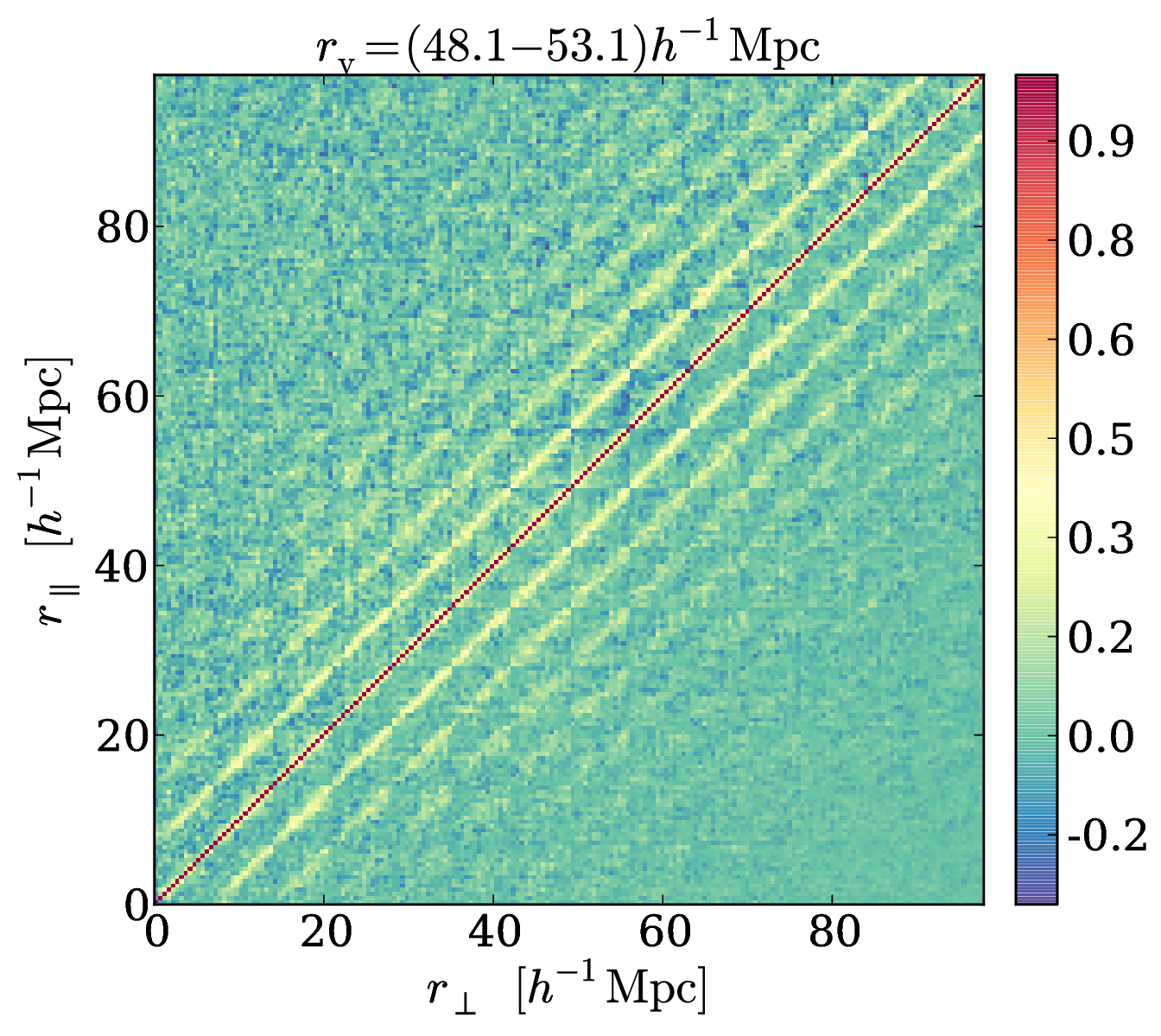}}
\resizebox{.75\hsize}{!}{
\includegraphics[trim=0 0 0 0,clip]{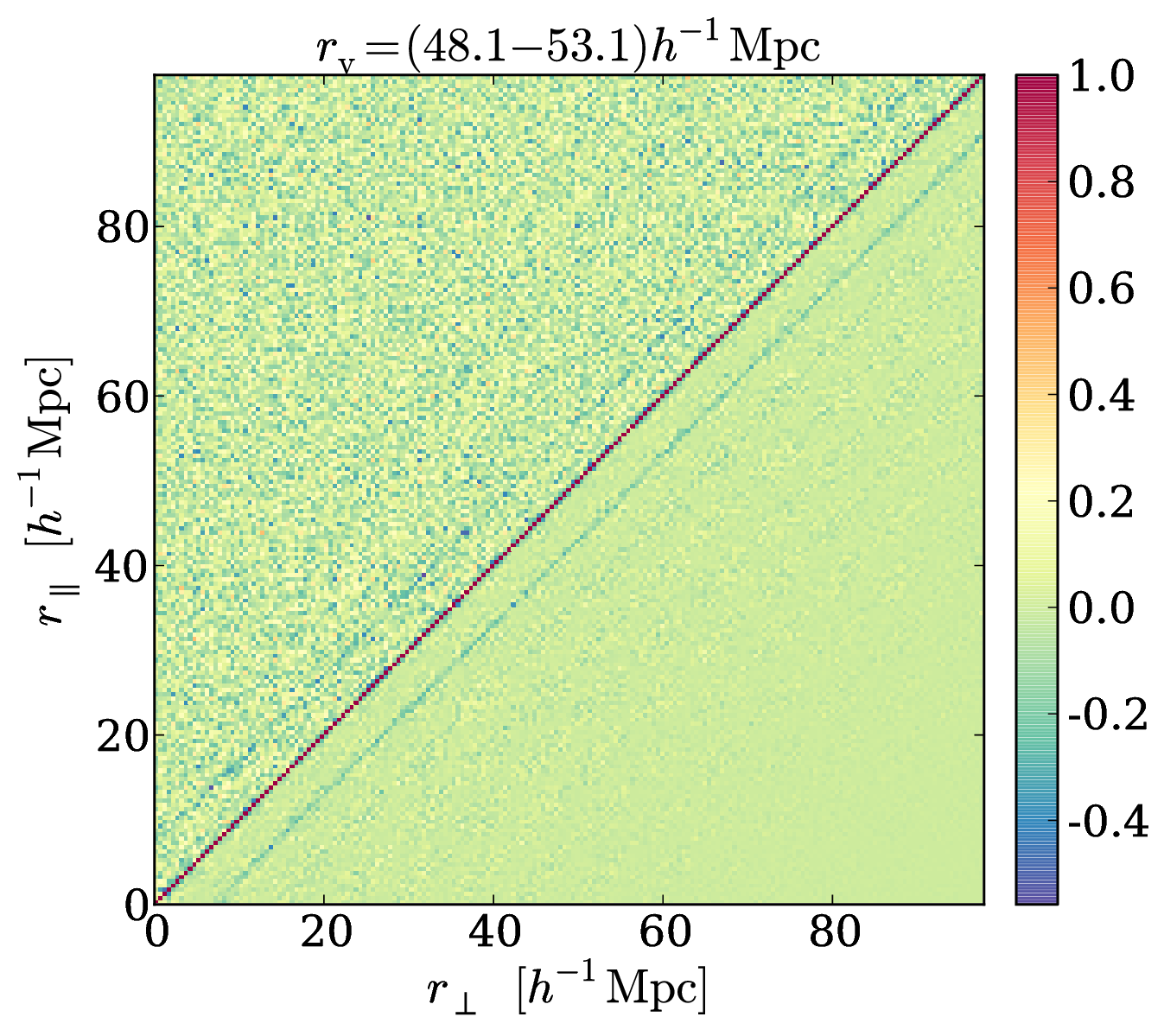}}
\caption{Covariance matrix (top) and its inverse (bottom) of a sparse mock-galaxy void stack with radius range $r_\void=(48.1-53.1)\hMpc$ at $z=0.5$. Each matrix is normalized by its diagonal, the lower triangular parts have been obtained using the tapering technique as described in the text.}
\label{Cvg}
\end{figure*}

In contrast, the value of $b\simeq1.8$ for our sparse mock-galaxy catalog at higher redshift leads to a smaller effective growth rate of $f/b=\Omega_\matter^{0.55}(z=0.5)/b\simeq0.58$, as can be seen in the posterior for the void radius range of $r_\void=(48.1-53.1)\hMpc$ in fig.~\ref{pdfg}. Although the high bias value leads to a steeper void-density profile, the lower galaxy number density and hence void abundance of this catalog results in wider parameter contours. Comparing the covariance matrices and their inverse in figs.~\ref{Cvt} and \ref{Cvg}, one can observe the stronger influence of the tapering procedure on the data from the sparse mock sample. Nevertheless, the posteriors for the logarithmic growth rate and the AP parameter nicely agree with their expected values within the $95.5\%$ contour region.

\section{Discussion\label{sec:discussion}}

\subsection{Joint constraints \label{sec:discussion:joint}}
So far we have only considered the parameter inference from a single void stack within a narrow effective radius range. What remains to be investigated is how the parameter constraints among stacks from different void radii compare and whether it is possible to combine them in order to achieve a higher precision.

\begin{figure*}[!p]
\centering
\vspace{-38pt}
\resizebox{.73\hsize}{!}{
\includegraphics[trim=0 38 0 0,clip]{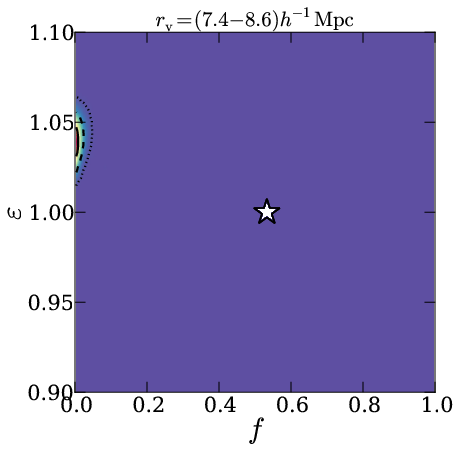}
\includegraphics[trim=27 38 0 0,clip]{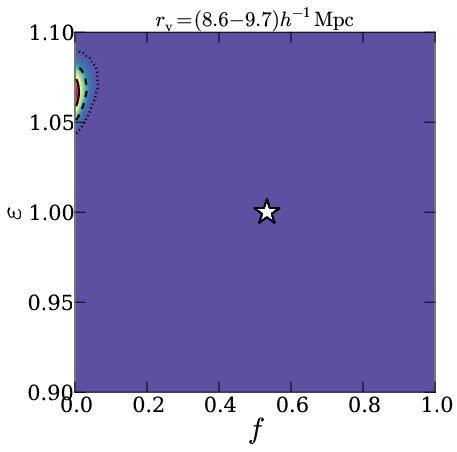}}
\resizebox{.73\hsize}{!}{
\includegraphics[trim=0 38 0 0,clip]{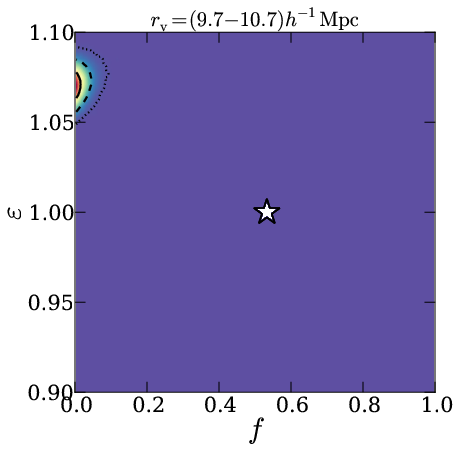}
\includegraphics[trim=27 38 0 0,clip]{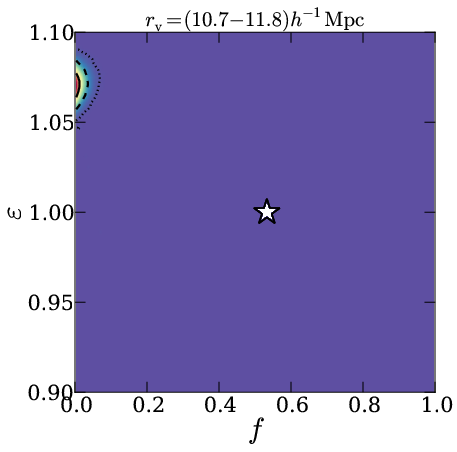}}
\resizebox{.73\hsize}{!}{
\includegraphics[trim=0 38 0 0,clip]{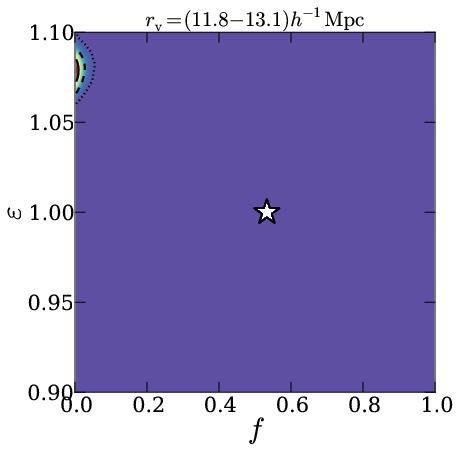}
\includegraphics[trim=27 38 0 0,clip]{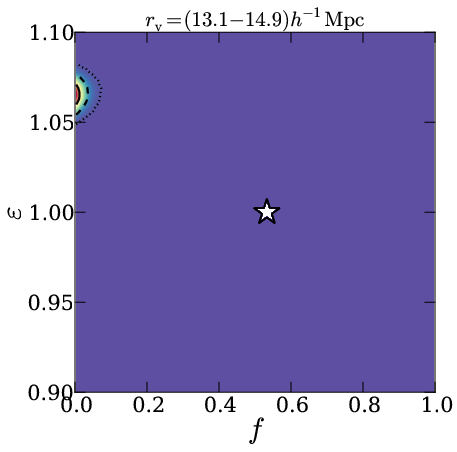}}
\resizebox{.73\hsize}{!}{
\includegraphics[trim=0 0 0 0,clip]{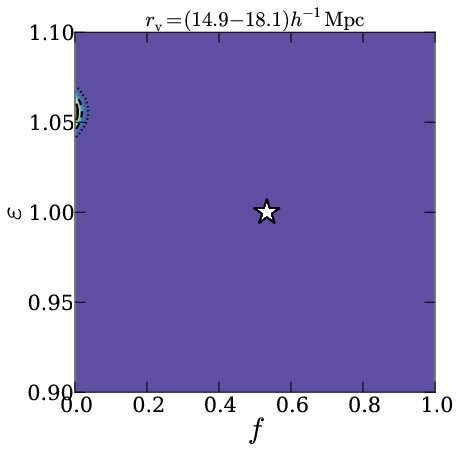}
\includegraphics[trim=27 0 0 0,clip]{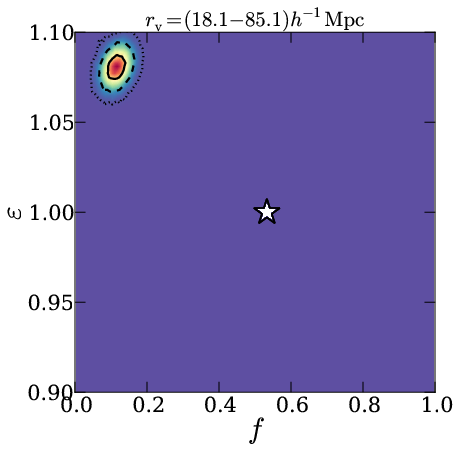}}
\caption{Constraints on growth rate $f$ and AP parameter $\varepsilon$ for each void stack in the dark matter. White stars show fiducial values of $f=\Omega_\matter^{0.55}(z=0)=0.53$ and $\varepsilon=1$. Confidence regions as shown only include statistical uncertainties and marginalize over the shape of the void density profile, the residual offset is due to modeling error (as discussed in the text).}
\label{pdfcm}
\end{figure*}

\begin{figure*}[!p]
\centering
\vspace{-38pt}
\resizebox{.73\hsize}{!}{
\includegraphics[trim=0 38 0 0,clip]{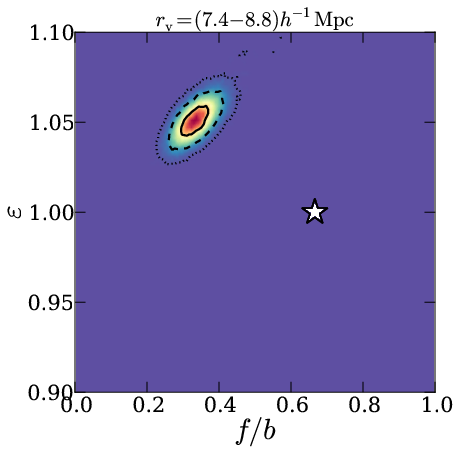}
\includegraphics[trim=27 38 0 0,clip]{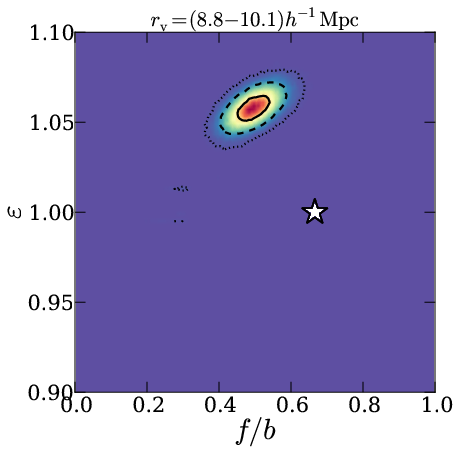}}
\resizebox{.73\hsize}{!}{
\includegraphics[trim=0 38 0 0,clip]{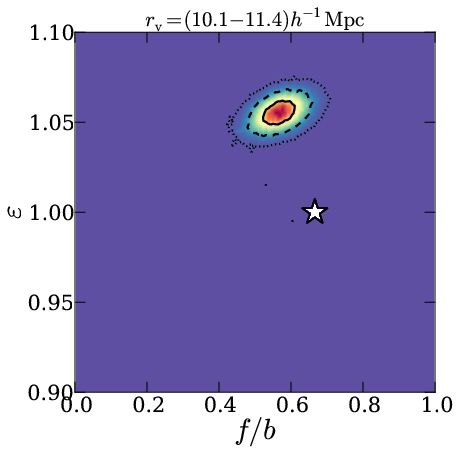}
\includegraphics[trim=27 38 0 0,clip]{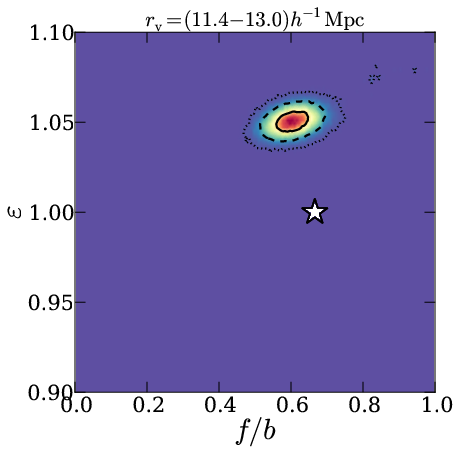}}
\resizebox{.73\hsize}{!}{
\includegraphics[trim=0 38 0 0,clip]{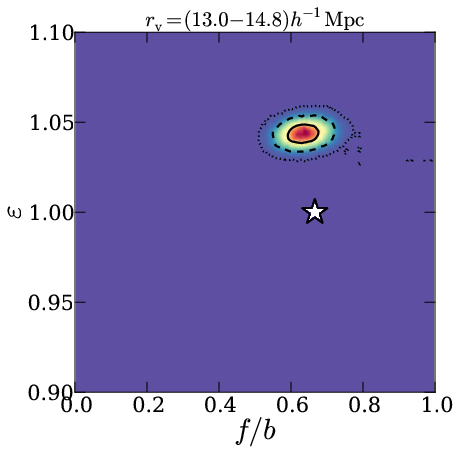}
\includegraphics[trim=27 38 0 0,clip]{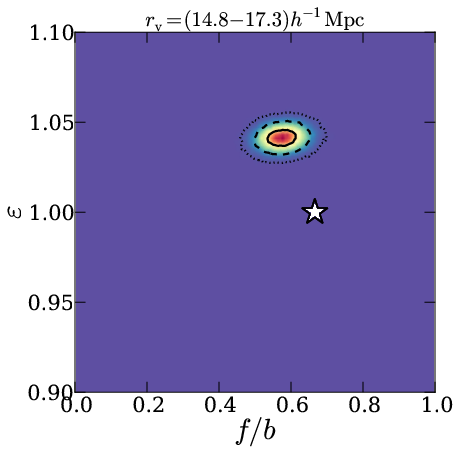}}
\resizebox{.73\hsize}{!}{
\includegraphics[trim=0 0 0 0,clip]{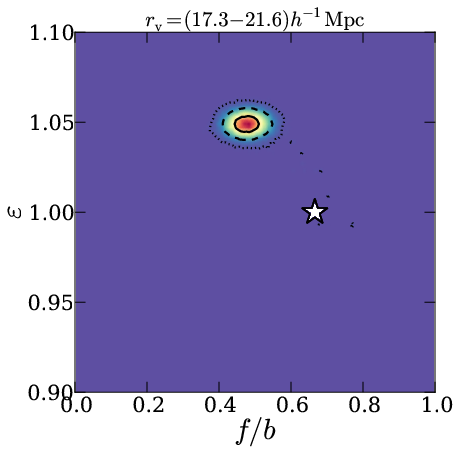}
\includegraphics[trim=27 0 0 0,clip]{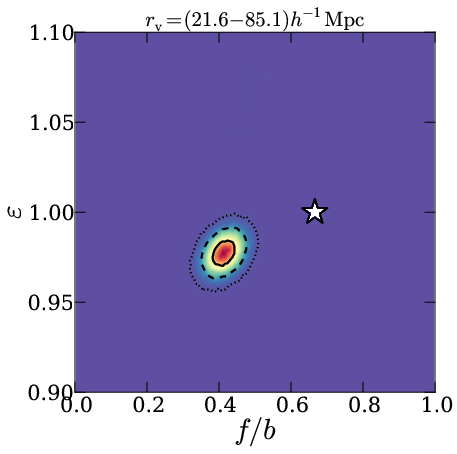}}
\caption{Constraints on growth rate $f/b$ and AP parameter $\varepsilon$ for each void stack in the dense mock-galaxy sample. White stars show fiducial values of $f/b=\Omega_\matter^{0.55}(z=0)/0.8=0.67$ and $\varepsilon=1$. Confidence regions as shown only include statistical uncertainties and marginalize over the shape of the void density profile, the residual offset is due to modeling error (as discussed in the text).}
\label{pdfct}
\end{figure*}

\begin{figure*}[!p]
\centering
\vspace{-38pt}
\resizebox{.73\hsize}{!}{
\includegraphics[trim=0 38 0 0,clip]{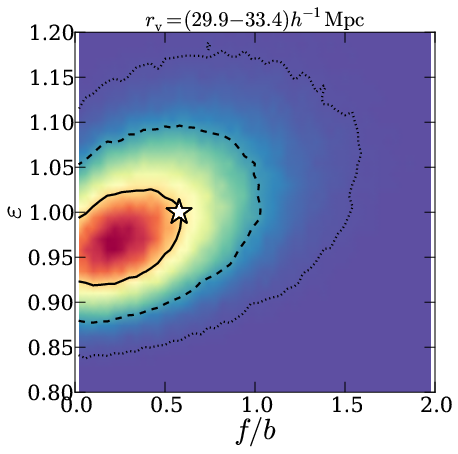}
\includegraphics[trim=27 38 0 0,clip]{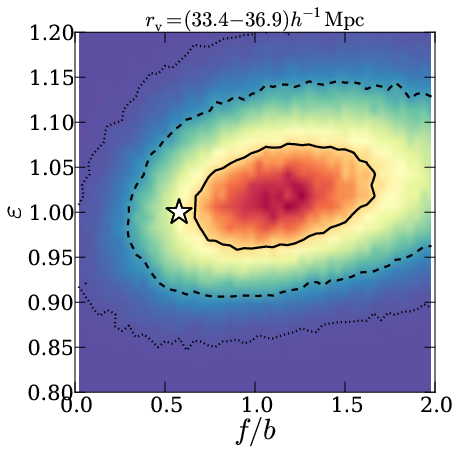}}
\resizebox{.73\hsize}{!}{
\includegraphics[trim=0 38 0 0,clip]{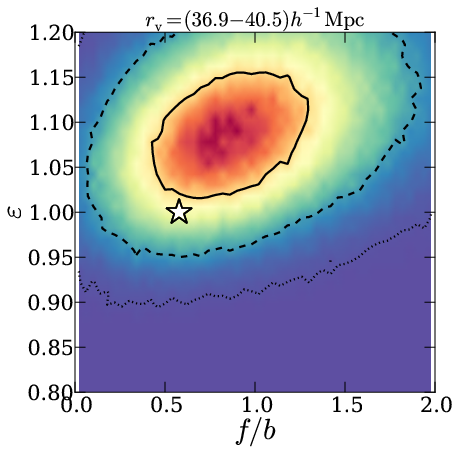}
\includegraphics[trim=27 38 0 0,clip]{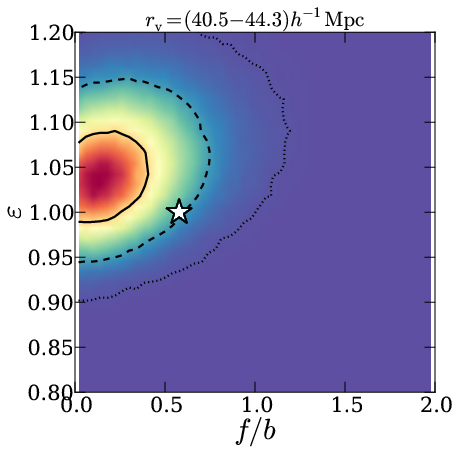}}
\resizebox{.73\hsize}{!}{
\includegraphics[trim=0 38 0 0,clip]{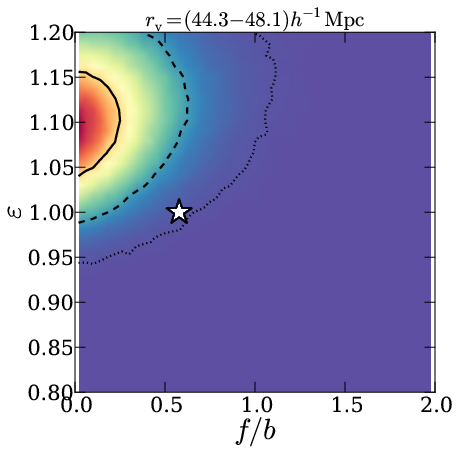}
\includegraphics[trim=27 38 0 0,clip]{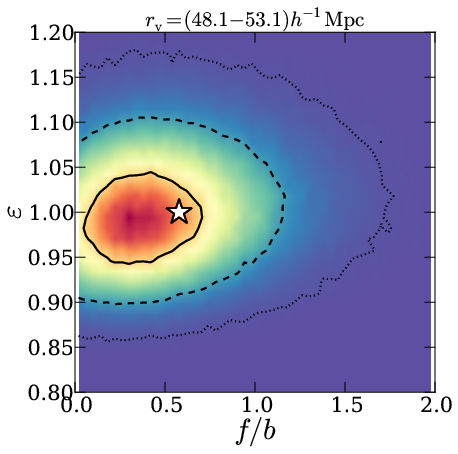}}
\resizebox{.73\hsize}{!}{
\includegraphics[trim=0 0 0 0,clip]{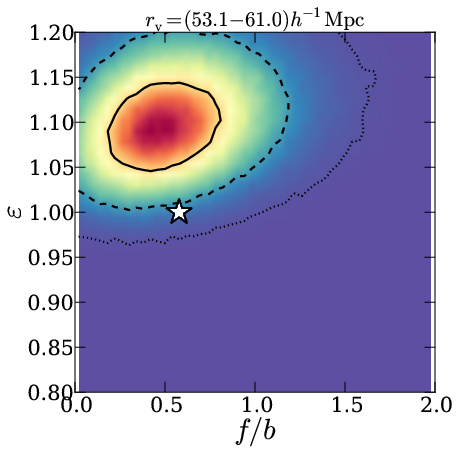}
\includegraphics[trim=27 0 0 0,clip]{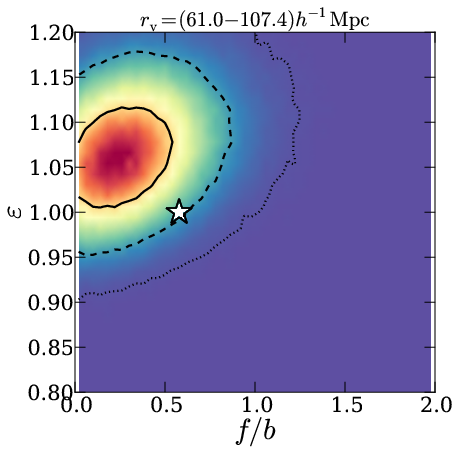}}
\caption{Constraints on growth rate $f/b$ and AP parameter $\varepsilon$ for each void stack in the sparse mock-galaxy sample. White stars show fiducial values of $f/b=\Omega_\matter^{0.55}(z=0.5)/1.8=0.58$ and $\varepsilon=1$. Confidence regions as shown only include statistical uncertainties and marginalize over the shape of the void density profile, the residual offset is due to modeling error (as discussed in the text).}
\label{pdfcg}
\end{figure*}

Figures~\ref{pdfcm}, \ref{pdfct} and \ref{pdfcg} show the posterior distributions of $f$ (respectively $f/b$) and $\varepsilon$ for each of our void stacks in dark matter, dense and sparse mock galaxies, respectively. Since these are the two main parameters of interest for the purpose of this paper, we do not investigate the remaining ones in detail anymore. As apparent from the different panels in these figures, the constraints on $f$ and $\varepsilon$ remain quite stable over the range of void sizes. While the Kaiser effect in dark matter voids can be only barely detected at the largest void radii, the growth rate inferred from voids in mock galaxies is quite consistent with its expected values from theory. For the sparse void sample, these values mostly fall within the $95.5\%$-confidence region of the posterior distribution. The smallest voids are more prone to nonlinear effects, as they exhibit the highest surrounding ridges~\cite{Hamaus2014b} that cause more extreme peculiar velocities (FoG)~\cite{Pisani2015b}. This can cause some discrepancy when using linear theory to model the void stacks in high-density tracers, as can be seen in fig.~\ref{pdfct}. We observe a slight bias of about $5\%$ overestimation of the AP parameter and a varying bias underestimating the growth rate, depending on void radius. The best agreement is achieved for intermediate-size voids, which also provide the tightest constraints. It is a reassuring result in view of the fact that the preferred range of effective void radii for this type of analysis lies within the sample of voids we can identify.

In order to combine the constraints from individual void stacks of a given tracer sample, we consider the joint likelihood as a product of the likelihoods of the individual stacks $i$,
\begin{equation}
 \mathcal{L}_\mathrm{tot}=\prod_i\mathcal{L}\left(\tilde{\xi}^{(i)}_{\void\tracer}|\boldsymbol{\theta}\right)\;.
\end{equation}
Note that we do not distinguish the model parameters $\boldsymbol{\theta} = (\rs,\dc,\alpha,\beta,\sigma_v,f,\varepsilon)$ between different stacks. As argued above, the parameters of interest $f$ and $\varepsilon$ do not depend on void size, so we can infer them from the joint dataset. This is of course not the case for the density-profile parameters $(\rs,\dc,\alpha,\beta)$, which all depend on void radius. However, we can simply marginalize over those, as we are not interested in their posterior distribution here.

\begin{figure*}[!t]
\centering
\resizebox{\hsize}{!}{
\includegraphics[trim=0 0 0 0,clip]{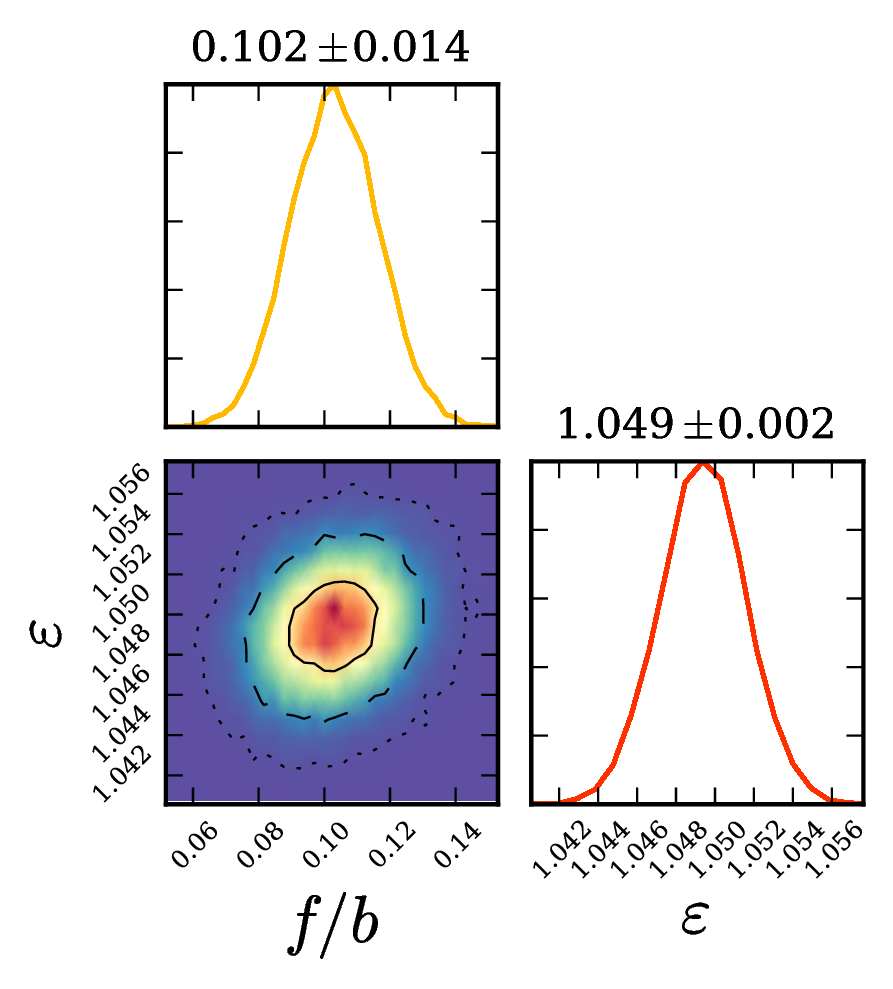}
\phantom{\includegraphics[trim=0 0 0 0,clip]{pdfcm_all.png}}}
\resizebox{\hsize}{!}{
\includegraphics[trim=0 0 0 0,clip]{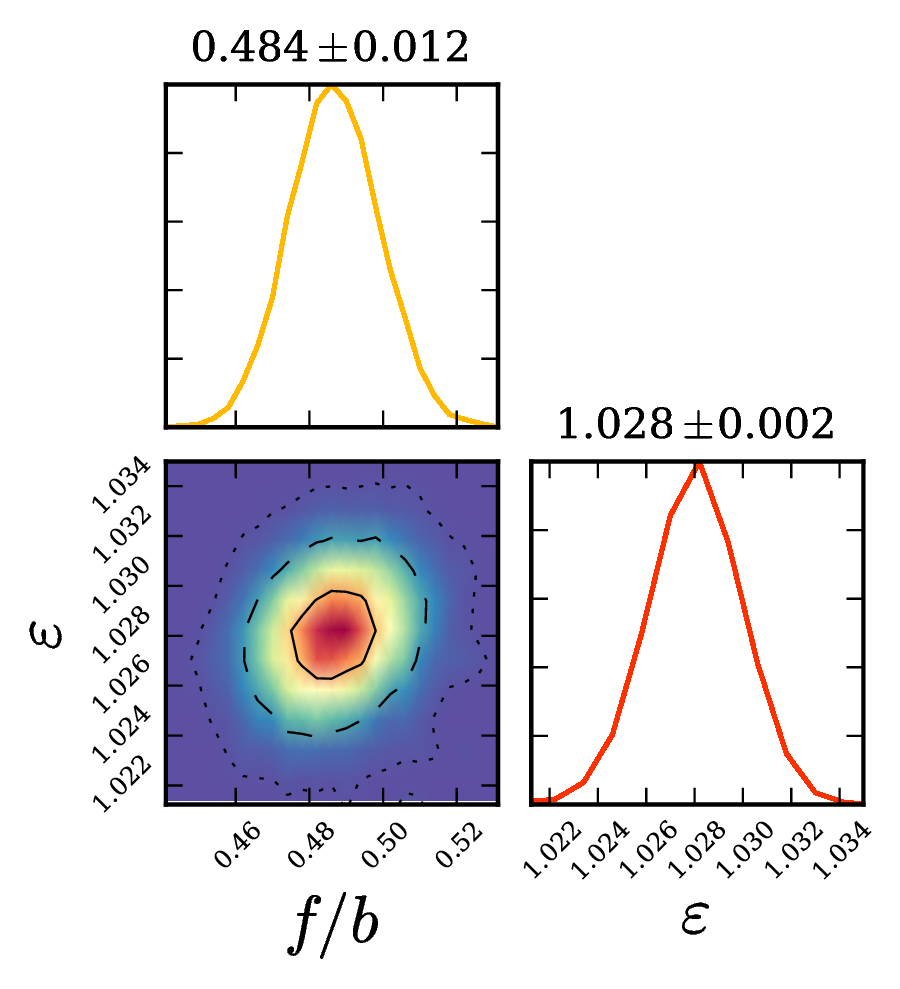}
\includegraphics[trim=0 0 0 0,clip]{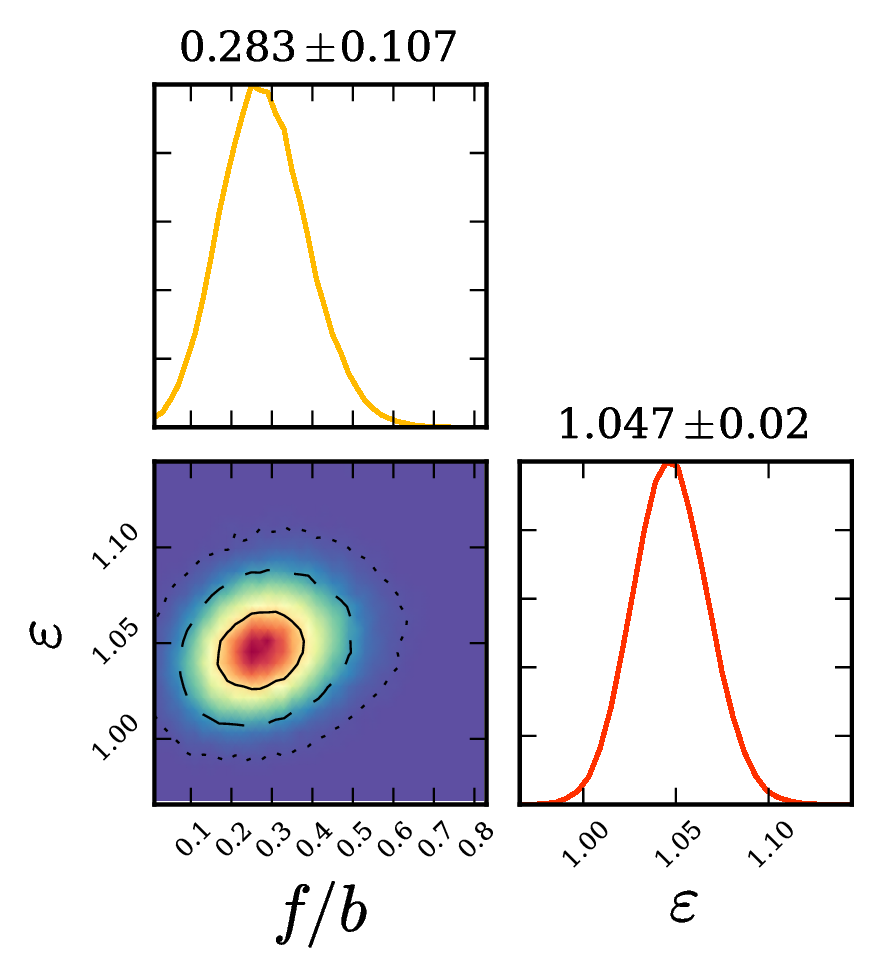}}
\caption{Joint constraints on growth rate $f$ (respectively $f/b$) and AP parameter $\varepsilon$ for \emph{all} void stacks from dark matter at $z=0$ (top), dense mock galaxies at $z=0$ (bottom left), and sparse mock galaxies at $z=0.5$ (bottom right). Parameter values at the maximum of the marginal posterior distribution and its standard deviation are given at the top of each column.}
\label{pdfc_all}
\end{figure*}

Figure~\ref{pdfc_all} shows the joint posterior distribution for $f$ and $\varepsilon$ from all void stacks obtained in our dark matter sample, as well as in our dense and sparse mock-galaxy catalogs. When compared to the constraints from the individual void stacks shown in figs.~\ref{pdfcm}, \ref{pdfct} and \ref{pdfcg}, we observe a considerable gain in precision on the parameters of interest. In each of the mock-galaxy samples the $1\sigma$ marginal error on $f/b$ and $\varepsilon$ is reduced by roughly a factor of three. It amounts to a relative $1\sigma$-uncertainty of $\sim2\%$ $(20\%)$ on $f/b$ and $\sim0.2\%$ $(2\%)$ on $\varepsilon$ (relative to their fiducial values) from the joint analysis of voids identified in the dense (sparse) mock-galaxy catalog. Thus, making use of the entirety of voids in a given catalog may significantly increase the information content that can be inferred via redshift-space distortions. However, using voids of all sizes in a single stack is suboptimal, as the redshift-space distortion signal from different physical scales is blurred out that way.

\subsection{Model details \label{sec:discussion:model}}
In order to examine the sensitivity to our model assumptions we have conducted a number of tests. First of all, we tried replacing the Gaussian form of the pairwise velocity probability distribution function in eq.~(\ref{pdf}) with an exponential distribution, as suggested by some authors (e.g.~\cite{Fisher1994,Sheth1996,Landy1998,Cabre2009}). Besides longer computing times in the MCMC sampling process, we found no noticeable differences in the final parameter constraints and concluded that prior knowledge on the exact shape of this function is of minor importance for the presented analysis.

Similarly, we tested how sensitive our results are affected by the velocity dispersion model adopted in eq.~(\ref{sv_model}), which depends on distance $r$ from the void center, but ignores the direction dependence from eq.~(\ref{sv}). By adding an additional parameter, we reran our MCMC analysis to constrain the two velocity dispersion amplitudes $\sigma_\parallel$ and $\sigma_\perp$ separately, and fixed the width $\omega$ of the Lorentzian in eq.~(\ref{sv_model}) for each of its measured values along and perpendicular to the line of sight. In addition, we repeated the entire inference process with simply a constant $\sigma_v$, neither dependent on scale, nor on direction. Apart from yielding slightly different best-fit values for $\sigma_v$, respectively $\sigma_\parallel$ and $\sigma_\perp$, these modification did not alter our parameter constraints in any notable fashion. This argues for the fact that the anisotropic shape of the void-tracer cross-correlation function is mainly determined through the mean of the pairwise velocity distribution function in eq.~(\ref{pdf}), and not its dispersion.

Also note that the parameter space used in the presented inference process can be reduced. In ref.~\cite{Hamaus2014b} it was found that the void density profile slopes $\alpha$ and $\beta$ are strongly correlated with the scale radius $\rs$ for voids of different size. A simple parametrization from a fit to simulations was proposed to reduce the effective number of parameters by two. We implemented this approach into our MCMC analysis, but found almost no difference in the final constraints on neither $f$, nor $\varepsilon$. This result is entirely expected, given the fact that we found no strong degeneracies between the void-density profile parameters and the cosmological parameters in sec.~\ref{sec:analysis:MCMC:matter}.

\section{Conclusion\label{sec:conclusion}}
In this paper we presented a first attempt to analyze and model the effects of redshift-space distortions around cosmic voids with a focus on cosmological parameter inference. To this end we generated catalogs of dark matter particles and realistic mock galaxies and used them to identify and stack voids in redshift space. In order to interpret the simulated mock data, we adopt the Gaussian streaming model for voids. We find that it successfully describes the void-tracer cross-correlation function in redshift space, given a realistic fitting function for the nonlinear real-space density profile of voids found in ref.~\cite{Hamaus2014b}, and its linear-theory relation to the corresponding velocity profile. All of these ingredients are then implemented into a Bayesian MCMC framework to infer the posterior parameter distribution of the model. The main conclusions drawn from this analysis are:
\begin{itemize}
 \item The values for the void density-profile parameters $(\rs,\dc,\alpha,\beta)$ inferred from the two-dimensional void-tracer cross-correlation function in redshift space $\tilde{\xi}_{\void\tracer}(\tilde{r}_\parallel,\tilde{r}_\perp)$ are consistent with the values found in a dedicated one-dimensional analysis of the real-space void-density profile~\cite{Hamaus2014b}.
 \item Void density-profile parameters $(\rs,\dc,\alpha,\beta)$ and cosmological parameters $(\sigma_v,f,\varepsilon)$ show no strong degeneracies, because the former set describes the isotropic, the latter set the anisotropic part of $\tilde{\xi}_{\void\tracer}(\tilde{r}_\parallel,\tilde{r}_\perp)$.
 \item The growth rate $f$ (respectively $f/b$) and the AP parameter $\varepsilon\propto 1/D_AH$ does not depend on void effective radius $r_\void$, in contrast to the void density-profile parameters. This allows to place joint constraints on these parameters from the entire available range of void sizes, yielding improvements in precision by factors of a few.
 \item The relative uncertainties on $f/b$ and $\varepsilon$ achievable in a survey volume of $V=1h^{-3}{\rm Gpc}^3$ with the presented mock galaxies range between $\sigma_{f/b}\left/(f/b)\right.\sim0.02-0.2$ and $\sigma_\varepsilon/\varepsilon=\sigma_{D_AH}/D_AH\sim0.002-0.02$. Increasing the survey volume, or adding independent information from the CMB and the BAO, for example, will further reduce these numbers.
 \item The $15\%$ systematic error in the conventional AP analysis~\cite{Lavaux2012,Sutter2012b} can be reduced to about $5\%$ when peculiar velocities are taken into account. This type of measurement potentially contains a tremendous amount of information, consistent with earlier estimates~\cite{Lavaux2012}. However, the level of precision achievable, in particular for dense catalogs, falls below the model accuracy provided by linear theory. This argues for an extension of the current framework to the nonlinear regime with a more sophisticated model.
\end{itemize}
It is important to note that the range of scales considered in this analysis is below the BAO scale, in a regime where theoretical models of the galaxy two-point correlation function fail (below $30\hMpc$ to $50\hMpc$, depending on redshift and galaxy bias, see e.g.~\cite{Samushia2014,Sanchez2014,Anderson2014,Song2014,Beutler2014,Alam2015}). We find that void-galaxy cross-correlations have the potential to unlock this scale range for precision cosmology. The principal advantages of void-galaxy cross-correlations are twofold: Firstly, even linear theory is able to describe the relation between density and velocity fluctuations around voids down to scales of $\sim10\hMpc$ or lower~\cite{Hamaus2014b}, and secondly the contributions from redshift-space distortions only enter this statistic linearly, not squared, as the motion of void centers can be neglected~\cite{Sutter2014c}. In addition to that, the void auto-correlation function can be used to independently perform an AP test and further improve the constraints on $\varepsilon$~\cite{Hamaus2014c}. These advantages are, however, somewhat mitigated by the relatively low number statistics of voids compared to galaxies. 

Nevertheless, the analysis of void redshift-space distortions has the potential to contribute independent cosmological information from galaxy surveys that has so far remained untapped. This method not only permits to improve our knowledge about the standard model of cosmology and gravity, but promises to test its alternatives. In particular, theories of modified gravity predict deviations from general relativity to be most pronounced in unscreened low-density environments~\cite{Clifton2012}, making voids a smoking gun for the detection of fifth forces~\cite{Martino2009,Li2011,Li2012,Clampitt2013,Cai2015,Zivick2015,Lam2015,Barreira2015}. Redshift-space distortions of tracers around voids not only probe these low densities, they also contain information about the tracer velocities in that regime, which are directly related to the gravitational potential. Existing public void catalogs~\cite{Pan2012,Sutter2012a,Nadathur2014,Sutter2013b,Leclercq2015} already allow testing these scenarios, and future data from ongoing and upcoming galaxy surveys will extend the scope of this method. Combined with measurements of weak lensing to probe the underlying mass distribution inside and around voids~\cite{Melchior2014,Clampitt2014}, such studies promise to shed more light on our concepts of dark energy, dark matter, and the laws of gravity. We will make our analysis tools publicly available by incorporating them into the VIDE pipeline on this website: \url{https://bitbucket.org/cosmicvoids/vide_public/}.

\begin{acknowledgments}
We thank Michael Warren for providing his $N$-body simulation and Alice Pisani, Roman Scoccimarro, Martin White and Donghui Jeong for useful discussions. NH is particularly grateful to Dante Paz, Adam Hawken and Yan-Chuan Cai for inspiring conversations at the 2015 LSS conference held at ESO in Garching. This work made in the ILP LABEX (under reference ANR-10-LABX-63) was supported by French state funds managed by the ANR within the Investissements d'Avenir program under reference ANR-11-IDEX-0004-02. This work was also partially supported by NSF AST 09-08693 ARRA. PMS is supported by the INFN IS PD51 ``Indark''. BDW is supported by a senior Excellence Chair by the Agence Nationale de Recherche (ANR-10-CEXC-004-01) and a Chaire Internationale at the Universit\'e Pierre et Marie Curie.
\end{acknowledgments}

\bibliography{ms.bib}
\bibliographystyle{JHEP.bst}

\end{document}